\documentclass{bmcart}
\usepackage[utf8]{inputenc}
\usepackage{graphicx}
\usepackage{comment}
\usepackage{multirow}

\usepackage{braket}

\usepackage{bm}
\usepackage{amsmath}
\usepackage{subfigure}
\usepackage{amsfonts,amsmath,amssymb,mathrsfs,bm,amsthm}
\usepackage{nccmath}
\usepackage{array}
\usepackage{tabu}
\usepackage{dcolumn}
\usepackage{float}
\usepackage{url}
\usepackage{graphics}
\usepackage{hyperref}
\usepackage{ragged2e}
\usepackage{cite}

\begin{document}

%%% Start of article front matter
\begin{frontmatter}

\begin{fmbox}
\dochead{Research}


\title{Measuring the stability of fundamental constants with a network of clocks}

%%%%%%%%%%%%%%%%%%%%%%%%%%%%%%%%%%%%%%%%%%%%%%
%%                                          %%
%% Enter the authors here                   %%
%%                                          %%
%% Specify information, if available,       %%
%% in the form:                             %%
%%   <key>={<id1>,<id2>}                    %%
%%   <key>=                                 %%
%% Comment or delete the keys which are     %%
%% not used. Repeat \author command as much %%
%% as required.                             %%
%%                                          %%
%%%%%%%%%%%%%%%%%%%%%%%%%%%%%%%%%%%%%%%%%%%%%%

\author[
  addressref={aff1},                   % id's of addresses, e.g. {aff1,aff2}
  corref={aff1},                       % id of corresponding address, if any
  email={g.barontini@bham.ac.uk}   % email address
]{\fnm{G.} \snm{Barontini}}
\author[
  addressref={aff2},                   % id's of addresses, e.g. {aff1,aff2}
]{\fnm{L.} \snm{Blackburn}}
\author[
  addressref={aff1},                   % id's of addresses, e.g. {aff1,aff2}
]{\fnm{V.} \snm{Boyer}}
\author[
  addressref={aff4, aff12},                   % id's of addresses, e.g. {aff1,aff2}
]{\fnm{F.} \snm{Butuc-Mayer}}
\author[
  addressref={aff2},                   % id's of addresses, e.g. {aff1,aff2}
]{\fnm{X.} \snm{Calmet}}
\author[
  addressref={aff7},                   % id's of addresses, e.g. {aff1,aff2}
]{\fnm{J. R.} \snm{Crespo L\'opez-Urrutia}}
\author[
  addressref={aff4},                   % id's of addresses, e.g. {aff1,aff2}
]{\fnm{E. A.} \snm{Curtis}}
\author[
  addressref={aff10},                   % id's of addresses, e.g. {aff1,aff2}
]{\fnm{B.} \snm{Darqui\'e}}
\author[
  addressref={aff2},                   % id's of addresses, e.g. {aff1,aff2}
]{\fnm{J.} \snm{Dunningham}}
\author[
  addressref={aff3},                   % id's of addresses, e.g. {aff1,aff2}
]{\fnm{N. J.} \snm{Fitch}}
\author[
  addressref={aff1},                   % id's of addresses, e.g. {aff1,aff2}
]{\fnm{E. M.} \snm{Forgan}}
\author[
  addressref={aff1},                   % id's of addresses, e.g. {aff1,aff2}
]{\fnm{K.} \snm{Georgiou}}
\author[
  addressref={aff4},                   % id's of addresses, e.g. {aff1,aff2}
]{\fnm{P.} \snm{Gill}}
\author[
  addressref={aff4},                   % id's of addresses, e.g. {aff1,aff2}
]{\fnm{R. M.} \snm{Godun}}
\author[
  addressref={aff1},                   % id's of addresses, e.g. {aff1,aff2}
]{\fnm{J.} \snm{Goldwin}}
\author[
  addressref={aff1},                   % id's of addresses, e.g. {aff1,aff2}
]{\fnm{V.} \snm{Guarrera}}
\author[
  addressref={aff4},                   % id's of addresses, e.g. {aff1,aff2}
]{\fnm{A. C.} \snm{Harwood}}
\author[
  addressref={aff4},                   % id's of addresses, e.g. {aff1,aff2}
]{\fnm{I. R.} \snm{Hill}}
\author[
  addressref={aff4},                   % id's of addresses, e.g. {aff1,aff2}
]{\fnm{R. J.} \snm{Hendricks}}
\author[
  addressref={aff1},                   % id's of addresses, e.g. {aff1,aff2}
]{\fnm{M.} \snm{Jeong}}
\author[
  addressref={aff4},                   % id's of addresses, e.g. {aff1,aff2}
]{\fnm{M. Y. H.} \snm{Johnson}}
\author[
  addressref={aff2},                   % id's of addresses, e.g. {aff1,aff2}
]{\fnm{M.} \snm{Keller}}
\author[
  addressref={aff8,aff9},                   % id's of addresses, e.g. {aff1,aff2}
]{\fnm{L. P.} \snm{ Kozhiparambil Sajith}}
\author[
  addressref={aff2},                   % id's of addresses, e.g. {aff1,aff2}
]{\fnm{F.} \snm{Kuipers}}
\author[
  addressref={aff4},                   % id's of addresses, e.g. {aff1,aff2}
]{\fnm{H. S.} \snm{Margolis}}
\author[
  addressref={aff1},                   % id's of addresses, e.g. {aff1,aff2}
]{\fnm{C.} \snm{Mayo}}
\author[
  addressref={aff1},                   % id's of addresses, e.g. {aff1,aff2}
]{\fnm{P.} \snm{Newman}}
\author[
  addressref={aff4},                   % id's of addresses, e.g. {aff1,aff2}
]{\fnm{A. O.} \snm{Parsons}}
\author[
  addressref={aff1},                   % id's of addresses, e.g. {aff1,aff2}
]{\fnm{L.} \snm{Prokhorov}}
\author[
  addressref={aff4},                   % id's of addresses, e.g. {aff1,aff2}
]{\fnm{B. I.} \snm{Robertson}}
\author[
  addressref={aff3},                   % id's of addresses, e.g. {aff1,aff2}
]{\fnm{J.} \snm{Rodewald}}
\author[
  addressref={aff5},                   % id's of addresses, e.g. {aff1,aff2}
]{\fnm{M. S.} \snm{Safronova}}
\author[
  addressref={aff3},                   % id's of addresses, e.g. {aff1,aff2}
]{\fnm{B. E.} \snm{Sauer}}
\author[
  addressref={aff4},                   % id's of addresses, e.g. {aff1,aff2}
]{\fnm{M.} \snm{Schioppo}}
\author[
  addressref={aff2},                   % id's of addresses, e.g. {aff1,aff2}
]{\fnm{N.} \snm{Sherrill}}
\author[
  addressref={aff6, aff11},                   % id's of addresses, e.g. {aff1,aff2}
]{\fnm{Y. V.} \snm{Stadnik}}
\author[
  addressref={aff4},                   % id's of addresses, e.g. {aff1,aff2}
]{\fnm{K.} \snm{Szymaniec}}
\author[
  addressref={aff3},                   % id's of addresses, e.g. {aff1,aff2}
]{\fnm{M. R.} \snm{Tarbutt}}
\author[
  addressref={aff3},                   % id's of addresses, e.g. {aff1,aff2}
]{\fnm{R. C.} \snm{Thompson}}
\author[
  addressref={aff4, aff3},                   % id's of addresses, e.g. {aff1,aff2}
]{\fnm{A.} \snm{Tofful}}
\author[
  addressref={aff4},                   % id's of addresses, e.g. {aff1,aff2}
]{\fnm{J.} \snm{Tunesi}}
\author[
  addressref={aff1},                   % id's of addresses, e.g. {aff1,aff2}
]{\fnm{A.} \snm{Vecchio}}
\author[
  addressref={aff3},                   % id's of addresses, e.g. {aff1,aff2}
]{\fnm{Y.} \snm{Wang}}
\author[
  addressref={aff1, aff8, aff9},                   % id's of addresses, e.g. {aff1,aff2}
]{\fnm{S.} \snm{Worm}}

%%%%%%%%%%%%%%%%%%%%%%%%%%%%%%%%%%%%%%%%%%%%%%
%%                                          %%
%% Enter the authors' addresses here        %%
%%                                          %%
%% Repeat \address commands as much as      %%
%% required.                                %%
%%                                          %%
%%%%%%%%%%%%%%%%%%%%%%%%%%%%%%%%%%%%%%%%%%%%%%

\address[id=aff1]{%                           % unique id
  \orgdiv{School of Physics and Astronomy},             % department, if any
  \orgname{University of Birmingham},          % university, etc
  \city{Edgbaston, Birmingham, B15 2TT},                              % city
  \cny{UK}                                    % country
}

\address[id=aff2]{%                           % unique id
  \orgdiv{Department of Physics and Astronomy},             % department, if any
  \orgname{University of Sussex},          % university, etc
  \city{Brighton BN1 9QH},                              % city
  \cny{UK}                                    % country
}

\address[id=aff4]{%  
          % department, if any
  \orgname{National Physical Laboratory},          % university, etc
  \city{ Hampton Road, Teddington TW11 0LW},                              % city
  \cny{UK}                                    % country
}

\address[id=aff12]{%                           % unique id         % department, if any
  \orgdiv{Department of Physics},% unique id            % department, if any
  \orgname{University of Oxford},          % university, etc
  \city{Clarendon Laboratory, Parks Road, Oxford OX1 3PU},                              % city
  \cny{UK}                                     % country
}

\address[id=aff7]{%                           % unique id         % department, if any
  \orgname{Max-Planck-Institut f\"ur Kernphysik},          % university, etc
  \city{Saupfercheckweg 1, 69117 Heidelberg},                              % city
  \cny{Germany}                                    % country
}

\address[id=aff10]{%                           % unique id
  \orgdiv{Laboratoire de Physique des Lasers},             % department, if any
  \orgname{CNRS, Universit\'e Sorbonne Paris Nord},          % university, etc
  \city{99 avenue Jean-Baptiste Cl\'ement, 93430 Villetaneuse},                              % city
  \cny{France}                                    % country
}

\address[id=aff3]{%                           % unique id
  \orgdiv{Blackett Laboratory},             % department, if any
  \orgname{Imperial College London},          % university, etc
  \city{Prince Consort Road, London SW7 2AZ},                              % city
  \cny{UK}                                    % country
}
\address[id=aff8]{%                           % unique id
  \orgdiv{Institut f\"ur Physik},             % department, if any
  \orgname{Humboldt-Universit\"at zu Berlin},          % university, etc
  \city{Newtonstra{\ss}e 15, 12489 Berlin},                              % city
  \cny{Germany}                                    % country
}

\address[id=aff9]{%                           % unique id            % department, if any
  \orgname{Deutsches Elektronen-Synchrotron (DESY) },          % university, etc
  \city{Platanenallee 6, D-15738 Zeuthen},                              % city
  \cny{Germany}                                    % country
}

\address[id=aff5]{%                           % unique id
  \orgdiv{Department of Physics and Astronomy},             % department, if any
  \orgname{University of Delaware},          % university, etc
  \city{Newark, Delaware 19716},                              % city
  \cny{USA}                                    % country
}

\address[id=aff6]{%                           % unique id
  \orgdiv{Kavli Institute for the Physics and Mathematics of the Universe (WPI), The University of Tokyo Institutes for Advanced Study},             % department, if any
  \orgname{University of Tokyo},          % university, etc
  \city{Kashiwa, Chiba 277-8583},                              % city
  \cny{Japan}                                    % country
}

\address[id=aff11]{%                           % unique id
  \orgdiv{School of Physics},             % department, if any
  \orgname{University of Sydney},          % university, etc
  \city{NSW 2006},                              % city
  \cny{Australia}                                    % country
}

\end{fmbox}% comment this for two column layout

%%%%%%%%%%%%%%%%%%%%%%%%%%%%%%%%%%%%%%%%%%%%%%%
%%                                           %%
%% The Abstract begins here                  %%
%%                                           %%
%% Please refer to the Instructions for      %%
%% authors on https://www.biomedcentral.com/ %%
%% and include the section headings          %%
%% accordingly for your article type.        %%
%%                                           %%
%%%%%%%%%%%%%%%%%%%%%%%%%%%%%%%%%%%%%%%%%%%%%%%

\begin{abstractbox}

\begin{abstract} % abstract
\justifying The detection of variations of fundamental constants of the Standard Model would provide us with compelling evidence of new physics, and could lift the veil on the nature of dark matter and dark energy. In this work, we discuss how a network of atomic and molecular clocks can be used to look for such variations with unprecedented sensitivity over a wide range of time scales. This is precisely the goal of the recently launched QSNET project: A network of clocks for measuring the stability of fundamental constants. QSNET will include state-of-the-art atomic clocks, but will also develop next-generation molecular and highly charged ion clocks with enhanced sensitivity to variations of fundamental constants. We describe the technological and scientific aims of QSNET and evaluate its expected performance. We show that in the range of parameters probed by QSNET, either we will discover new physics, or we will impose new constraints on violations of fundamental symmetries and a range of theories beyond the Standard Model, including dark matter and dark energy models.   
\end{abstract}

%%%%%%%%%%%%%%%%%%%%%%%%%%%%%%%%%%%%%%%%%%%%%%
%%                                          %%
%% The keywords begin here                  %%
%%                                          %%
%% Put each keyword in separate \kwd{}.     %%
%%                                          %%
%%%%%%%%%%%%%%%%%%%%%%%%%%%%%%%%%%%%%%%%%%%%%%


\end{abstractbox}
\end{frontmatter}

\section{Introduction}

The Standard Model of particle physics and the Standard Model of cosmology form the current foundation of fundamental physics. The cosmological model introduces two new forms of energy: dark matter and dark energy. Astrophysical observations suggest that these two forms of energy account for 95\% of the energy balance of our universe \cite{Zyla:2020PDG}, with only the remaining 5\% described by the Standard Model of particle physics. Dark matter is understood to be a non-relativistic form of matter not accounted for by the Standard Model of particle physics and that is believed to play a crucial role in the dynamics of galaxies. Dark energy, usually in the form of a cosmological constant, is instead postulated to explain the observed accelerated expansion of the universe. The precise natures of both dark matter and dark energy remain an open question. 

The two Standard Models rely on a large number of \emph{fundamental constants}. 
Crucially, in these models all fundamental constants are assumed to be immutable in space and time and to have had the same value throughout the history of the universe. Challenging this central assumption could be the key to solving the dark matter and dark energy enigmas, and also to understand how to unify particle physics and gravity into a unified theory of nature. Many models of physics beyond these Standard Models lead to a cosmological time evolution of physical constants \cite{khoury04, Avelino06, Dvali02, Banks02,Taylor:1988dilaton,gambini03,taveras08} and, in many of these models, all constants vary if one does \cite{Uzan15}.
In other models with ultra light new particles, e.g., models of ultra light dark matter, fundamental constants can have an effective space-time dependence due to the interactions between these ultra light particles and those of the Standard Model of particle physics \cite{Tilburg:2015DM,Stadnik:2015DM-LI,Stadnik:2015DM-VFCs,Arvanitaki16,Hees:2018DM}. 

On the opposite end of the energy spectrum with respect to the theories just mentioned, quantum technologies allow us to perform extremely precise measurements. It has recently been realised that such an exceptional precision is a formidable tool for performing tests of fundamental physics \cite{RevModPhys.90.025008}. Atomic clocks in particular can now reach uncertainties as low as 1 part in 10$^{18}$ and below~\cite{Brewer2019, Oelker2019}, and this has been exploited to provide some of the tightest constraints on present-day temporal variations of the fine structure constant, $\alpha$, and the electron-to-proton mass ratio, $\mu$, two of the fundamental constants of the Standard Model of particle physics \cite{Godun14,Huntemann14,Lange21,BACON:2021DM}. 
Furthermore, the networking of clocks for detection of dark matter or dark energy signatures is emerging as an effective way to increase detection sensitivity and confidence, and also to expand the range of dark sector phenomena that can be probed \cite{Derevianko:2014TDM,Derevianko:2018stochastic,Roberts17,Wcislo18, Roberts20,Stadnik:2020TDM}. 

In this article, we present the science case of the recently launched QSNET project: `A network of clocks for measuring the stability of fundamental constants' \cite{QSNET_SPIE}. The project represents a multidisciplinary effort, bringing together theoretical and experimental physicists from a wide range of research communities. It is our primary aim here to review and summarise the physics and phenomenology linked to variations of fundamental constants. At the same time, we describe how atomic and molecular clocks can be used to measure variations of fundamental constants, and discuss the typical range of parameters in which these clocks operate.  

\begin{figure}
	\centering
		\includegraphics[width=0.9\textwidth]{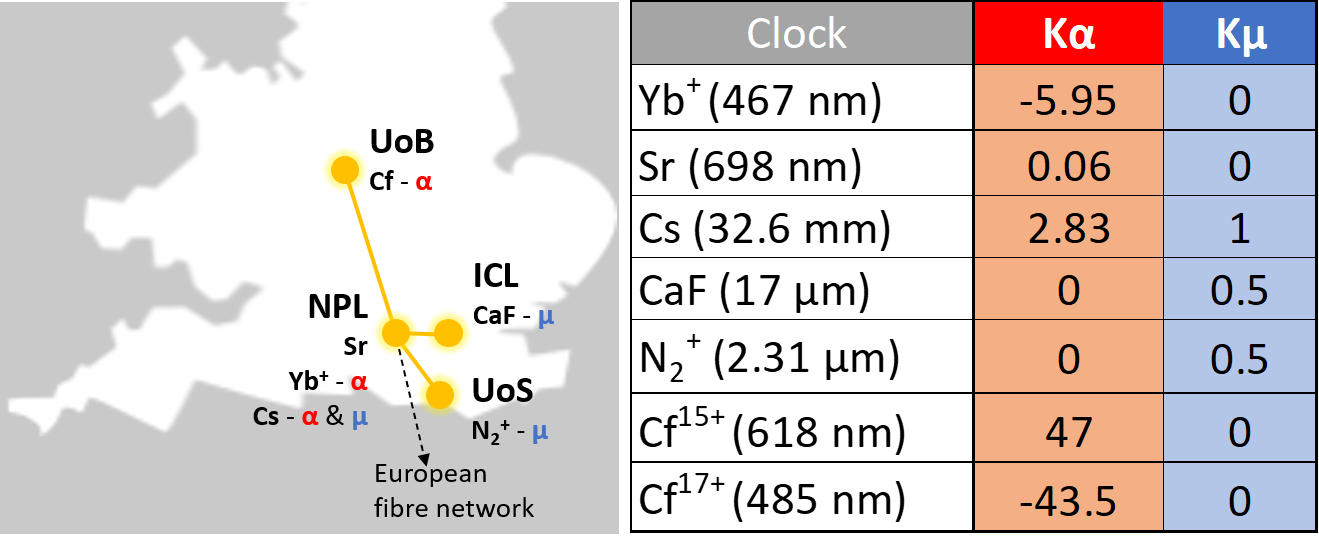}
	\caption{The QSNET network, including the University of Birmingham (UoB), the National Physical Laboratory (NPL), the Imperial College London (ICL) and the University of Sussex (UoS). For each node we indicate the atomic or molecular clock used and to which fundamental constant variations they are sensitive. In the table, we report the wavelength of the clock transitions and the sensitivity coefficients $K_\alpha$ \cite{Flambaum2009,Cfclock}
 and $K_\mu$, as defined by Eqs. (\ref{eq_Ka}) and (\ref{eq_Km}). 
	}
\label{figQSNET}
\end{figure}

QSNET will network a range of state-of-the-art clocks and next-generation clocks that feature enhanced sensitivity to variations of fundamental constants. The initial stage of QSNET is summarised in Fig.~\ref{figQSNET}; it includes existing Sr, Yb$^+$ and Cs atomic clocks at the National Physical Laboratory (NPL) in London and  several new clocks currently being developed: a N$_2^+$ molecular ion clock at the University of Sussex, a CaF molecular optical lattice clock at Imperial College London, and a Cf highly charged ion clock at the University of Birmingham. As the project progresses, this  national network can be expanded and linked with other clocks across the globe. We evaluate the potential sensitivities of these clocks to variations of $\alpha$ and $\mu$ over different timescales. We then estimate the impact of the QSNET performance on specific dark matter and dark energy models, soliton models, and violations of fundamental symmetries. 

This article is organised as follows: in Sec.~\ref{Sec:theory}, we review the theory about variations of fundamental constants. In Sec.~\ref{Sec:clock_network}, we describe how atomic and molecular clocks can be used to detect variations of fundamental constants, and how networking clocks brings both scientific and technological advantages. We then provide a description of the QSNET clocks and derive their expected performance. In Sec.~\ref{Sec:pheno}, we consider how frequency comparisons between pairs of these clocks can probe the parameter space of specific dark matter, dark energy and soliton models. Additionally, we discuss how QSNET can perform tests of violations of fundamental space-time symmetries, grand unification theories and quantum gravity. 
 Sec.~\ref{Sec:outlook} is finally devoted to the conclusions. 

%%%%%%%%%%%%%%%%
\section{Theoretical framework}
\label{Sec:theory}
The study of a possible time variation of the parameters of the fundamental laws of nature, such as the fine structure constant, has a long history.  In particular, there were early speculations about a cosmological time change of Newton's constant $G_N$ \cite{Dirac:1937ti,Dirac:1938mt,Milne,Jordan1,Jordan2}. Dirac's {\em large numbers hypothesis} is based on his comparison of the strength of gravity to those of other forces of nature. Indeed, for elementary particles, gravity is about 30 to 40 orders of magnitude weaker than electromagnetism, for example. Dirac speculated that this may not be a mere coincidence but instead could imply a cosmology with some unusual features. He speculated that the strength of gravity, which is fixed by Newton's constant, is inversely proportional to the age of the universe: $G_N \propto 1/t$; the mass of the universe would be proportional to the square of the universe's age $M \propto t^2$. Therefore what we consider physical constants would not actually be constant, but their values could depend on the age of the universe. Dirac's hypothesis has been ruled out by cosmological observations, but the interest of the physics community in a possible cosmological time evolution of physical constants has not ceased \cite{Uzan:2010pm}, further motivated by possible astrophysical observations of a cosmological time evolution of fundamental constants; see e.g.~\cite{Webb:2000mn,Chand:2004ct}.

The Standard Model of particle physics (Standard Model hereafter) and the  Standard Model of cosmology ($\Lambda$CDM model hereafter, where $\Lambda$ stands for a cosmological constant and CDM for cold dark matter) are extremely successful at describing all particle physics experiments performed thus far on Earth and all observations in cosmology and astrophysics. There are, however, serious limitations to these models. The most obvious one is that the $\Lambda$CDM model posits that 85$\%$ of all matter of the universe is described by cold dark matter, which is not accounted for by the Standard Model. Furthermore, the $\Lambda$CDM model assumes that General Relativity is the correct theory for gravity on all scales. It is however well known that General Relativity is not easy to reconcile with quantum field theory, which is the mathematical framework used to formulate the Standard Model.

 Another issue is that both General Relativity and quantum field theories are not very predictive in the following sense: there is no fundamental guiding principle to tell us which fields to introduce in our models or which gauge symmetries to impose. Furthermore, while we may have the correct differential equations to describe the evolution of the universe and the interactions between particles, we do not have a theory of initial conditions. 
 
 There is also no specific reason to have three generations of particles or why leptons and quarks need to be introduced. These particles are introduced because they are found in nature, but without experimental guidance theorists would not have been able to determine how many fields to introduce or how to gauge these particles. There is also no basic reason why Lorentz invariance and its local version which lead to General Relativity are symmetries of space-time. One could have imagined other fundamental symmetries of space and time. 

Furthermore, these models have a number of fundamental constants that cannot be calculated from first principles. Within the Standard Model alone, there are 28 such fundamental parameters (note that 22 of these parameters are needed to describe fermion masses). In these 28 constants, we include also Newton's constant which describes the strength of the gravitational interactions. The number of fundamental constants is even larger if we include the speed of light $c$, the Planck constant $h$, and other cosmological parameters such as the cosmological constant or the non-minimal coupling of the Higgs boson to the Ricci scalar. 
The absence of a theory of fundamental constants is a problem with our current understanding of nature. Indeed, modern theories of nature are based on renormalisable quantum field theories. Within this mathematical framework, it is impossible to calculate the value of the coupling constants from first principles. There is only one class of quantum-field-theoretical models with no free parameters \cite{tHooft:2011aa}, but these models are an exception and they are far from describing the real world. Additionally, all such {fundamental} constants are taken to be invariable in space and time. However, as discussed here below, many theoretical frameworks that attempt to describe physics beyond the Standard Model predict or allow variations of the fundamental constants.  

Because quantum gravity does not seem to be described by a renormalizable quantum field theory, extensions such as non-commutative geometry \cite{Connes:1994yd} or string theory \cite{Polchinski:1998rq,Polchinski:1998rr} have been considered. While it is difficult to make the link between these models and the real world, they are interesting because in principle they allow us to calculate some, if not all, fundamental constants. In particular, within string theory, coupling constants are fixed by the expectation values of moduli, which are scalar fields, and are thus calculable, at least in principle. 

Some extensions of the Standard Model, and in particular models that couple gravity to the Standard Model, such as inflationary models or quintessence, require or allow a time dependence of the parameters of the model. In models with extra-dimensions, such as Kaluza-Klein models or string theory, fundamental constants are often given by the expectation values of moduli fields which depend on the size of compactified extra-dimensions. As the size of these extra-dimensions could vary during the cosmological evolution of the universe, fundamental constants could also have a cosmological time evolution \cite{Marciano:1983wy}. Probing the cosmological time evolution of fundamental constants is therefore important, because this tests the validity of models with extra-dimensions and may help us develop a theory of fundamental constants. 

Recently, it has been realised that other physical phenomena could mimic a time-variation of fundamental constants. For example, very light scalar fields which could account for dark matter could lead to an effective time variation of fundamental constants. A cosmological time evolution or time dependence of the coupling constants of the Standard Model can be parametrised by a scalar field $\phi$ which couples to the electron $\psi_e$ (with mass $m_e$), light quarks ($u$, $d$ and $s$-quarks) $\psi_q$ (with mass $m_q$), the photon $A_\mu$ or gluons $G^a_\mu$ according to
\begin{equation}
\label{general_linear_interactions}
	\mathcal{L}=\kappa  \phi \left (\frac{d^{(1)}_e}{4} F_{\mu\nu}F^{\mu\nu} -d^{(1)}_{m_e} m_e \bar \psi_e \psi_e \right ) 
	+ \kappa  \phi \left (\frac{d^{(1)}_g}{4} G_{\mu\nu}G^{\mu\nu} -d^{(1)}_{m_q} m_q \bar \psi_q \psi_q \right ),
\end{equation}
 with $\kappa=\sqrt{4 \pi G_N}$,  $F_{\mu\nu}=\partial_\mu A_\nu-\partial_\nu A_\mu$ and $G_{\mu\nu}=\partial_\mu G_\nu-\partial_\nu G_\mu-i g_s [G_\mu,G_\nu]$, where $g_s$ is the QCD coupling constant. The $d^{(i)}_j$ are numerical constants which determine the strength of the interactions between the scalar field and Standard Model particles, which can be stronger than the gravitational one if $d^{(i)}_j>1$ or weaker if $d^{(i)}_j<1$. We could also add to Eq.~(\ref{general_linear_interactions}) couplings to the field strength of the neutrinos, heavier leptons and quarks, electroweak gauge bosons of the Standard Model and to the Higgs boson, but these particles usually do not play an important role for very low energy tabletop experiments such as clocks and we shall thus not include these couplings at this stage. Note that these operators are dimension-5 operators as they are suppressed by one power of the reduced Planck scale $M_P=1/\sqrt{8 \pi G_N}\,$.  

For some applications, it might be necessary to consider scalar fields that transform under some discrete, global or gauge symmetry, in which case the simplest coupling to matter is given by dimension-6 operators 
\begin{equation}
\label{general_quadratic_interactions}
	\mathcal{L}=\kappa^2  \phi^2 \left (\frac{d^{(2)}_e}{4} F_{\mu\nu}F^{\mu\nu} -d^{(2)}_{m_e} m_e \bar \psi_e \psi_e \right )  + \kappa^2  \phi^2 \left (\frac{d^{(2)}_g}{4} G_{\mu\nu}G^{\mu\nu} -d^{(2)}_{m_q} m_q \bar \psi_q \psi_q \right).
\end{equation}
In other words, the interactions of the scalar field with stable matter are suppressed by two powers of the reduced Planck scale.

These Lagrangians can account for a variety of physical phenomena:
\begin{itemize}
\item Scalar dark matter models in which case the magnitude of $\phi$ is related to the density of dark matter, see section \ref{Sec:DM}. 
\item Quintessence-like models, see section \ref{Sec:DE}.
\item A generic hidden sector scalar field \cite{Calmet:2019frv}.
\item Kaluza-Klein models/moduli models \cite{Marciano:1983wy}. In these models the size of compactified extra-dimensions can be described by a generic scalar field (the moduli field); if the size of extra-dimensions changes with cosmological time, the scalar field would have a cosmological time evolution. Generically speaking, in string theory coupling constants are moduli fields. Each coupling constant has its moduli field and its expectation value fixes the value of the coupling constant. 
\item Dilaton field models, see e.g.  \cite{Polchinski:1998rq,Polchinski:1998rr}. These are similar to moduli models, but dilaton fields are expected to couple universally to matter, like gravity. These models include Brans-Dicke fields and also scalar fields that are coupled non-minimally to the Ricci scalar $R$.
\item Soliton models, transient phenomena, cosmic strings, domain walls, and kink solutions would also be accounted for by a simple scalar field; see section \ref{Sec:solitons}.
\end{itemize}

One expects on very general grounds that quantum gravity will generate an interaction between any scalar field $\phi$ and regular matter with $d_j^{(i)}\sim {\cal O}(1)$, whether such a coupling exists or not when gravity decouples \cite{Calmet:2019jyz,Calmet:2019frv,Calmet:2020rpx,Calmet:2020pub,Calmet:2021iid}. However, very light scalar fields coupling linearly to regular matter (i.e. dimension-5 operators) are essentially ruled out by the E\"ot-Wash torsion pendulum experiment \cite{Kapner:2006si,Hoyle:2004cw,Adelberger:2006dh,Lee:2020zjt} for $d_j^{(1)}\sim {\cal O}(1)$. Indeed, E\"ot-Wash's data imply that if $d_j^{(1)}\sim 1$, the mass of the scalar field must satisfy $m_\phi>10^{-2}$~eV.  If a neutral scalar field with $m_\phi<10^{-2}$~eV and a linear coupling to regular matter was found by some experiment, we would learn that dimension-5 operators are not generated by quantum gravity.  On the other hand, non-linear couplings are far less constrained by current experiments (see section \ref{Sec:unifiedtheory}). This provides a very important test of quantum gravity \cite{Calmet:2019jyz,Calmet:2019frv,Calmet:2020rpx,Calmet:2020pub,Calmet:2021iid}.

Besides tests of quantum gravity, a time variation of fundamental parameters  would enable tests of grand unified theories \cite{Calmet:2001nu,Calmet:2002ja,Calmet:2002jz,Langacker:2001td,Campbell:1994bf,Olive:2002tz,Dent:2001ga,Dent:2003dk,Landau:2000cc,Wetterich:2003jt,Flambaum:2006ip,Calmet:2014qxa}, because in grand unified models, shifts in, e.g., the fine structure constant $\alpha$ and the coupling constant of quantum chromodynamics $\alpha_s$ are related. The same can apply also to shifts in lepton and quark masses. In grand unified theories, the relations between the different fundamental parameters are strongly model dependent. This is why very low-energy measurements can be used to probe very high energy theories (see section \ref{Sec:unifiedtheory}). Space-time variations of the fine structure constant have also been studied in the context of violations of fundamental symmetries \cite{Kostelecky:2002ca,Bertolami:2003qs,Ferrero:2009jb}.
More generally, similar experimental techniques for probing the stability of fundamental constants have led to stringent constraints on violations of Lorentz and CPT invariance, diffeomorphism invariance and the equivalence principle \cite{datatables} (see section \ref{Sec:symmetries}). Definitions of various space-time transformations can be found in, e.g.
Ref.~\cite{Kostelecky:2021xhb}.
 
We hope this short introduction to the theoretical framework underlying QSNET will have convinced the reader of the richness of the science that can be investigated by probing the time variation of fundamental constants. We can study cosmology by looking for a field responsible for the expansion of the universe, but also astrophysics by searching for extremely light dark matter. We can also probe fundamental high energy theories of particle physics and the symmetries of an ultra-violet complete theory of everything.  Indeed, in the next sections, we will show that one can probe fundamental physics at the Planck scale and test grand unification physics, quantum gravity, and the fundamental symmetries of nature with tabletop experiments. 

%%%%%%%%%%%%%%%%%%%%%
\section{Clock Network}
\label{Sec:clock_network}

In this section we describe how atomic and molecular clocks can be used to detect variations of fundamental constants, and we discuss the advantages of a networked approach. We then describe the clocks of the QSNET network and derive the performance that could be achieved in terms of sensitivities to variations of $\alpha$ and $\mu$. 

\subsection{Clocks and fundamental constants}
\label{Sec:Clock_and_FCs}

All atomic and molecular energy spectra depend on the fundamental constants of the Standard Model. For example, the scale of atomic transitions is set by the Rydberg constant, which can be written as 
\begin{ceqn}
\begin{equation}
    R_\infty=\frac{c}{4\pi\hbar}\alpha^2m_e \, ,
\end{equation}
\end{ceqn}
with both the electron mass $m_e$ and the fine structure constant $\alpha$ being fundamental constants of the Standard Model. 
It follows that, if these constants vary either in space or time, then so do atomic and molecular spectra. Clocks based on atoms or molecules rely on using the frequency of a spectral line to set the rate at which a clock `ticks’.  Changes in the spectra will therefore result in changes in the clock frequencies.  The narrower the spectral line, the more precisely the clock frequency can be determined and the better the resolution for detecting any changes.  The most favourable species to be used for clocks are therefore those which possess transitions that are narrow in frequency (forbidden at least to first order) and not easily perturbed by changes in background electric or magnetic fields.  Optical atomic clocks have already been demonstrated to achieve fractional frequency instabilities and inaccuracies at the level of $10^{-18}$ and below~\cite{Brewer2019, Oelker2019}, making them among the most precise measurement instruments ever built.  High-precision spectroscopy with atomic clocks has therefore provided some of the tightest constraints on variations of $\alpha$ and the electron-to-proton mass ratio $\mu=m_e/m_p$ \cite{Godun14,Huntemann14,Lange21,BACON:2021DM}, with 
$m_p$ the proton mass. 

Depending on the nature of the transition employed, different clocks are more or less sensitive to variations of specific fundamental constants. To illustrate this, let us express the frequency of clocks employing optical transitions as 
\begin{ceqn}
\begin{equation}
    \nu_\textrm{opt}=A \cdot F_\textrm{opt}(\alpha) \cdot cR_\infty \, ,
    \label{eq_opt}
\end{equation}
\end{ceqn}
with $A$ a constant depending on the specific atomic species and transition and $F_\textrm{opt}(\alpha)$ describing the relativistic correction to the specific transition. In contrast, microwave (MW) clocks utilise transitions between hyperfine energy levels, whose frequency can be written as
\begin{ceqn}
\begin{equation}
    \nu_\textrm{MW}=B \cdot \alpha^2 F_\textrm{MW}(\alpha) \cdot \mu \cdot cR_\infty \, ,
    \label{eq_MW}
\end{equation}
\end{ceqn}
where $B$ is a constant that depends on the specific atomic species and transition and $F_\textrm{MW}(\alpha)$ is the relativistic correction to the specific MW transition. 
Finally, the frequency of molecular clocks based on vibrational transitions can be expressed as
\begin{ceqn}
\begin{equation}
    \nu_\textrm{vib}=C \cdot \mu^{1/2} \cdot cR_\infty \, ,
    \label{eq_mol}
\end{equation}
\end{ceqn}
with $C$ a constant depending on the specific molecule and transition used.  

The sensitivity of a certain atomic or molecular transition $\nu_i$ to variations of a fundamental constant $X=\{\alpha,\mu\}$ is 
characterised by a \emph{sensitivity coefficient} $K_X$, which we define as
\begin{ceqn}
\begin{equation}
    K_X=\frac{\partial \ln\left(\frac{\nu_i}{cR_\infty} \right)}{\partial\ln X} \, . 
    \label{eq_K}
\end{equation}
\end{ceqn}
The larger the value of $K_X$, the more sensitive a specific transition is to variations of $X$. From Eq.~({\ref{eq_K}}), and using Eqs.~(\ref{eq_opt}), (\ref{eq_MW}) and (\ref{eq_mol}), it follows that 
\begin{ceqn}
\begin{equation}
K_\alpha=\begin{cases}
\partial \ln F_\textrm{opt}/\partial\ln\alpha & \text{for optical transitions}  \\  
2+\partial \ln F_\textrm{MW}/\partial\ln\alpha & \text{for MW transitions}  \\   
0 & \text{for vibrational transitions}  \\ 
\end{cases}    
\label{eq_Ka}
\end{equation}
\end{ceqn}
The sensitivities of optical and MW transitions to variations of $\alpha$ are calculated by numerically varying the value of $\alpha$ in the computation of atomic spectra \cite{Flambaum:2009VFCs,Flambaum:2014VFCs}. Note that the magnitude of $K_\alpha$ generally increases in heavier atomic systems due to increased relativistic effects \cite{Flambaum:1999VFCs-A,Flambaum:1999VFCs-B,Holliman22}. Similarly, we obtain
\begin{ceqn}
\begin{equation}
K_\mu=\begin{cases}
0 & \text{for optical transitions}  \\  
1 & \text{for MW transitions}  \\   
1/2 & \text{for vibrational transitions}  \\ 
\end{cases}    
\label{eq_Km}
\end{equation}
\end{ceqn}

To measure variations of fundamental constants using atomic or molecular clocks, the frequency of one clock relative to another, i.e. their frequency ratio, needs to be measured over time.  As dimensionless quantities, frequency ratios also avoid any ambiguities of whether the system of units employed in the measurements is varying over time or not. For a given frequency ratio $R=\nu_1/\nu_2$, the sensitivity to variations of a certain fundamental constant $X$ is proportional to the difference between the sensitivity coefficients, i.e.: 
\begin{ceqn}
\begin{equation}
    \frac{dR}{R}=\left[K_{X,1}-K_{X,2}\right]\frac{dX}{X} \, . 
    \label{eq_ratio}
\end{equation}
\end{ceqn}
Ratios of transitions with similar sensitivity coefficients are therefore almost insensitive to variations of $X$, whereas comparing two transitions with large magnitudes of $K_X$ and of opposite signs greatly boosts the sensitivity. In the case of combinations of transitions that are sensitive to variations of more than one fundamental constant, the different contributions are weighted with the corresponding values of $K_{X,1}-K_{X,2}$. 

From Eqs.~(\ref{eq_Ka}) and (\ref{eq_Km}), it follows that the ratio between the transition frequencies of two optical or two microwave clocks is only appreciably sensitive to variations of $\alpha$ via the transition-specific relativistic correction factors $F(\alpha)$. However, a ratio between an optical clock and a microwave clock is sensitive both to variations of $\alpha$ and $\mu$ as is a frequency ratio between a molecular vibrational transition and an optical atomic transition.     

%%%%%%
\subsection{Networks of clocks for measuring the stability of fundamental constants}
\label{Sec:clock_network_stability}

To perform the clock-to-clock comparisons needed to measure variations of fundamental constants, clocks can be networked \emph{online} using either satellite \cite{Roberts17} or fibre links \cite{Roberts20}, or \emph{offline} via time-stamping of  measurements \cite{Wcislo18}. 

In QSNET, the clocks in the network will be linked with optical fibres.  Over distances of several hundred kilometres, the frequency uncertainties introduced by phase noise in the optical fibre links can be cancelled to a level well below the measurement uncertainties from the clocks themselves.  The signal that is sent through the fibres will be an optical carrier frequency from a transfer laser, around $1.5~\mu{\rm m}$ in wavelength.  A frequency comb at each end of the fibre will be used to measure the local clock frequency relative to the transfer laser, and the exact frequency of the transfer laser cancels out in the ratio of the clock frequencies.  Large networks of telecom fibres already exist; however, it is important to cancel the phase noise that is picked up by the light along the fibre route.  This is standard practice in optical frequency transfer for precision metrology.  The phase noise is monitored by setting up an interferometer with a fraction of light that has been back-reflected from the far end of the fibre being compared against the incident light.  A correction signal can then be applied to an acousto-optic modulator to cancel the noise. 

{Existing telecom fibres can be used for the networking, but some modifications are required to ensure the infrastructure is suitable for transferring ultrastable optical frequencies between the clocks.  Most notably, any telecom equipment that relies on optical-to-electrical conversion along the fibre route must be replaced or circumvented by all-optical amplifiers to avoid scrambling the phase of the transfer laser.  The all-optical amplifiers must also be bi-directional to allow back-reflected light to be used in the interferometer for phase-noise cancellation.  Changing the amplifiers affects all channels on a fibre and, for this reason, amplifiers cannot be switched out on fibres carrying live telecom signals to other users.  It is therefore necessary to use ‘dark fibres’, i.e. existing telecom fibres that have no other users at present.  Dark fibre routes are available between all the partners in QSNET.  A recent demonstration, transferring ultrastable frequencies along a 2,220-km dark fibre link, has shown that noise from the link can be cancelled to the level of $10^{-16}$ in 10 s \cite{Schioppo22} and much lower with longer averaging times.  This is already sufficient to avoid degrading comparisons between the pairs of clocks proposed in QSNET and strategies exist to reduce the link noise even further, as clocks improve in the future.}   

This networked approach is an important aspect of the QSNET project. While each clock pair is an excellent detector for searches of variations in fundamental constants, by combining them into a network, better sensitivities, very high detection confidence, and new capabilities can be achieved. The networked approach brings both technological and scientific advantages:
\vspace{0.3 cm}

\noindent\textbf{Technological advantages}
\begin{itemize}
\item A network makes it possible to compare two clocks in different locations, thus exploiting the resources and expertise spread across different institutes. No single institution has the range of expertise required to run a sufficiently large and diverse set of clocks with different sensitivities to variations of fundamental constants. A network makes it easy to compare a diverse range of systems, such as highly charged ion clocks, molecular clocks and nuclear clocks, as well as more standard atomic clock systems.
\item Validation of the results can be achieved with simultaneous measurements from multiple pairs of clocks. Ideally, the clocks should be in different environments and with different sensitivities to systematic frequency offsets.
\end{itemize}

\noindent\textbf{Physics advantages} 
\begin{itemize}

\item Networks enable probing of space-time correlations \cite{Derevianko:2018stochastic}. These correlations increase the detection confidence and provide added information, such as the speed and directionality of the oscillating dark matter fields discussed in Subsection \ref{Sec:DM}. 
\item Networking is the only possibility of detecting transient events linked to macroscopic dark objects, such as topological defects, solitons, Q-balls and dark stars. This is discussed in Subsection \ref{Sec:solitons}. Correlation functions across the network are important in discriminating noise from transient effects linked to dark-sector fields \cite{Pustelny2013}.

\item Networking is ideal for implementing dark matter detection using both variations of $\alpha$ and $\mu$, which is instrumental in discriminating between models that predict a variation of the unified coupling constant or a time-variation of the unification scale, or both; see the discussion in Subsection \ref{Sec:unifiedtheory} and e.g. \cite{Calmet2002,Calmet2002_2,Calmet2006} and references therein. 
\item Similarly, using a network of different types of sensors can veto noise and make it possible to determine the origin of a signal. For example, in a clock network combined with the Global Network of Optical Magnetometers for Exotic Physics (GNOME), correlation of read-outs from optical and non-optical magnetometers can be used to rule out magnetic artifacts, such as solar wind  \cite{Pustelny2013}.
\item Having $N$ pairs of clocks within the coherence length of oscillating dark-sector fields can improve the limit from a single pair of clocks by a factor $\sqrt{N}$ \cite{Derevianko:2018stochastic}, increasing the signal to noise ratio.
\item {It has been suggested that a global network of entangled clocks could give an international time scale with unprecedented stability and accuracy \cite{Komar2014}. There are also suggestions of how to use entanglement in local networks to measure non-commuting observables, e.g. different components of a field  \cite{Baumgratz2016}, or to deal with nuisance parameters \cite{Kok2017}. Those ideas for using quantum correlations across a network might, in the future, be used to improve the precision when estimating non-local parameters, i.e. ones that are a function of the values at the different spatially separated sensors \cite{Proctor2018}. The question of how to measure such correlations through an optical fibre network is a good topic for future research.}
\end{itemize}

%%%%%%
\subsection{The QSNET network}
\label{Sec:QSNET_network}

In this subsection we discuss the clocks being developed within QSNET. As summarised in Fig.~\ref{figQSNET}, these include Yb$^+$ and Cf highly charged ion clocks with enhanced sensitivity to variations of $\alpha$; Cs, N$_2^+$ and CaF clocks, that are most sensitive to variations of $\mu$; and a Sr clock, that has $K_{\mu}=0$ and $K_{\alpha}$ close to zero. For the established clocks, we discuss the state-of-the-art  and provide the current measurement limits set by systematic uncertainties. For the next-generation molecular and highly charged ion clocks, we instead predict their systematic uncertainties based on realistic assumptions. More details are provided in Appendix~\ref{Appendix:QSNET_clocks}.

\subsubsection{Established standards}

Atomic clocks are already very well-developed on certain atomic transitions, particularly those which are used as frequency standards in national measurement institutes. In QSNET, three established clocks will be used in the network: a microwave $^{133}$Cs clock, and optical clocks based on $^{87}$Sr and $^{171}$Yb$^+$.  More details are given in Appendix~\ref{Appendix:standards} about these systems and the current state-of-the-art performances that have been achieved. 

\begin{figure}
		\includegraphics[width=0.98\textwidth]{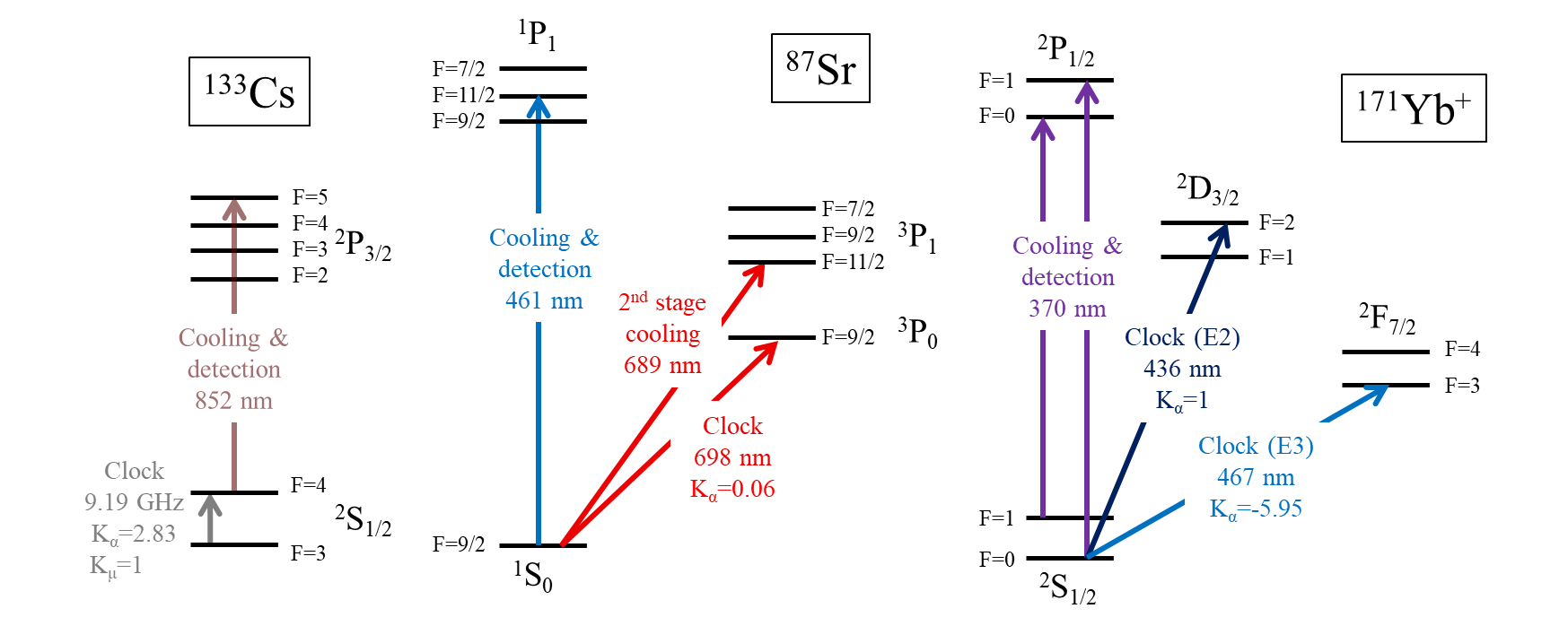}
	\caption{Relevant energy levels for the QSNET atomic clocks. The wavelengths and sensitivity coefficients of the clock transitions are also specified.}
\label{Fig:NPLtransitions}
\end{figure}

Of all the species used for atomic clocks, caesium is the most common.  This is because the definition of the SI second is based on fixing the value of the transition frequency between the $^{133}$Cs hyperfine ground states to be exactly $9\,192\,631\,770$~Hz. See Figure~\ref{Fig:NPLtransitions} for the relevant energy levels involved in this microwave clock transition.  The highest accuracy Cs clocks rely on laser cooling the atoms in a magneto-optical trap and then launching them vertically in a `fountain' configuration.  These fountain clocks allow the atoms to pass twice through a microwave cavity: once on the way up and then a second time on the way down. The two interactions with the microwave cavity constitute separated Ramsey pulses that drive the clock transition.  The Ramsey dark time is $\sim 1$~s, leading to Ramsey fringes with a linewidth at the Hz level. There are several state-of-the art Cs fountain clocks around the world~\cite{Weyers2018,Heavner2014,Guena2012,Szymaniec2016,Levi2014}, all operating with fractional frequency uncertainties at the level of $1$--$2 \times 10^{-16}$ from systematic shifts; see Appendix~\ref{Appendix:Cs_clock}.  A caesium clock is useful in the QSNET network because its transition frequency is sensitive to changes in both the fine structure constant ($K_{\alpha} = 2.83$) and the electron-to-proton mass ratio ($K_{\mu} = 1$). 

The fermionic isotope of strontium, $^{87}$Sr, is one of the most commonly used species for optical atomic clocks. The `forbidden' $^1{\rm S}_0 \rightarrow\,^3{\rm P}_0$ transition at 698~nm provides a suitable clock transition with a linewidth of 2$\pi \times 1$~mHz, as shown in Figure~\ref{Fig:NPLtransitions}.  The cooling transitions are readily accessible with diode lasers and clouds of about $10^4$ atoms can be trapped in an optical lattice potential, formed from a standing wave of laser light.  The standing wave is tuned close to a `magic wavelength' at 813~nm, where the differential polarisability between ground and excited states is zero, leading to cancellation of the AC Stark shift.  The $^{87}$Sr clock has very small sensitivity factors, $K_{\alpha} = +0.06$ and $K_{\mu} = 0$.  This is useful when comparing against clocks with larger sensitivities because it allows the corresponding frequency ratio to have a large differential sensitivity to variations in the constants.  The current state-of-the-art for the $^{87}$Sr optical lattice clock has an estimated fractional frequency uncertainty from systematic shifts of $2.0 \times 10^{-18}$~\cite{Bothwell2019}. The uncertainty budget is discussed in Appendix~\ref{Appendix:Sr_clock}.  

Finally, QSNET will exploit singly-charged ytterbium, $^{171}$Yb$^+$, which has two optical clock transitions that are used as frequency references, as shown in Figure~\ref{Fig:NPLtransitions}.  One is based on the electric quadrupole (E2) transition $^2{\rm S}_{1/2} \rightarrow \ ^2{\rm D}_{3/2}$ at 436~nm with a linewidth = $2\pi\times 3$~Hz, while the other is based on the electric octupole (E3) transition $^2{\rm S}_{1/2} \rightarrow \ ^2{\rm F}_{7/2}$ at 467~nm with a linewidth $\sim$~nHz.  Both of these transitions have a good sensitivity to changes in the fine structure constant, but the octupole transition is significantly more sensitive with $K_{\alpha}^{E3} = -5.95$, compared to the electric quadrupole transition with $K_{\alpha}^{E2} = +1.00$~\cite{Flambaum2009}.  The E3 transition is also less perturbed than the E2 transition by background electric and magnetic fields, giving rise to a lower uncertainty contribution from systematic shifts for the E3 transition frequency. The current state-of-the-art for the E3 transition of $^{171}$Yb$^+$ has an estimated fractional frequency uncertainty from systematic shifts of $2.7 \times 10^{-18}$~\cite{Sanner2019}.  The uncertainty budget is discussed in Appendix~\ref{Appendix:Yb_clock}.

\subsubsection{Molecular clocks}

\begin{figure}
		\includegraphics[width=0.98\textwidth]{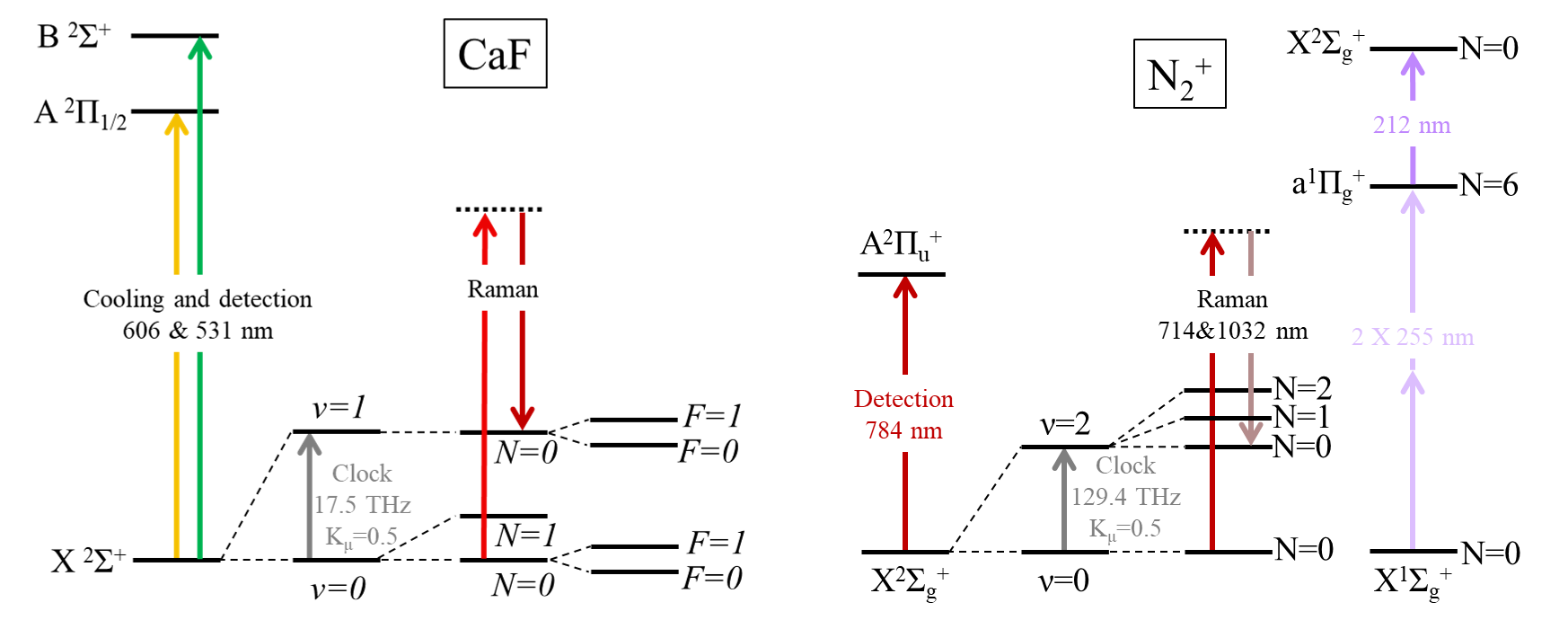}
	\caption{Relevant energy levels for the QSNET molecular clocks. The wavelengths and sensitivity coefficients of the clock transitions are also specified.}
\label{Fig:mol_transitions}
\end{figure}

We propose a molecular lattice clock based on the fundamental vibrational transition in CaF, which has a frequency $f_0 = 17.472$~THz, an estimated linewidth of 0.7~Hz, and $K_\mu=0.5$. The main ideas for such a clock were presented in \cite{Kajita2018}. Figure \ref{Fig:mol_transitions} shows the relevant energy levels of CaF. The electronic transitions $A ^{2}\Pi_{1/2}-X^{2}\Sigma^{+}$ and $B ^{2}\Sigma^{+}-X^{2}\Sigma^{+}$ are used to laser-cool the molecules to a few $\mu$K using the methods described in \cite{Truppe2017, Truppe2017b, Williams2017, Williams2018, Caldwell2019}. They can then be loaded into an optical dipole trap~\cite{Anderegg2018} or an optical lattice. For the clock transition, we choose to drive the Raman transition from $\ket{v,N,F,m} = \ket{0,0,F,m}$ to $\ket{1,0,F,m}$. All four choices of $(F,m)$ are useful for controlling systematic shifts. By choosing a transition that leaves $N$ unchanged, the upper and lower states have almost identical properties, so Zeeman, DC Stark and AC Stark shifts all cancel to high accuracy. Furthermore, because $N=0$ in both states, problematic tensor Stark shifts are eliminated. In the optical lattice, the molecules are deep in the Lamb-Dicke regime which eliminates first-order Doppler shifts. A 3D lattice also eliminates collisional shifts. Furthermore, the $N=0-1$ rotational transition at $20.5$~GHz can be used to measure the lattice intensity and the local electric and magnetic fields to high accuracy, further improving the accuracy of the clock. In Appendix~\ref{Appendix:CaF_clock} we discuss a set of parameters that will allow us to reach a fractional frequency uncertainty from systematic shifts of $\simeq7.5\times 10^{-18}$.

The QSNET network will include a second molecular clock with comparable sensitivity to variation of $\mu$ but different systematic uncertainties. Recently, molecular nitrogen ions have been proposed as a candidate for precision spectroscopy \cite{Kajita}. The vibrational clock transition at $129.4$~THz has an estimated linewidth on the order of nHz and a sensitivity of K$_\mu = 0.5$\footnote{From a detailed analysis of the molecular structure we found that the exact value is 0.49 \cite{Kajita}.}. The systematic shifts are predicted to be comparable with the current best optical clocks, facilitating frequency measurements at an uncertainty below $10^{-18}$. Within nitrogen molecular ions, there are two promising transitions with small systematic shifts and high sensitivity to changes in the electron-to-proton mass ratio, as shown in Fig.~\ref{Fig:mol_transitions}. The direct transition between the ro-vibrational ground state $\nu=0, N=0$ and the $\nu=2, N=2$ state is electric-quadrupole allowed and has been demonstrated by Germann \textit{et al.} \cite{Germann}. The transition between the $\nu=0, N=0$ and the $\nu=2, N=0$ states with a transition frequency of $129.4$~THz has not been investigated yet but was proposed in \cite{Kajita} and exhibits smaller systematic shifts for the $I=0$ isotopomer. N$_2^+$ has a X$^2\Sigma_g^+$ ground state described by Hund's case (b) with the angular momentum being solely the electron’s spin. Hence, the interaction with external magnetic fields is dominated by the interaction with the electron's magnetic moment. While there is a strong linear Zeeman shift between the magnetic sub-levels of $1.4$~MHz/$\mu$T, the differential shift of the two clock levels is exactly zero. Also, the quadratic Zeeman shift is zero, which makes transitions between the states with the same magnetic quantum number insensitive to external magnetic fields even at the second order. Due to the $\Sigma$ character of the electronic ground state, there is no permanent quadrupole moment which can interact with the trap electric field in the rotational ground state. Hence, the X$^2\Sigma_g^+, N=0$ ground state does not show a quadrupole shift. This is particularly important because, to conduct high resolution spectroscopy, molecular ions need to be trapped alongside atomic ions for sympathetic cooling and state detection. This inevitably results in a local electric field gradient that would cause a quadrupole shift. In Appendix~\ref{Appendix:N2_clock} we evaluate that the fractional frequency uncertainty from systematic shifts for N$_2^+$ is $\simeq3.9\times 10^{-18}$ under conditions that can be easily reached in current experiments.

\subsubsection{Highly charged ion clocks}

Novel technological platforms are currently being explored to develop clocks with enhanced sensitivity to variations of $\alpha$. There are two promising avenues that could be followed. One regards the development of a nuclear clock \cite{Peik_2003, Flambaum_Th_2006}. The other involves the use of various transitions in highly charged ions (HCI) \cite{berengut, derevianko, kozlov}. The estimates of potential accuracy of nuclear and HCI clocks are similar, but HCI do not have the complication of 150 nm clock transition wavelength that also imply the design of new frequency combs with sufficient power.
 Recent experimental breakthroughs \cite{Schmoger1233,HCIRSI} allow the HCIs to be cooled to temperatures below 1~mK, where high resolution spectroscopy and advanced techniques such as quantum logic spectroscopy can be used \cite{Schmoger1233,Micke}. 

\begin{figure}
		\includegraphics[width=0.98\textwidth]{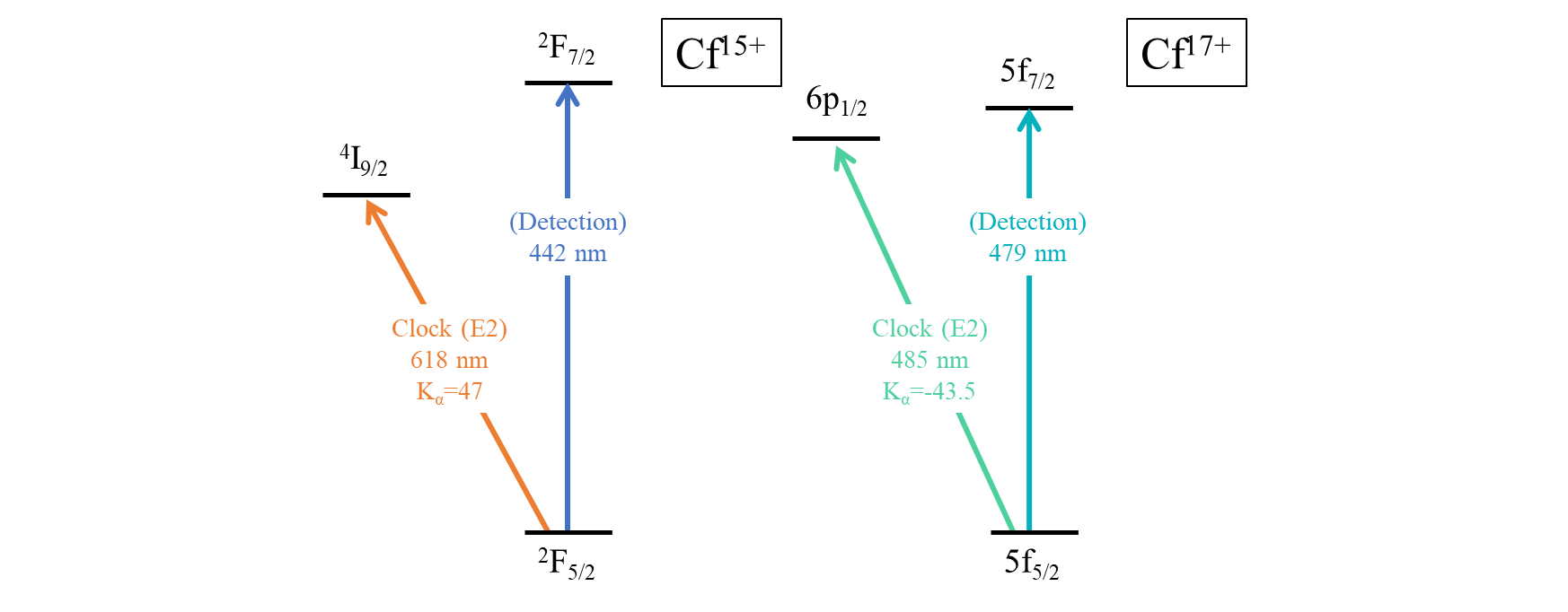}
	\caption{Relevant energy levels for the QSNET highly charged ion clocks. The wavelengths and sensitivity coefficients of the clock transitions are also specified.}
\label{Fig:transitions}
\end{figure}

HCIs are good candidates for searching for variation of the fine structure constant $\alpha$: they are less sensitive to external perturbations due to their compact electronic cloud and their sensitivity to variation of $\alpha$ is enhanced due to larger relativistic effects. {A vast number of highly charged ions can exist in stable form},  however to reach the level of accuracy and stability needed to measure variations of $\alpha$, it is required that they: i) feature optical transitions ($\simeq$200-1000 nm); ii) have lifetimes of the excited clock state between 1 and $10^4$~s; iii) have high sensitivities to variations of $\alpha$. Additionally, it is desirable that other strong transitions could be used for cooling and/or detection, and that the clock levels are suitable for cancellation of various systematic shifts. Another desirable feature is a relatively simple electronic structure to enable precision calculation of the atomic properties and significantly easier spectra identification. {A handful of HCIs which satisfy all of these requirements have been identified, among them the californium ionisation states} Cf$^{15+}$ and Cf$^{17+}$ that have been chosen for the QSNET project. The low-lying level structures for Cf$^{15+}$ and Cf$^{17+}$ are shown in Fig.~\ref{Fig:transitions}~\cite{Cfclock}, together with the corresponding $K$ enhancement factors \cite{kozlov}. The Cf$^{15+}$ HCI features a clock transition at 618 nm with $K_\alpha=+47$ and a predicted linewidth of order mHz, while the Cf$^{17+}$ HCI has a clock transition at 485 nm with $K_\alpha=-43.5$ and a linewidth of $\simeq0.5$~Hz. In Appendix~\ref{Appendix:HCI_clock} we evaluate that under realistic conditions, it would be possible to reach fractional frequency uncertainty on the order of $10^{-19}$ for both ionisation states. Additionally, the possibility of realising a dual clock co-trapping Cf$^{15+}$ and Cf$^{17+}$ is particularly appealing due to the opposite sign of the large $K$ coefficients of the clock transitions and the cancelling of any residual common systematic effect. Finally, both ionisation states of Cf feature a relatively strong M1 transition that could be used for direct detection. 

%%%%%%%%%%%%%
\subsection{Predicted performance}
\label{Sec:predicted_performance}

In this subsection we estimate the sensitivity of the QSNET clocks to variations of $\alpha$ and $\mu$ on different timescales. The fractional frequency uncertainty with which the clocks can be operated will depend on both the inaccuracy, as determined by the systematic uncertainties described in the subsection above and Appendix~\ref{Appendix:QSNET_clocks}, and the instability. Ignoring additional factors of order 1, the fractional frequency instability for the different clocks can be estimated according to~\cite{Ludlow2015}:
\begin{ceqn}
\begin{equation}
    \sigma_y(\tau) \sim \frac{\Delta \nu}{\nu_0}\frac{1}{\sqrt{N}}\sqrt{\frac{T_c}{\tau}} \, ,
\end{equation}
where $\Delta \nu$ and $\nu_0$ are the linewidth and frequency of the probed transition, $N$ is the number of atoms, $T_c$ is the cycle time {of the experiment} and $\tau$ is the total measurement time.  Under the assumption that the probe time $T_p$ completely dominates the measurement cycle time, then $T_c = T_p$. Futhermore, assuming the natural linewidth of the atomic transition is sufficiently small, then the resolved linewidth will be determined by the Fourier transform limit of the probe time, $\Delta \nu \sim 1/T_p$. Accordingly, the fractional frequency instability can be expressed as
\begin{ceqn}
\begin{equation}
    \sigma_y(\tau) \sim \frac{1}{\nu_0 \sqrt{N T_p \tau}} \quad.
\end{equation}
\end{ceqn}
\end{ceqn}

\begin{figure}
\begin{center}
		\includegraphics[width=0.5\textwidth]{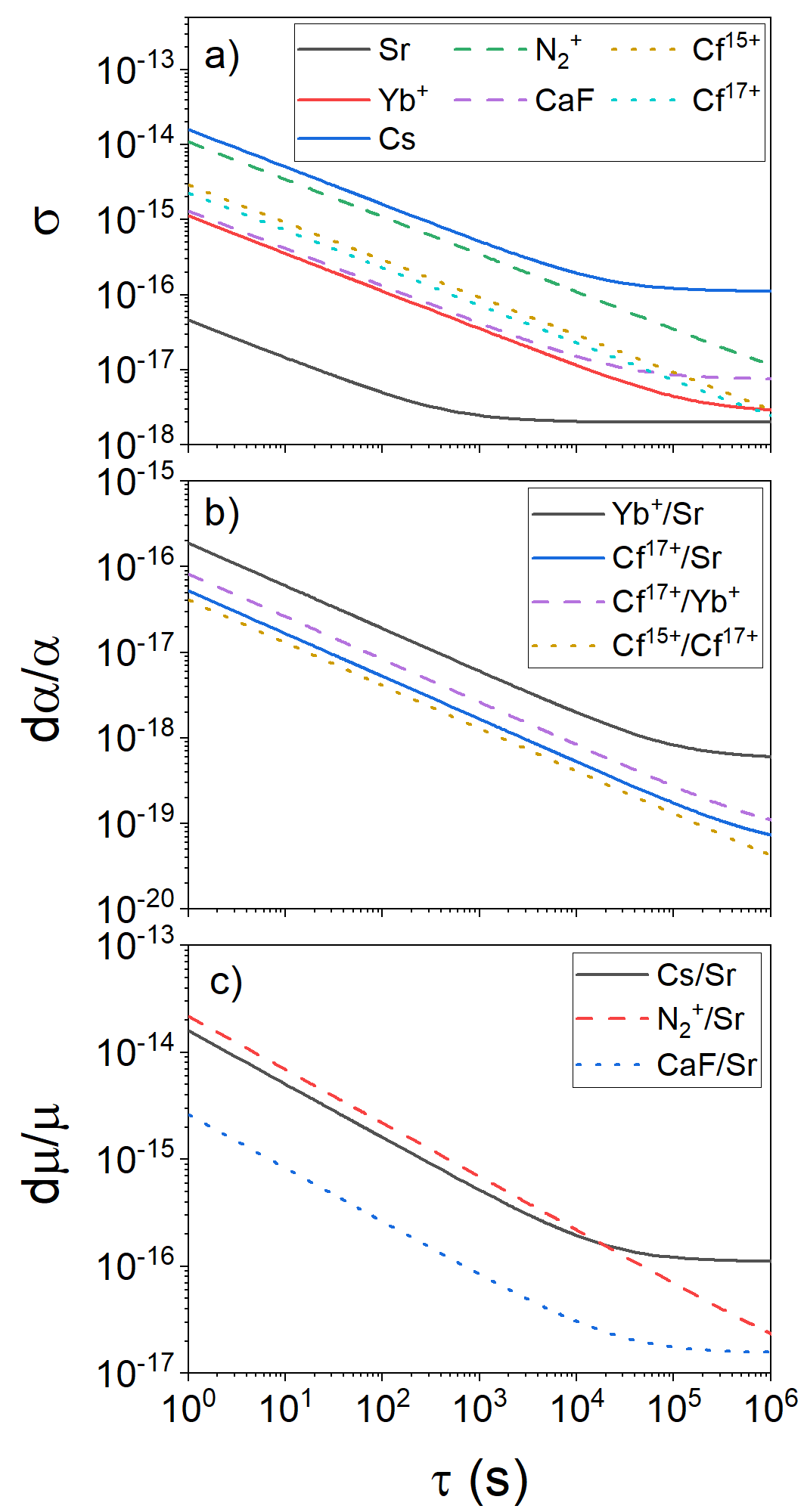}
\end{center}
	\caption{a) Fractional uncertainty as a function of the integration time $\tau$ evaluated for the QSNET clocks. The solid lines are for atomic clocks, dashed lines for molecular clocks and dotted lines for highly charged ion clocks. b) and c) Predicted sensitivities to temporal variations of $\alpha$ and $\mu$ obtained by comparing different clocks.}
\label{figsigma}
\end{figure}

For each clock, we calculate the fractional uncertainty as 
\begin{ceqn}
\begin{equation}
    \sigma(\tau)=\sqrt{\sigma_y^2(\tau)+\sigma_i^2},
\end{equation}
\end{ceqn}
where $\sigma_i$ is the inaccuracy for the specific clock, as reported in Subsection \ref{Sec:QSNET_network} and calculated in Appendix~\ref{Appendix:QSNET_clocks}. For CaF and Cs  we set $T_p=190$ and 1000~ms, respectively\footnote{For Cs $T_p$ has been chosen to match the current state-of-the-art experiments, for CaF $T_p$ is limited by the natural linewidth of the transition.}, while for all the other clocks $T_p=500$ ms. For Sr and Cs clocks the number of atoms are $N\simeq10^4$ and $5\times10^6$ respectively, while for the CaF clock we aim to use $N=10^4$ molecules. For all the ion clocks $N=1$. The resulting fractional uncertainties as a function of the integration time $\tau$ are reported in Fig. \ref{figsigma}a). From the values of $\sigma$ for each clock it is then possible to estimate the sensitivity to variations of $\alpha$ or $\mu$ using Eq.~(\ref{eq_ratio}). The results are reported in Fig. \ref{figsigma}b) and c) for the clock-to-clock comparisons that deliver the best performance.

{It is a well-known consequence of general relativity that time runs at different rates in different gravity potentials, where the term ‘gravity potential’ includes both gravitational and centrifugal components.  If a clock on the surface of the Earth experiences a height change of 1 cm, this will change the frequency of the clock by one part in $10^{18}$.  When looking for variations in fundamental constants at this level of precision, care must therefore be taken to avoid confusion with any frequency changes brought about by gravitational effects. 
For two clocks operated in the same location (and at the same height), gravitational effects are common-mode and will not alter the frequency ratio between the clocks.  However, if two clocks are in different locations, as proposed in QSNET, they can be subject to gravity potential differences.  Any static difference in the gravity potential (for example due to a constant height offset between the clocks) will lead to a constant offset in the frequency ratio and is irrelevant for QSNET, which is concerned only with \emph{temporal changes} in frequency ratios as a signature of \emph{changes} in fundamental constants. It is therefore only time-varying differences in the gravity potential that might need to be taken care of.  Such time-varying effects can arise from a variety of sources including solid Earth tides, ocean tides and non-tidal mass redistributions in the atmosphere and oceans \cite{ Voigt_2016}.  By far the largest contribution to time-varying gravity potentials is from solid Earth tides, which can create fractional changes in individual clock frequencies with peak-to-peak variations up to 5$\times10^{-17}$ over timescales of several hours.  However, the effect on the frequency ratio between two clocks depends on the gravity potential difference between them.  The clocks involved in QSNET are all located in the UK, within 2 degrees of longitude of each other, thus sharing a large common-mode component in the time-varying potential.  This means that the largest time-varying effect from gravity potentials on the frequency ratios involved in QSNET is at or below the level of 10$^{-18}$ over several hours.
Both solid Earth tides and ocean tides are highly predictable and their effects can be modelled and subtracted from the data.  In particular, they are oscillatory in nature with characteristic timescales of several hours so are easily distinguished from oscillations at all other timescales.  Furthermore, in the case of an oscillating dark-matter field (see Sec. \ref{Sec:DM}), the signal is expected to be nearly monochromatic with a quality factor of $Q \sim 10^6$, which is higher than that for oscillatory changes in Earth's gravity potential.   
Far smaller frequency offsets can also arise from non-tidal mass redistributions, which can include gravity potential effects that are non-periodic and hard to predict, such as the influence of local weather conditions.  Disturbances from these non-periodic effects on an individual clock frequency would produce transient offsets, similar to those caused by any other localised disturbances in a specific laboratory.  The networked approach, however, is resilient to these local disturbances.  Transient effects are disregarded unless they are correlated with other independent frequency ratios that have been measured in different locations.  This avoids mistaking technical noise in a given system for evidence of new physics on a global scale. 
In summary, careful attention should be paid to the frequency offsets created by time-varying gravity potentials but they do not prevent the searches for new physics using the network of clocks proposed in QSNET.}

%%%%%%%%%%%%%%%%%%%%%%%%%%
\section{Phenomenology}
\label{Sec:pheno}

In this section, we evaluate how QSNET can probe some specific dark matter and dark energy models, namely ultra light dark matter models in Subsection~\ref{Sec:DM}, quintessence dark energy models in Subsection~\ref{Sec:DE}, and solitonic dark sector models in Subsection~\ref{Sec:solitons}. Additionally, we show in Subsection~\ref{Sec:symmetries} that QSNET can probe fundamental space-time symmetries. Finally, in Subsection~\ref{Sec:unifiedtheory} we discuss how QSNET can provide stringent tests of grand unification theories and quantum gravity. In this section, unless explicitly stated otherwise, we employ the natural units $\hbar = c = 1$.

%%%
\subsection{Dark matter}
\label{Sec:DM}

Dark matter comprises about $85\%$ of the total matter content of the universe. To date, all available evidence for dark matter has involved the gravitational effects of dark matter on ``visible'' ordinary matter. Nevertheless, it is generally expected that dark matter should also interact non-gravitationally with ordinary matter, albeit feebly. Over the past few decades, the majority of experimental efforts in direct dark matter searches with terrestrial detectors have focused on weakly-interacting massive particles (WIMPs) in the $\sim \textrm{GeV} - \textrm{TeV}$ mass range \cite{Baudis:2016WIMP}. In light of the continued absence of strong experimental evidence for WIMPs, there has been growing interest over the past decade in performing searches for dark-matter candidates other than WIMPs, in particular for feebly-interacting, \emph{low-mass bosonic particles} \cite{Jaeckel:2010WISPs,Irastorza:2018axions,FIPs:2021}. The axion (a pseudoscalar particle) is the leading candidate to explain the strong \textit{CP} problem of quantum chromodynamics \cite{Peccei:1977hh,Peccei:1977ur,Weinberg:1977ma,Wilczek:1977pj,Kim:1979if,Shifman:1979if,Zhitnitsky:1980tq,Dine:1981rt}, whilst also being an excellent candidate for cold dark matter \cite{Preskill:1982cy,Abbott:1982af,Dine:1982ah}. 

Low-mass spinless bosons may be produced non-thermally via the ``vacuum misalignment'' mechanism \cite{Preskill:1982cy,Abbott:1982af,Dine:1982ah} and subsequently form a coherently oscillating classical field, which in the rest frame is given by: 
\begin{ceqn}
\begin{equation}
\label{oscillating_DM_field}
\phi(t) \approx \phi_0 \cos( m_\phi c^2 t / \hbar )  \, , 
\end{equation}
\end{ceqn}
which occurs, for example, in the case of the harmonic potential $V(\phi) = m_\phi^2 \phi^2 / 2$ when $m_\phi \gg H$, where $m_\phi$ is the boson mass and $H$ is the Hubble parameter that describes the relative rate of expansion of the universe. The scalar field in Eq.~(\ref{oscillating_DM_field}) carries an energy density, averaged over a period of oscillation, of $\left< \rho_\phi \right> \approx m_\phi^2 \phi_0^2 / 2$. Since present-day dark matter must be cold, all of the boson energies satisfy $E_\phi \approx m_\phi c^2$, which implies that the oscillations of the field in Eq.~(\ref{oscillating_DM_field}) are temporally coherent on sufficiently small time scales (and are also spatially coherent on sufficiently small length scales). Over time, the galactic dark matter is expected to have become virialised, attaining a root-mean-square speed of $\sim 10^{-3} c$ in our local galactic region. The typical spread in the boson energies is hence $\Delta E_\phi / E_\phi \sim \left< v_\phi^2 \right> / c^2 \sim 10^{-6}$, implying a coherence time of $\tau_\textrm{coh} \sim 2 \pi / \Delta E_\phi \sim 10^6 T_\textrm{osc}$, where $T_\textrm{osc} \approx 2 \pi / m_\phi$ is the period of oscillation. In other words, the oscillations of the bosonic field are practically monochromatic, with an associated quality factor of $Q \sim 10^6$. The spatial gradients associated with the field $\phi(t,\boldsymbol{x}) \approx \phi_0 \cos( m_\phi t - m_\phi \boldsymbol{v}_\phi \cdot \boldsymbol{x} )$ give rise to a coherence length of $\lambda_\textrm{coh} \sim 2\pi / (m_\phi \sqrt{\left< v_\phi^2 \right>})$, which is $\sim 10^3$ times the Compton wavelength. On time and length scales exceeding the coherence time and coherence length, respectively, the stochastic nature of the bosonic field needs to be taken into account \cite{Derevianko:2018stochastic,Safdi:2018stochastic}. In particular, the bosonic field amplitudes $\phi_0$ are expected to follow a Rayleigh-type distribution, while the bosonic particle velocities $\boldsymbol{v}_\phi$ are expected to follow a Maxwell-Boltzmann-type distribution.  

In Appendix~\ref{Appendix:DM_mass_range}, we discuss the relevant range of bosonic dark-matter particle masses, which is $10^{-21}~\textrm{eV} \lesssim m_\phi \lesssim 1~\textrm{eV}$, corresponding to oscillation frequencies in the range $10^{-7}~\textrm{Hz} \lesssim f \lesssim 10^{14}~\textrm{Hz}$. The lower end of this frequency range corresponds to periods of oscillation of the order of a month and the upper end corresponds to the infra-red region of the electromagnetic spectrum. Searching for possible particle-like signatures of such low-mass dark matter is practically impossible, since the kinetic energies of non-relativistic very-low-mass particles are extremely small and typically many orders of magnitude below the energy thresholds of conventional WIMP detectors. On the other hand, it may be possible to take advantage of wave-like signatures of low-mass bosonic dark matter due to the large number density of the bosons. While searches for wave-like signatures of the oscillating classical field in Eq.~(\ref{oscillating_DM_field}) via its gravitational effects (e.g., using pulsar timing methods \cite{Khmelnitsky:2014pulsar,Porayko:2018pulsar}) are limited to the lowest allowable dark-matter particle masses, much larger ranges of dark-matter particle masses may be probed if the bosonic field interacts non-gravitationally with fields from the Standard Model sector. 

In particular, atomic clocks are excellent detectors to search for ``scalar-type'' interactions, either via comparisons with other clocks \cite{Tilburg:2015DM,Stadnik:2015DM-LI,Stadnik:2015DM-VFCs,Leefer:2015DM,Hees:2016DM,Stadnik:2016DM-clocks,BACON:2021DM} or via referencing against cavities \cite{Stadnik:2015DM-LI,Wcislo18,Stadnik:2016cavity,Kennedy:2020cavity}. 
The QSNET clock network is particularly well-suited to search for scalar-type interactions of a spinless dark-matter field with the electromagnetic field and electron. From Eq.~(\ref{general_linear_interactions}), the lowest order of such interactions involves the first power of the scalar field $\phi$: 
\begin{ceqn}
\begin{equation}
\label{linear_DM_interactions}
\mathcal{L} = \frac{\phi}{\Lambda_\gamma} \frac{F_{\mu \nu} F^{\mu \nu}}{4} - \frac{\phi}{\Lambda_e} m_e \bar{\psi} \psi  \, , 
\end{equation}
\end{ceqn}
where we have defined $1/\Lambda_\gamma \longleftrightarrow \kappa d_e^{(1)}$ and $1/\Lambda_e \longleftrightarrow \kappa d_{m_e}^{(1)}$. The parameters $\Lambda_{\gamma,e}$ denote the effective new-physics energy scales of the underlying model; higher energy scales correspond to feebler interactions (for comparison, the effective energy scale associated with the usual gravitational interaction is set by the reduced Planck scale). 

Comparing the interactions in Eq.~(\ref{linear_DM_interactions}) with the corresponding terms in the Standard Model Lagrangian, $\mathcal{L}_\textrm{Standard Model} \supset - F_{\mu \nu} F^{\mu \nu} / 4 - e J_\mu A^\mu - m_e \bar{\psi} \psi$, where $-e$ is the electric charge carried by an electron, $J^\mu = \bar{\psi} \gamma^\mu \psi$ is the electromagnetic 4-current and $A^\mu$ is the electromagnetic 4-potential, shows that the oscillating field in Eq.~(\ref{oscillating_DM_field}) induces the following apparent oscillations of the electromagnetic fine structure constant $\alpha$ and the electron mass \cite{Stadnik:2015DM-LI}: 
\begin{ceqn}
\begin{equation}
\label{linear-in-phi_VFCs}
\frac{d \alpha}{\alpha} \approx \frac{\phi_0 \cos (m_\phi t)}{\Lambda_\gamma} , \quad \frac{d m_e}{m_e} \approx \frac{\phi_0 \cos (m_\phi t)}{\Lambda_e} \, . 
\end{equation}
\end{ceqn}

If the linear-in-$\phi$ interactions in Eq.~(\ref{linear_DM_interactions}) are precluded, e.g.~if the underlying model admits a $Z_2$ symmetry under the $\phi \to - \phi$ transformation, then the lowest-order interactions would involve the second power of the scalar field $\phi$, as described by Eq.~(\ref{general_quadratic_interactions}). It follows that Eqs.~(\ref{linear_DM_interactions}) and (\ref{linear-in-phi_VFCs}) are modified according to \cite{Stadnik:2015DM-LI,Stadnik:2015DM-VFCs}: 
\begin{ceqn}
\begin{equation}
\label{quadratic_DM_interactions}
\mathcal{L} = \frac{\phi^2}{(\Lambda'_\gamma)^2} \frac{F_{\mu \nu} F^{\mu \nu}}{4} - \frac{\phi^2}{(\Lambda'_e)^2} m_e \bar{\psi} \psi  \, , 
\end{equation}
\begin{equation}
\label{quadratic-in-phi_VFCs}
\frac{d \alpha}{\alpha} \approx \frac{\phi_0^2 \cos^2 (m_\phi t)}{(\Lambda'_\gamma)^2} , \quad \frac{d m_e}{m_e} \approx \frac{\phi_0^2 \cos^2 (m_\phi t)}{(\Lambda'_e)^2}  \, . 
\end{equation}
\end{ceqn}

The apparent oscillations of the fundamental constants in Eqs.~(\ref{linear-in-phi_VFCs}) and (\ref{quadratic-in-phi_VFCs}) would cause atomic and molecular transition frequencies to undergo small oscillations about their mean values. Therefore, from Eq.~(\ref{eq_ratio}) we obtain that the comparison of two atomic transition frequencies in time would also undergo small oscillations: 
\begin{ceqn}
\begin{equation}
\label{clock_DM_signal}
\frac{d R}{R} = \left( K_{X,1} - K_{X,2} \right) \frac{d X}{X} \propto \left( K_{X,1} - K_{X,2} \right) \cos(2\pi f_\textrm{signal}t)  \, ,
\end{equation}
\end{ceqn}
where the signal frequency is given by $f_\textrm{signal} \approx m_\phi c^2 / h$ in the case of linear-in-$\phi$ interactions (\ref{linear_DM_interactions}) or $f_\textrm{signal} \approx 2 m_\phi c^2 / h$ in the case of quadratic-in-$\phi$ interactions (\ref{quadratic_DM_interactions}).

\begin{figure*}[t!]
	\centering
	\includegraphics[width=0.8\textwidth]{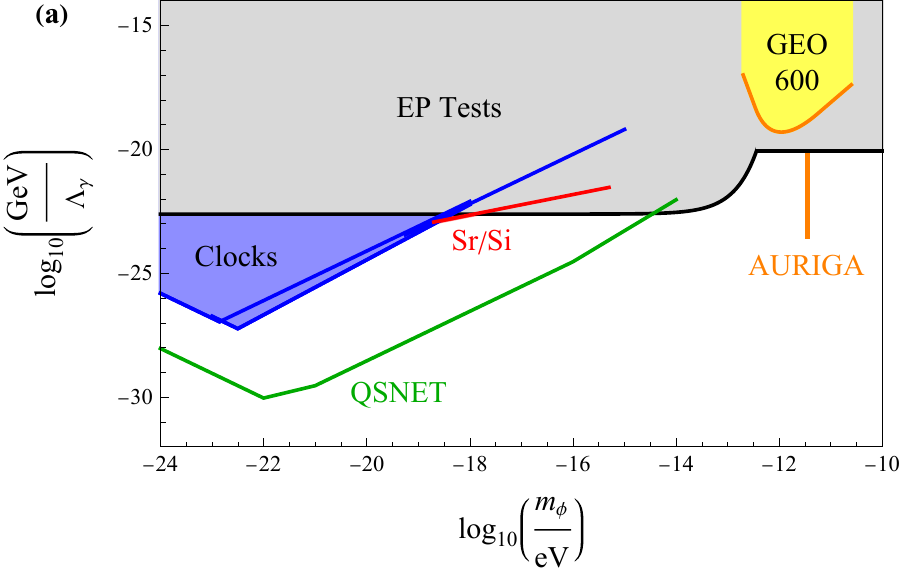}
    \par \medskip
    \includegraphics[width=0.8\textwidth]{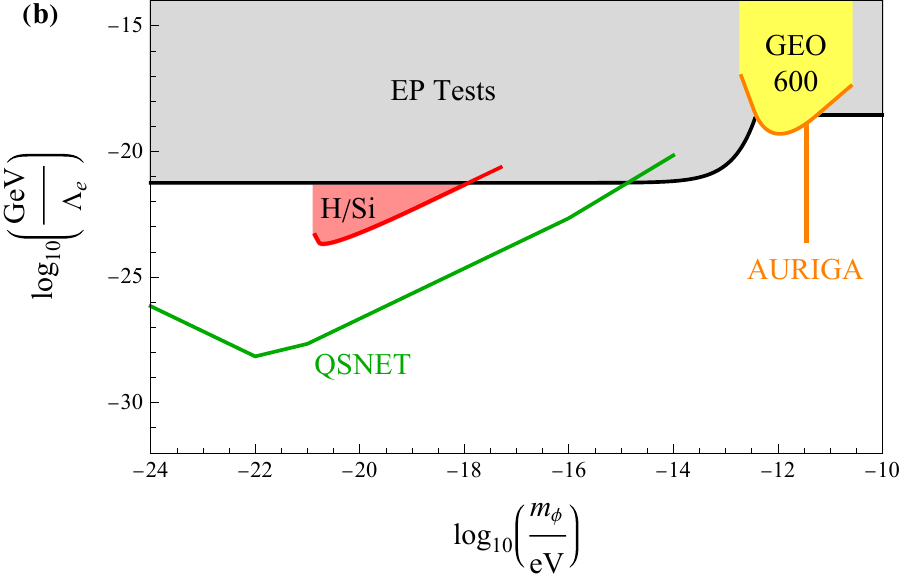}
	\caption{
	Parameter spaces for a model of an oscillating scalar dark-matter field interacting with \textbf{(a)} the electromagnetic field and \textbf{(b)} the electron via the linear-in-$\phi$ interactions in Eq.~(\ref{linear_DM_interactions}). 
	The estimated sensitivities of clocks within the QSNET network to apparent oscillations of the fundamental constants are shown by the green lines. 
	Also shown are existing limits from previous searches for oscillations of the fundamental constants via clock comparisons (blue region) \cite{Leefer:2015DM,Hees:2016DM,BACON:2021DM}, clock-cavity comparisons (red regions) \cite{Kennedy:2020cavity}, optical interferometry (yellow regions) \cite{GEO600:2021DM} and a resonant-mass detector (narrow orange regions) \cite{AURIGA:2017DM}, as well as complementary bounds from searches for equivalence-principle-violating forces (grey regions) \cite{RotWash:1999,EotWash:2008,MICROSCOPE:2017A,MICROSCOPE:2017B}. 
	See the main text for more details. 
	}
\label{Fig:DM_exclusion_plots_linear}
\end{figure*}

\begin{figure*}[t!]
	\centering
	\includegraphics[width=0.8\textwidth]{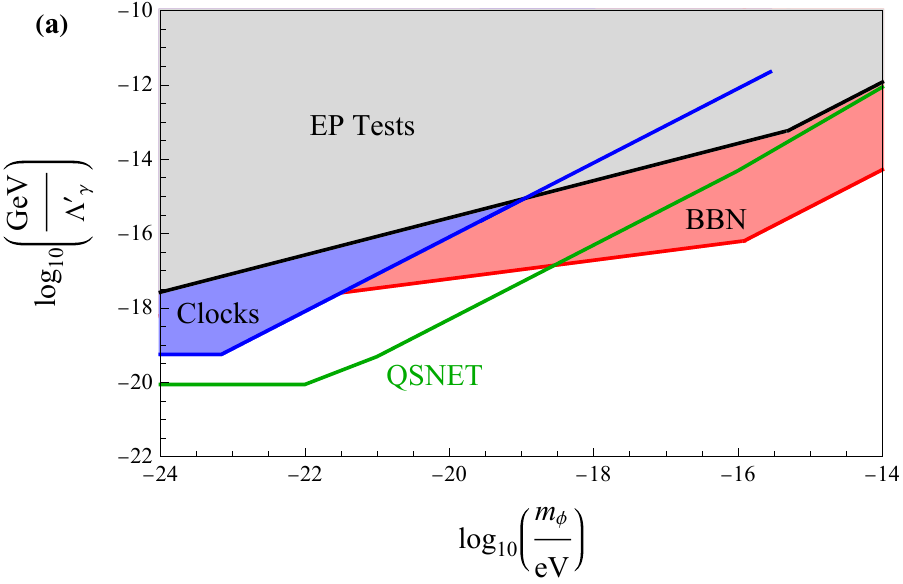}
    \par \medskip
	\includegraphics[width=0.8\textwidth]{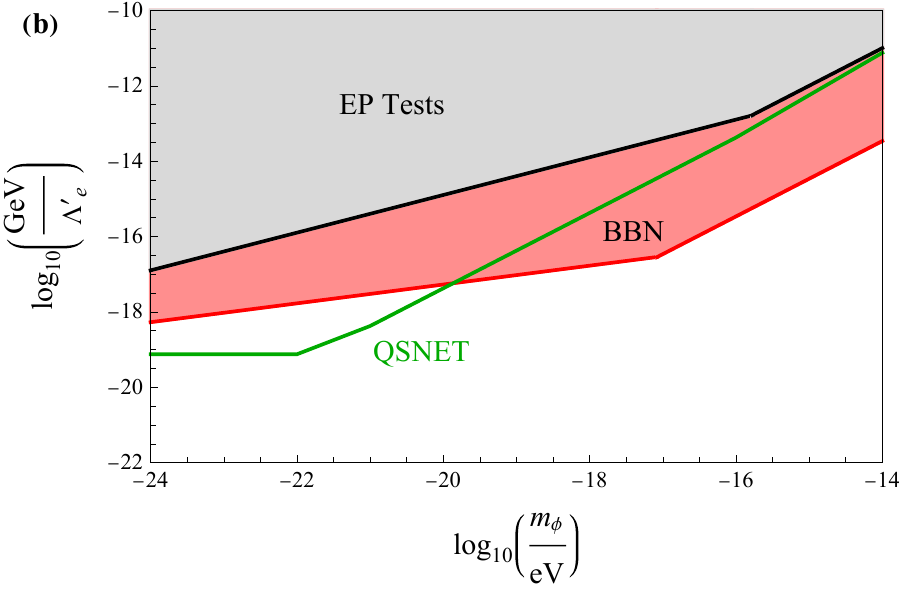}
	\caption{
	Parameter spaces for a model of an oscillating scalar dark-matter field interacting with \textbf{(a)} the electromagnetic field and \textbf{(b)} the electron via the quadratic-in-$\phi$ interactions in Eq.~(\ref{quadratic_DM_interactions}). 
	The estimated sensitivities of clocks within the QSNET network to apparent oscillations of the fundamental constants are shown by the green lines. 
	Also shown are existing limits from previous searches for oscillations of the fundamental constants via clock comparisons (blue region) \cite{Stadnik:2015DM-VFCs,Stadnik:2016DM-clocks}, as well as complementary types of bounds from searches for equivalence-principle-violating forces (grey regions) \cite{Hees:2018DM} and measurements and calculations pertaining to big bang nucleosynthesis (red regions) \cite{Stadnik:2015DM-VFCs}. 
	See the main text for more details. 
	}
\label{Fig:DM_exclusion_plots_quadratic}
\end{figure*}

As discussed in Sec.~\ref{Sec:Clock_and_FCs}, the sensitivity coefficients $K_X$ in Eq.~(\ref{clock_DM_signal}) depend on the specific transitions under consideration. All of the transitions that will be utilised within the QSNET network are summarised in Fig.~\ref{figQSNET}. 
%Currently, the most sensitive (in absolute terms) pairs of transitions within QSNET to variations of the fundamental constants are Yb$^+$/Sr and Cs/Sr, for which $K_\alpha (\textrm{Yb}^+) - K_\alpha (\textrm{Sr}) \approx -6.0$ and $K_{m_e} (\textrm{Cs}) - K_{m_e} (\textrm{Sr}) \approx +1$. 
%In the longer term, the most promising pairs of transitions within QSNET to variations of the fundamental constants appear to be Cf$^{15+}$/Cf$^{17+}$ and CaF/Sr, for which $K_\alpha (\textrm{Cf}^{15+}) - K_\alpha (\textrm{Cf}^{17+}) \approx +90$ and $K_{m_e} (\textrm{CaF}) - K_{m_e} (\textrm{Sr}) \approx +0.5$. 
The estimated performances of QSNET are discussed in Subsection \ref{Sec:predicted_performance}, and  are reported in Fig.~\ref{figsigma}. From these, we can estimate the sensitivity of the QSNET network to apparent oscillations of the fundamental constants induced by an oscillating dark-matter field via linear-in-$\phi$ and quadratic-in-$\phi$ interactions. The results are reported in  Figs.~\ref{Fig:DM_exclusion_plots_linear} and \ref{Fig:DM_exclusion_plots_quadratic}, respectively. In this case, the signal-to-noise ratio (SNR) generally improves with the total measurement time $t_\textrm{int}$ as $\textrm{SNR} \propto t_\textrm{int}^{1/2}$. We assume $t_\textrm{int} = 1~\textrm{year}$, with individual data points sampled every $\tau=1~\textrm{s}$. Additional details are discussed in Appendix~\ref{Appendix:DM_exclusion_plots}.

%%%

%%%

In  Figs.~\ref{Fig:DM_exclusion_plots_linear} and \ref{Fig:DM_exclusion_plots_quadratic}, we assume that the ultra-low-mass bosons saturate the observed dark matter abundance when possible. For the boson masses $m_\phi \lesssim 10^{-21}~\textrm{eV}$, for which bosons cannot comprise the entirety of the observed dark matter, we assume that such bosons make up a maximally allowable fraction of the dark matter, which is $\mathcal{O}(10\%)$ in this case (see also Appendix~\ref{Appendix:DM_mass_range}). Since the new-physics energy scales appear in the clock-based observables in the combination $\sqrt{\rho_\phi} / \Lambda_X$ or $\sqrt{\rho_\phi} / \Lambda'_X$, the sensitivity to $\Lambda_X$ or $\Lambda'_X$ is only weakened by a factor of $\approx 3$ in this case, compared to the case when dark matter is saturated entirely by these bosons. 

In the case of linear-in-$\phi$ interactions, we also show existing limits from previous searches for oscillating dark-matter signatures via clock comparisons \cite{Leefer:2015DM,Hees:2016DM,BACON:2021DM}, clock-cavity comparisons \cite{Kennedy:2020cavity}, optical interferometry \cite{GEO600:2021DM} and a resonant-mass detector \cite{AURIGA:2017DM}, as well as complementary bounds from searches for equivalence-principle-violating forces \cite{RotWash:1999,EotWash:2008,MICROSCOPE:2017A,MICROSCOPE:2017B}. In the case of the clock- and cavity-based bounds, we have taken into account a degradation factor of $\approx 3$ due to only partial sampling of the distribution of stochastically-fluctuating scalar-field amplitudes \cite{centers2020stochastic}. In the case of quadratic-in-$\phi$ interactions, we also present existing limits from previous searches for oscillating dark-matter signatures via clock comparisons \cite{Stadnik:2015DM-VFCs,Stadnik:2016DM-clocks}, as well as complementary types of bounds from searches for equivalence-principle-violating forces \cite{Hees:2018DM} and measurements and calculations pertaining to big bang nucleosynthesis \cite{Stadnik:2015DM-VFCs}. The limits on linear-in-$\phi$ interactions from searches for equivalence-principle-violating forces do not involve any assumptions about the possible contribution of the $\phi$-bosons towards the observed dark-matter abundance, whereas all of the other types of limits do. We remark that, in the case of $\phi^2$ couplings, the bosonic dark-matter field can be screened near Earth's surface; however, such screening is negligible for the majority of the parameter space that is relevant for the QSNET clock experiments \cite{Hees:2018DM}. 

From our analysis, we see that the QSNET network has the potential to probe large regions of previously unexplored parameter space in models of oscillating scalar dark-matter fields.

%%%
\subsection{Dark energy}
\label{Sec:DE}

It is now well established that our universe has been undergoing accelerated expansion over the past several billion years. A cosmological constant in Einstein's equations that is characterised by an energy scale of $\Lambda \sim 10^{-3}$~eV can account for this observed acceleration. Together with the cold dark matter, this corresponds to the $\Lambda$CDM cosmological model. While current cosmological data are fully consistent with the presence of a cosmological constant, the tiny energy scale $\sim 10^{-3}$~eV is unexplained and much smaller than other relevant energy scales in nature, such as the Planck scale in gravity or the electroweak scale in particle physics. This observation has motivated a class of models known as quintessence, in which the cosmological constant is replaced by a dynamical scalar field; see, e.g., \cite{Martin:2008qp} for a review. 
In quintessence models, gravity is described by General Relativity and the matter content of the universe consists of radiation, dark matter, visible matter and quintessence, which is a scalar field $\phi$ that evolves on a cosmological time scale. If the quintessence field couples to visible matter, fundamental constants could be slowly evolving with cosmological time. In particular, if the quintessence field couples linearly to matter as in Eq.~(\ref{general_linear_interactions}), then slow changes in the values of the fundamental constants may leave an imprint on clock experiments. 

The classical equation of motion for a scalar field $\phi$ with potential $V(\phi)$ in an expanding universe reads as follows: 
\begin{ceqn}
\begin{equation}
\label{KGE_expanding_general}
\ddot \phi+ 3 H \dot \phi + \frac{\partial V}{\partial \phi} = \frac{\partial \mathcal{L}_\textrm{int}}{\partial \phi}  \, , 
\end{equation}
\end{ceqn}
where $H(t)$ is the Hubble parameter that describes the relative rate of expansion of the universe and the Lagrangian $\mathcal{L}_\textrm{int} (\phi)$ encodes the non-gravitational interactions of the scalar field with ordinary matter. 
A wide variety of potentials that admit a slow evolution of $\phi$ at the present day are possible; see, e.g., Refs.~\cite{Wetterich:1994cosmon,Amendola:1999A-quint,Amendola:1999B-quint,Dvali:2002quintessence,Chiba:2002quintessence,Veneziano:2002A-quint,Veneziano:2002B-quint,Wetterich:2003quintessence,Goldberg:2003quintessence,Copeland:2004quintessence,Lee:2004quintessence,Marra:2005quintessence,Lee:2007quintessence}. 
The key features of quintessence models can be understood by considering the simple example of the harmonic potential $V (\phi) = m_\phi^2 \phi^2 / 2$ and the limiting case of sufficiently feeble interactions, in which case the classical equation of motion (\ref{KGE_expanding_general}) simplifies to: 
\begin{ceqn}
\begin{equation}
\label{KGE_expanding_harmonic}
\ddot \phi+ 3 H \dot \phi + m_\phi^2 \phi \approx 0  \, , 
\end{equation}
\end{ceqn}
which represents the equation of motion for a damped harmonic oscillator. 
In the strongly overdamped regime when $m_\phi \ll 3H/2$, $\phi$ remains approximately constant over time and so does not appreciably affect the values of the fundamental constants. 
On the other hand, in the strongly underdamped regime when $m_\phi \gg 3H/2$, the scalar field begins oscillating well before the present day, similarly to the case of scalar-field dark matter discussed in Sec.~\ref{Sec:DM}, and so does not explain the observed dark energy. 
Appreciable changes in the scalar field $\phi$ and consistency with dark energy therefore can only occur when $m_\phi \sim H_0 \sim 10^{-33}~\textrm{eV}$, where $H_0$ is the present Hubble scale; in this case, the apparent values of the fundamental constants would change slowly and linearly with time during the present epoch. 

The most stringent bound on linear drifts in $\alpha$ at redshifts $z \lesssim 0.5$ comes from optical clock comparison measurements with $\textrm{Yb}^+$ and is at the level $|d\ln(\alpha)/dt| \lesssim 10^{-18}~\textrm{yr}^{-1}$ \cite{Lange21}. 
This bound is one and two orders of magnitude, respectively, more stringent than bounds pertaining to the Oklo phenomenon \cite{Shlyakhter:1976Oklo,Dyson:1996Oklo,Fujii:2000Oklo,Petrov:2006Oklo} and meteorite dating \cite{Olive:2002meteorites}; see Table~\ref{tab:quintessence_drifts_limits} for a summary of the bounds. The bounds from two optical clock measurements separated in time can be improved with longer time intervals between the measurements.  In QSNET, there are clocks with differential sensitivities to $\alpha$ that are $\simeq 10$ times greater than those used in present slow-drift constraints.  %We thus see that clocks within the QSNET network have the potential to probe unconstrained regions of parameter space in dark-energy-type models involving scalar fields.
Meanwhile, the most stringent limit on $d_e^{(1)}$ via searches for equivalence-principle-violating forces comes from the MICROSCOPE mission \cite{MICROSCOPE:2017A} and is at the level $|d_e^{(1)}| \lesssim 10^{-4}$~\cite{Hees:2018DM,MICROSCOPE:2017B}. The quantity $d\ln(\alpha)/dt$ and parameter $d_e^{(1)}$ can be related within specific scalar-field models; e.g., in the model considered in Refs.~\cite{Veneziano:2002A-quint,Veneziano:2002B-quint}, $|d\ln(\alpha)/dt|$ at the present day can vary from $\sim 10^{-17}~\textrm{yr}^{-1}$ to $\sim 10^{-23}~\textrm{yr}^{-1}$ when taking into account the current MICROSCOPE bound. We thus see that clocks within the QSNET network have the potential to probe unconstrained regions of parameter space in dark-energy-type models involving scalar fields. 

\begin{table}
\centerline{\begin{tabular}{|c |c| c|}
\hline
 Measurement type  &   $|d\ln(\alpha)/dt|/\textrm{yr}$  & Reference         \\
\hline
Yb$^+$ clocks    & $\sim 10^{-18}$       & \cite{Lange21} \\
Oklo phenomenon      & $\sim 10^{-17}$    & \cite{Shlyakhter:1976Oklo,Dyson:1996Oklo,Fujii:2000Oklo,Petrov:2006Oklo} \\
Meteorite dating     & $\sim 10^{-16}$    & \cite{Olive:2002meteorites} \\
MICROSCOPE (indirect limits)         & $\sim 10^{-17} - 10^{-23}$       & \cite{Hees:2018DM,MICROSCOPE:2017A,MICROSCOPE:2017B} \\
\hline
\end{tabular}}
\caption{Summary of bounds on drifts of the electromagnetic fine structure constant $\alpha$ and indirect limits from searches for equivalence-principle-violating forces in the context of the model considered in \cite{Veneziano:2002A-quint,Veneziano:2002B-quint}.}
\label{tab:quintessence_drifts_limits}
\end{table}

It is worth mentioning however that there is already some tension between quintessence and constraints from the E\"ot-Wash and MICROSCOPE experiments. Indeed, quantum gravity predicts that linear coupling constants of order one will be generated \cite{Carroll:1998zi} between a quintessence field and visible matter. Linear couplings of the quintessence field with  $d^{(1)}_e$ or $d^{(1)}_{m_e}$ of order one are ruled out by E\"ot-Wash and MICROSCOPE for a field of mass $10^{-33}$ eV. In that sense, QSNET clocks will provide a direct test of quantum gravity and quintessence.

%%%%%%%%%%%%%%%%%%%%%
\subsection{Solitons}
\label{Sec:solitons}

Under certain conditions, spinless bosons may form structured ``dark objects'', such as solitons. Solitons can be either topological or non-topological in nature. Topological solitons are made up of one or more fields that acquire stability due to the presence of two or more vacua, which are energetically equivalent but topologically distinguishable (e.g., due to a difference in the sign or overall phase associated with a field at the positions of the vacua).  Such objects may be produced during a cosmological phase transition \cite{Vilenkin:1985defects}. Topological solitons may arise in a variety of dimensionalities, namely: zero-dimensional monopoles \cite{Hooft:1974monopole,Polyakov:1996monopole}, one-dimensional strings \cite{Abrikosov:1957string,Nielsen:1973string} and two-dimensional domain walls \cite{Zeldovich:1975wall}. Monopoles, being practically pressureless, are a good candidate to explain the observed dark matter, while strings and domain walls may only comprise a sub-dominant fraction of the dark components \cite{Spergel:1989walls,Urrestilla:2008strings}. A notable example of a non-topological soliton is the Q-ball \cite{Sirlin:1976Qball,Coleman:1985Qball}, a monopole-like soliton that is a good candidate for dark matter. 

Recent interest in utilising spatially-separated clocks to study dark sector phenomena has mainly focused on macroscopic scalar-field topological domain walls \cite{Derevianko:2014TDM,Roberts17,Wcislo18,Roberts20,Stadnik:2020TDM,Wcislo:2016TDM}. The simplest model admitting topological domain walls involves a single real scalar field $\phi$ with the following $\phi^4$ potential (see, e.g., \cite{Derevianko:2014TDM,Stadnik:2020TDM,Zeldovich:1975wall}): 
\begin{ceqn}
\begin{equation}
\label{DW_quartic_potential}
V ( \phi ) = \frac{\lambda}{4} ( \phi^2 - \phi_0^2 )^2  \, , 
\end{equation}
\end{ceqn}
where $\lambda$ is a dimensionless parameter. The potential in Eq.~(\ref{DW_quartic_potential}) admits two energetically equivalent, but topologically distinct minima at $\phi = \pm \phi_0$, separated by a potential barrier of height $\lambda \phi_0^4 / 4$. The $Z_2$ symmetry associated with the potential (\ref{DW_quartic_potential}) is spontaneously broken, since the vacuum states are not invariant under the $\phi \to -\phi$ transformation. If there exist two spatially separated regions of space with topologically distinct vacua, then a domain wall forms between the two vacua, with the following transverse ``kink'' profile (in the rest frame of the wall) \cite{Zeldovich:1975wall}: 
\begin{ceqn}
\begin{equation}
\label{DW_kink_solution}
\phi (x) = \phi_0 \tanh (x/d)  \, , 
\end{equation}
\end{ceqn}
where the transverse size of the wall is set by $d = \sqrt{2 / ( \lambda \phi_0^2)} = 1/m_{\phi,\textrm{eff}}$ ($m_{\phi,\textrm{eff}}$ is the effective scalar mass) and may in principle range in size from the microscopic scale up to a sizeable fraction of the observable universe. The regions on either side of the wall are referred to as domains, by analogy with the familiar ferromagnetic domains in condensed matter physics. The energy density inside a wall is given by $\rho_\textrm{inside} \sim \phi_0^2 / d^2$. For a network of walls with an average energy density of $\rho_\textrm{walls}$ and an average separation between adjacent walls of $L = v_\textrm{wall} \mathcal{T}$, there is a simple relation between the domain-wall parameters: 
\begin{ceqn}
\begin{equation}
\label{DW_parameters_relation}
\phi_0^2 \sim \rho_\textrm{walls} v_\textrm{wall} \mathcal{T} d  \, , 
\end{equation}
\end{ceqn}
where $v_\textrm{wall}$ is the typical speed of passage of domain walls through Earth and $\mathcal{T}$ is the average time between encounters of a wall with Earth. Domain walls with the potential (\ref{DW_quartic_potential}) have large spatial components in the associated energy-momentum tensor that give significant deviations from the equation of state for non-relativistic matter; numerical simulations indicate that such domain walls travel at semi-relativistic speeds $v_\textrm{wall} \sim c$ \cite{Spergel:1989walls,Avelino:2005wallsA,Avelino:2005wallsB}. Furthermore, the consideration of the gravitational effects of domain walls on photons originating from the cosmic microwave background constrains the present-day energy density stored in a network of domain walls to less than $\sim 10^{-5}$ times that of the present-day critical density of the universe \cite{Stadnik:2020TDM,Spergel:1989walls,Planck:2018};  i.e., $\sim 10^{-10}~\textrm{GeV/cm}^3$ or $\sim 10^{-10}$ times the local Galactic dark-matter energy density. Therefore, domain walls cannot account for all of the dark matter. 

If the scalar field $\phi$ interacts non-gravitationally with fields from the Standard Model sector, then there may be characteristic observable signatures in terrestrial experiments. Due to the smallness of the maximally allowed value of $\rho_\textrm{walls}$ and hence $\phi_0$, the scalar-type linear-in-$\phi$ interactions in Eq.~(\ref{linear_DM_interactions}) are strongly constrained by traditional searches for equivalence-principle-violating forces which do not depend on $\phi_0$ \footnote{Note that the situation in the case of an oscillating dark-matter scalar field of the type discussed in Sec.~\ref{Sec:DM} is different, since $\phi_0$ can be many orders of magnitude larger in that case.}. Therefore, we focus on the scalar-type quadratic-in-$\phi$ interactions in Eq.~(\ref{quadratic_DM_interactions}), which were previously considered in Refs.~\cite{Derevianko:2014TDM,Roberts17,Wcislo18,Roberts20,Stadnik:2020TDM,Wcislo:2016TDM}. Comparing the interactions in Eq.~(\ref{quadratic_DM_interactions}) with the corresponding terms in the Standard Model Lagrangian, $\mathcal{L}_\textrm{Standard Model} \supset - F_{\mu \nu} F^{\mu \nu} / 4 - e J_\mu A^\mu - m_e \bar{\psi} \psi$, we see that the apparent values of $\alpha$ and the electron mass are given by: 
\begin{ceqn}
\begin{equation}
\label{quadratic-in-phi_VFCs-TDM}
\alpha (\phi^2) \approx \alpha_0 \left[ 1 + \left( \frac{\phi}{\Lambda'_\gamma} \right)^2 \right]
 , \quad 
m_e (\phi^2) = m_{e,0} \left[ 1 + \left( \frac{\phi}{\Lambda'_e} \right)^2 \right]
 \, , 
\end{equation}
\end{ceqn}
where the subscript `0' refers to the local fundamental constant value when $\phi = 0$. The passage of a domain wall, e.g., with the transverse profile in (\ref{DW_kink_solution}), through a point or region of space is expected to induce transient changes in the apparent values of the fundamental constants  (the values of the fundamental constants are the same prior to and after the passage of a domain wall, and differ only in the central region of a wall during the wall's passage), which can be sought with clock- \cite{Derevianko:2014TDM,Roberts17,Roberts20} and cavity-based \cite{Stadnik:2015DM-LI,Wcislo18,Stadnik:2016cavity,Wcislo:2016TDM} measurements, ideally using a network of spatially-separated detectors. 

For a sufficiently small detector, the signal duration is given by $\Delta t \sim d / v_\textrm{wall}$, if a wall passes through the detector (and Earth) in an unperturbed manner. However, it has been pointed out in \cite{Stadnik:2020TDM} that the interactions in Eq.~(\ref{quadratic_DM_interactions}) cause ordinary matter (e.g., within Earth, Earth's atmosphere, and the apparatus itself) to create a repulsive potential that may affect the propagation of domain walls near bodies of ordinary matter. The precise outcome for a domain wall incident on a strongly repulsive potential requires further detailed investigation; for the purposes of the present discussion, we make the simple assumption adopted in Refs.~\cite{Derevianko:2014TDM,Roberts17,Wcislo18,Roberts20,Wcislo:2016TDM}, namely that the passage of a wall proceeds in an unperturbed fashion. When a domain wall of cosmological origin is far away from Earth, the ``back action'' of ordinary matter inside Earth on the scalar field via Eq.~(\ref{quadratic_DM_interactions}) induces quasi-static apparent variations of the fundamental constants with height above Earth's surface, which can be sought with clock comparison measurements at different heights; see \cite{Stadnik:2020TDM} for details. 

%%%
\begin{figure*}[t!]
	\centering
	\includegraphics[width=0.8\textwidth]{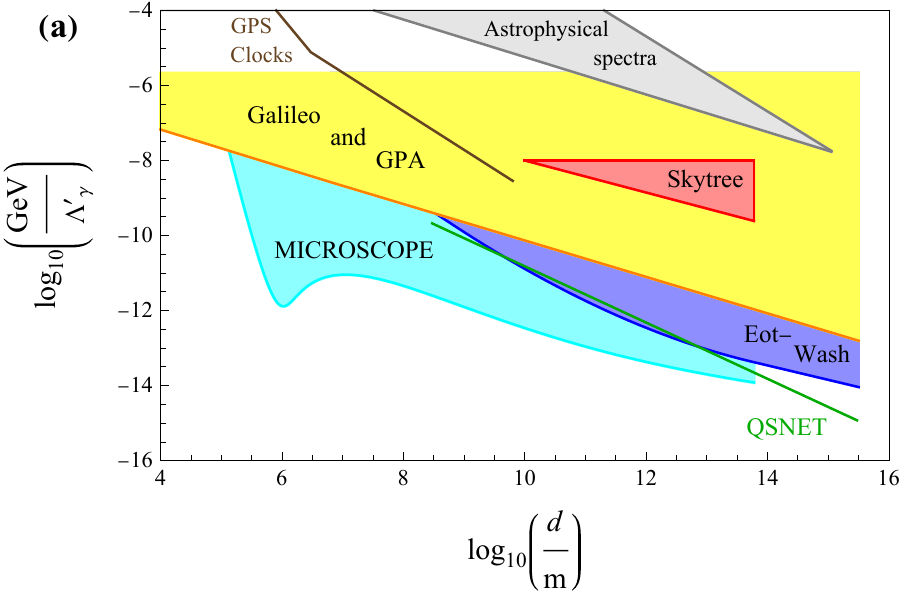}
    \par \medskip
	\includegraphics[width=0.8\textwidth]{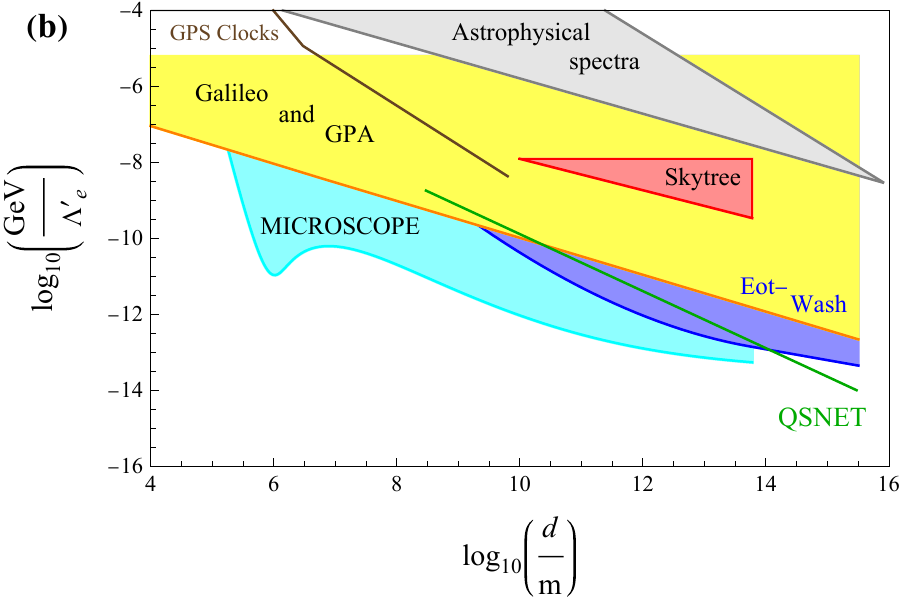}
	\caption{
	Parameter spaces for a model of a domain-wall scalar field with the potential (\ref{DW_quartic_potential}), interacting with \textbf{(a)} the electromagnetic field and \textbf{(b)} the electron via the quadratic-in-$\phi$ interactions in Eq.~(\ref{quadratic_DM_interactions}). We denote by $d$ the size of a domain wall of cosmological origin and $\Lambda'$ the energy scale associated with the quadratic-in-$\phi$ interactions between the scalar field and ordinary matter. The estimated sensitivities of clocks within the QSNET network to apparent transient variations of the fundamental constants induced by the passage of domain walls are shown by the green lines. Also shown are existing limits from previous searches for transient variations of the fundamental constants using the GPS network of clocks (brown lines) \cite{Roberts17}, and searches for quasi-static apparent variations of the fundamental constants with height above Earth's surface using clock comparison measurements at different heights (red and yellow regions) and accelerometers (blue and cyan regions) as well as comparisons of atomic and molecular spectra in ground-based laboratory and low-density astrophysical environments (light grey regions) \cite{Stadnik:2020TDM}. See the main text for more details. 
	}
\label{Fig:TDM_exclusion_plots}
\end{figure*}

In Figure~\ref{Fig:TDM_exclusion_plots}, we present the estimated sensitivity of the QSNET network to apparent transient variations of the fundamental constants induced by the passage of domain walls, assuming that data are continuously collected every $\sim 1~\textrm{s}$ over the course of $1~\textrm{year}$. We assume that domain walls make up a maximally allowable fraction of the dark components, which is $\sim 10^{-5}$ at the present day. We further assume that the average time between encounters of a wall with Earth is given by $\mathcal{T} = 1~\textrm{year}$, with adjacent walls well-separated ($\Delta t \ll \mathcal{T}$), that domain walls propagate at semi-relativistic speeds $v_\textrm{wall} \sim c$, and that the back-action effects of Earth and the apparatus on the incident domain walls can be neglected (transient signals may be qualitatively different when back-action effects are strong \cite{Stadnik:2020TDM}). 

For comparison, we also show limits from previous searches for transient variations of the fundamental constants using the GPS network of clocks \cite{Roberts17}, and searches for quasi-static apparent variations of the fundamental constants with height above Earth's surface using clock comparison measurements at different heights and accelerometers as well as comparisons of atomic and molecular spectra in ground-based laboratory and low-density astrophysical environments \cite{Stadnik:2020TDM}. Note that we have rescaled the limits in \cite{Roberts17} and \cite{Stadnik:2020TDM} to account for differences in the assumed values of $\rho_\textrm{walls}$ and $v_\textrm{wall}$ compared with the present article.

Our analysis shows that the QSNET network has the potential to probe unexplored regions of parameter space in domain-wall scalar-field models, specifically in regions of parameter space where the back-action effects of Earth and the apparatus are negligible \cite{Stadnik:2020TDM}. We note that QSNET may have more extensive reach in other models of solitons, such as monopoles \cite{Hooft:1974monopole,Polyakov:1996monopole,Sirlin:1976Qball,Coleman:1985Qball}; these possibilities require further detailed study. Additionally, exploiting the network configuration, QSNET could provide information about the speed and direction of propagation of the domain walls or other dark objects.

%%%
\subsection{Violation of fundamental symmetries}
\label{Sec:symmetries}

Fundamental symmetries are central concepts and guiding principles in physics. The study of space-time symmetries dates back hundreds of years to the insights of Galileo, Newton, and other contemporaries,
and forms the foundations of modern particle and gravitational theories. Lorentz invariance is a space-time symmetry at the heart 
of the Standard Model and General Relativity. It roughly states that 
physical laws are independent of the relative orientation 
and velocity of an experiment in space-time. As a consequence, 
the measurement outcomes of two otherwise identical experiments of 
distinct space-time orientation must be based on the same laws of physics, with results connected by the Lorentz transformation. 

Space-time symmetries have been studied in a number of new-physics scenarios, including string-based approaches \cite{Kostelecky:1988zi,Kostelecky:1994rn,Kostelecky:1991ak,Kostelecky:1995qk,Ellis:2008gg,Gliozzi:2011hj,Hashimoto:2012ig}, loop quantum gravity \cite{Gambini:2014kba,Rovelli:2010ed}, noncommutative field theory \cite{Carroll:2001ws,Carlson:2001sw,Calmet:2004ii,Calmet:2004dn,Bailey:2018ifc}, high-energy electrodynamics \cite{Carroll:1989vb, Coleman:1998ti}, and modified-gravity theories \cite{Kostelecky:2020hbb,Kostelecky:2021xhb,Kostelecky:2021tdf,deRham:2014zqa,Horava:2009uw,Bluhm:2004ep,Bluhm:2007bd,Bluhm:2014oua}. A subset of these works suggest Lorentz-violating effects may exist and be detectable in experiments with exceptional sensitivity, including those of QSNET. Indeed, recent years have witnessed a significant expansion in experimental searches for small violations of Lorentz invariance involving nearly every subfield of physics \cite{datatables}.
If such effects do exist, current constraints
suggest they are quite small. Given the present absence of any Lorentz-violating signal, a prudent approach is to use model-independent methods based on effective field theory (EFT) \cite{Weinberg:2009bg}. In the context of Lorentz violation, this framework exists and is known as the Standard-Model Extension (SME) \cite{Colladay:1996iz,Colladay:1998fq,Kostelecky:2003fs}\footnote{For accessible reviews of the SME see, e.g., Refs.~\cite{Bluhm:2005uj,Tasson:2014dfa}.}. The SME action includes all known physics in addition to possible Lorentz-violating and invariant terms,
\begin{ceqn}
\begin{equation}
\label{SMEaction}
S_{\text{SME}} = S_{\text{SM}} + S_{\text{EH}} + S_{\text{LV}},
\end{equation}
\end{ceqn}
where $S_{\text{SM}}$ is the Standard Model action, $S_{\text{EH}}$ is the Einstein-Hilbert action, and $S_{\text{LV}} \ll S_{\text{SM}}, S_{\text{EH}}$ includes in principle an infinite number of terms in EFT consistent with the choice of fields and preserved symmetries. Note that since CPT violation implies Lorentz violation in EFT, all possible CPT-violating operators in EFT are included in the SME by construction \cite{Colladay:1998fq}. Also note that violations of other fundamental symmetries, such as the weak equivalence principle, have also been developed within the SME framework \cite{Kostelecky:2010ze}. See, e.g., Ref.~\cite{Mewes:2020pvv} for a recent account of several applications. An example term modifying conventional quantum electrodynamics (QED) is 
\begin{ceqn}
\begin{equation}
\label{Lc}
\mathcal{L}_{c} \supset \tfrac{1}{2}c_{\mu\nu}\bar{\psi}\gamma^{\mu}
i\overset{\text{\tiny$\leftrightarrow$}}{\partial^{\nu}}\psi. 
\end{equation}
\end{ceqn}
The SME coefficient $c_{\mu\nu}$ is coupled to the fermion $\psi$ and controls the strength of Lorentz violation. In general, the inclusion of a nonzero $c_{\mu\nu}$ results in rich phenomenological signatures, including modified kinematic effects, quantum corrections, and shifts in atomic spectra \cite{Colladay:1996iz,Jackiw:1999yp,Bluhm:1998rk}. 
%\cite{Kostelecky:2010ze,Pihan-LeBars:2019qsd,Altschul:2014lua,Flambaum:2016tsn,Hohensee:2013cya,Hohensee:2013hra}.

Clock-comparison experiments\footnote{The phrase `clock-comparison experiment' originated in Ref.~\cite{Kostelecky:1999mr} and describes a broad class of experiments with atoms and ions. Experiments based on atomic clocks are a particular type of clock-comparison experiment.}, including those based on atomic clocks, are ideal systems for precision tests of Lorentz violation \cite{Kostelecky:1999mr,Bluhm:2001rw,Kostelecky:2018fmc,Vargas:2019swg}. Since these experiments involve comparatively low energies, operators
of lowest mass dimension $d = 3, 4$  involving the free propagation of electrons and nucleons are expected to represent the dominant experimental signals. This describes the extension of the free QED Lagrange density 
\begin{ceqn}
\begin{align}
\label{smeqed}
\mathcal{L} \supset \tfrac{1}{2}\bar{\psi}\Gamma_{\nu}
i\overset{\text{\tiny$\leftrightarrow$}}{\partial^{\nu}}\psi - 
\bar{\psi}M\psi,
\end{align}
\end{ceqn}
where $\psi = \{\psi_e, \psi_p, \psi_n\}$ stands for an electron, proton, or neutron field. The generalised kinetic and mass matrices are 
\begin{ceqn}
\begin{align}
\label{fermionterms}
& \Gamma_{\nu} = \gamma_{\nu} + c_{\mu\nu}\gamma^{\mu} + 
d_{\mu\nu}\gamma_{5}\gamma^{\mu} + e_{\nu} + if_{\nu}
\gamma_{5} + \tfrac{1}{2}g_{\lambda\mu\nu}\sigma^{\lambda\mu}, \nonumber\\
& M = m + a_{\mu}\gamma^{\mu} 
+ b_{\mu}\gamma_{5}\gamma^{\mu} + \tfrac{1}{2}H_{\mu\nu}\sigma^{\mu\nu},
\end{align}
\end{ceqn}
where the first two terms on the right-hand side of each equation --- $\gamma_\nu$ and $m$ --- are the conventional four-dimensional gamma matrices and fermion mass, respectively. Note the inclusion of the $c_{\mu\nu}$ coefficient from Eq.~\eqref{Lc}. The coefficients $c_{\mu\nu}, d_{\mu\nu}, e_\nu, f_\nu, g_{\lambda \mu\nu}$ have mass dimension zero and the coefficients $a_\mu, b_\mu, H_{\mu\nu}$ have mass dimension one. By convention, the units for SME coefficients of nonzero mass dimension are chosen to be powers of GeV. Additional technical details are provided in Appendix~\ref{ssec:fundsymmsdetails}.

Extractions of Lorentz-violating signals from atomic-clock tests have been performed with both nonrelativistic and relativistic methods. The starting point in either case is constructing the relativistic Hamiltonian stemming from Eq.~\eqref{smeqed}. For especially light systems, e.g. H, $\overline{\text{H}}$, He, Li, the nonrelativistic approach based on the Foldy-Wouthuysen sequence is expected to capture the dominant physics, where the relativistic-in-origin SME coefficients are treated as small corrections amenable to standard techniques of perturbation theory \cite{Foldy:1949wa,Kostelecky:1999zh}. Carrying out this procedure results in the full Lorentz-violating perturbation $\delta h_{\text{LV}}$ detailed in Appendix~\ref{ssec:fundsymmsdetails} \cite{Kostelecky:1999mr}. Of these renormalizable effects, the electron-sector $c$-type coefficients first introduced in Eq.~\eqref{Lc} represent a prime example of sensitive perturbations to atomic-clock transitions\footnote{Note that this class of fermion coefficients cannot be separated from the minimal CPT-even photon coefficients, as explained in Appendix~\ref{ssec:fundsymmsdetails}.}. The relevant
perturbing Hamiltonian for a bound electron of momentum $\vec{p}$ reads \cite{Kostelecky:2010ze,Hohensee:2013cya,Hohensee:2011wt}
\begin{ceqn}
\begin{equation}
\label{toth}
\delta h_{\text{LV}} = -\left(C_0^{(0)} - \frac{2U}{3c^2}c_{00}\right)\frac{\vec{p}^2}{2m_e} - \sum_{q=-2}^{2}\frac{(-1)^q}{6m_e}C_q^{(2)}T_{-q}^{(2)},
\end{equation}
\end{ceqn}
where $C_0^{(0)}, C_q^{(2)}$ are linear combinations of $c_{\mu\nu}$ matrix elements, $T_{-q}^{(2)}$ are spherical tensor operators, $U$ is the Newtonian gravitational potential and $c$ the speed of light.
The shifted spectra are obtained by calculating the matrix elements of $\delta h_{\text{LV}}$ between the unperturbed atomic states of interest. In most applications the total atomic angular momentum $F$ and its spin projection $m_F$ are conserved and may be used to label the states. The projection quantum number $m_F$ typically defines the laboratory $z$ axis and is identified with the direction of a uniformly applied external magnetic field. In this scenario, the conventional energy levels are shifted by 
\begin{ceqn}
\begin{equation}
\label{deltaE}
\delta E_{\text{LV}} = \braket{F,m_F|\delta h_{\text{LV}}|F,m_F}.
\end{equation}
\end{ceqn}
The shifted laboratory frequencies are simply proportional to the difference between the relevant energy shifts. Note that in general, and especially for systems based on heavy atoms or ions where relativistic corrections can be large, intrinsically relativistic methods, starting from solving the Dirac-Hartree-Fock equations,  are used to obtain the electronic wave functions used in Eq.~(\ref{deltaE}) to accurately capture the effects of the $c$-type coefficients  \cite{Dzuba:2015xva,SAFRONOVA2008191,Shaniv:2017gad,Dzuba:2016aiy,Pruttivarasin:2014pja,Hohensee:2013cya,Harabati:2015xea}. Relativistic analogues of the momentum-space matrix elements, e.g. $\braket{\vec{p}^2} \rightarrow \braket{c\gamma^0\gamma^j p_j}$ have also been calculated, though the differences thus far considered were found to be negligible  \cite{Pruttivarasin:2014pja,Dzuba:2015xva}. Additional studies have recently extended some of these approaches into the nonrenormalizable $d\geq 5$ sector \cite{Kostelecky:2018fmc,Vargas:2019swg}. 

A variety of clock-comparison experiments have produced leading constraints on fermion-sector SME coefficients. As detailed in Sec.~\ref{Sec:clock_network} and Appendix~\ref{Appendix:QSNET_clocks}, the established standards of QSNET are based on a Cs fountain clock and Sr and Yb$^+$ optical clocks. Systems based on a subset of these types of clocks currently hold leading constraints on minimal Lorentz violation affecting electrons and nucleons. Caesium parity-violation experiments have constrained the timelike component of the electron-sector $b$-type coefficient $|b_0| < 2\times 10^{-14}$\;GeV \cite{Roberts:2014dda,Roberts:2014cga}. In the nucleon sector, similar techniques have placed constraints $|b_0| \lesssim 10^{-8}$--$10^{-7}$\;GeV \cite{Roberts:2014dda,Roberts:2014cga,Stadnik:2014xja}, in addition to the timelike $d$-type coefficient $d_{00} \lesssim 10^{-8}$--$10^{-7}$ \cite{Roberts:2014cga,Stadnik:2014xja}. In the proton sector, fountain-based Cs clocks have stringently constrained linear combinations of all components of the $c$-type coefficients at levels $\lesssim 10^{-25}$--$10^{-16}$ \cite{Wolf:2006uu,Pihan-LeBars:2016pjg,Bars:2017yil}. Optical clocks based on ytterbium have also produced the current best limits $\lesssim 10^{-21}$--$10^{-16}$ on the electron sector $c$-type coefficients \cite{Sanner:2018atx}. Several other competitive constraints have been placed on minimal SME coefficients using a wide variety of clock-comparison experiments \cite{LCPT1,LHunterLCPT1,LCPT4,KornackLCPT4,Berglund:1995zz,Dzuba:2016aiy,Megidish:2018sey,Pruttivarasin:2014pja,Botermann:2014wua,Matveev:2013orb,Hohensee:2013cya,Hohensee:2011wt,Muller:2003qgj,Muller:2004zp,Muller:2007zz,Stadnik:2014xja,Peck:2012pt,Brown:2010dt,Humphrey:2001wm,Phillips:2000dr,Smiciklas:2011xq,Flambaum:2016dwc,Flambaum:2016tsn,Bars:2017yil,Allmendinger:2013eya,Gemmel:2010ft,KTullney,Altarev:2009wd,Flambaum:2009mz,Altschul:2009iz,Cane:2003wp}. In addition, the first constraints in the nonminimal $5 \leq d \leq 8$ electron and nucleon sectors were recently placed using H masers and the 1S-2S transition \cite{Kostelecky:2015nma} and H, $\overline{\text{H}}$ spectroscopy \cite{Kostelecky:2018fmc}.

The Yb$^+$ E3 clock has the highest sensitivity to Lorentz violation among all presently operating clocks, with the reduced matrix element of the $T^{(2)} $ operator $|\langle J\| T^{(2)}\| J\rangle|=145$~a.u. (atomic units) for the upper clock state \cite{ Shaniv:2017gad}. Cf$^{15+}$ and Cf$^{17+}$ have similar sensitivities, with respective matrix elements of  112~a.u. and 144~a.u. for ions in the ground state \cite{ Shaniv:2017gad}.  There are two types of measurements that the QSNET network can perform to search for Lorentz invariance violation in the electron-photon sector. First, one can follow an approach of the PTB Yb$^+$ Lorentz-violation experiment, comparing frequencies of two co-located Yb$^+$ clocks with different magnetic field orientations \cite{Sanner2019}. One has to investigate if such a scheme can be adapted for two different clocks. Such a method has the advantage of using usual clock-comparison metrological protocols. One can use clock-comparison data obtained for the dark matter searches, for example. However, the limits set by PTB already used a clock comparison at the $10^{-18}$ level, and higher accuracy will be required for an improvement of $c_{IJ}$ and $c_{TJ}$ coefficients (indices $T, I, J$ denote Sun-centred frame indices---see Appendix~\ref{ssec:fundsymmsSCF}). The $c_{TT}$ coefficient was not considered in \cite{Sanner2019} and significant improvement is possible. In addition, one can constrain nonminimal coefficients in such experiments. In the second class of experiments, one uses a dynamic decoupling proposal of \cite{ Shaniv:2017gad} to monitor the splitting between different Zeeman substates as Earth rotates around its axis and around the Sun placing a bound on $C_0^{(2)}$. This method would use Zeeman multiplets of either upper Yb$^+$ clock states or ground states of Cf highly charged ion clocks. We note that this method does require operating a Cf HCI clock (no need for a clock laser) but only the ability to carry out the dynamic decoupling sequence for the ground state, which does not involve optical transitions. This method can drastically, by orders of magnitude, improve Lorentz-violation tests for all $c_{\mu\nu}$ coefficients for electrons.

Atomic-clock experiments have demonstrated exceptional sensitivity to electron- and proton-sector Lorentz violation. As described in Appendix~\ref{ssec:fundsymmsSCF}, the SME coefficients depend on the choice of observer frame. This implies that every experiment is sensitive to a unique linear combination of SME coefficients and that dedicated studies investigating the performance of QSNET relative to existing experiments must ultimately be performed. As discussed in Sec.~\ref{Sec:theory}, since space-time variations of $\alpha$ have been associated with violations of Lorentz and CPT invariance \cite{Kostelecky:2002ca,Bertolami:2003qs,Ferrero:2009jb}, it is conceivable that increased constraints on $\alpha$ variations from QSNET could be translated into new constraints on Lorentz violation.
To summarise, given the existing capabilities and projected limits of QSNET described in Sec.~\ref{Sec:predicted_performance}, it is reasonable to suggest new, competitive and potentially leading constraints on violations of fundamental symmetries are within reach. 

\subsection{Tests of unification and quantum gravity}
\label{Sec:unifiedtheory}

\subsubsection{Tests of unification}
A discovery of a time variation of $\mu$ or $\alpha$ could be used to probe very high energy theories such as models of grand unification \cite{Fritzsch:1974nn,Georgi:1974sy}. Grand unified theories  are a natural extension of the Standard Model. The idea that all the forces of nature can be unified in one fundamental force is very attractive to theoretical physicists, as such models have the potential to reduce the number of fundamental constants in the model.

To define a grand unified theory, we need to decide which unification group to consider. Well studied examples of gauge groups are, e.g., SU(5), SO(10) or E$_6$, but other groups are possible. The minimal requirement for such a group is that the Standard Model gauge groups SU(3)$\times$SU(2)$\times$U(1) can be embedded in the grand unified theory group. The coupling constants of the gauge groups of the standard model, $\alpha_1$ for U(1), $\alpha_2$ for SU(2), and $\alpha_3$ for SU(3), are assumed to all reach the same value $\alpha_u$ (the unified coupling constant) at some unification scale $\Lambda_u$. At energies larger than $\Lambda_u$, the gauge symmetries of the unification group are manifest, while those of the Standard Model are manifest at energies below $\Lambda_u$. Besides the gauge group, one must decide which Higgs field and fermion representations to introduce in the model and how to couple the Higgs fields to fermions or themselves. This generically introduces Higgs boson masses and Yukawa couplings and thus a number of fundamental constants.

Measurements of variations of $\mu$ can be used to probe grand unified theories~\cite{Calmet:2001nu,Calmet:2002ja,Calmet:2002jz,Langacker:2001td,Campbell:1994bf,Olive:2002tz,Dent:2001ga,Dent:2003dk,Landau:2000cc,Wetterich:2003jt,Flambaum:2006ip,Calmet:2014qxa}. Ignoring possible cosmological time variations of Yukawa couplings and of Higgs boson masses \cite{Calmet:2017czo}, and working at the one loop level,  we only have two parameters: the unification scale $\Lambda_u$ and  the unified coupling constant $\alpha_u$. As  the proton mass is mainly determined by the QCD scale, quark masses can be neglected. We  focus on the QCD scale $\Lambda_{\rm QCD}$ and extract its value from the Landau pole of the renormalization group equation for the QCD coupling constant:
\begin{ceqn}
\begin{eqnarray} \label{running}
  \alpha^{\rm SM}_3(\mu_R)^{-1}
 &=&
\frac{1}{\alpha^{\rm SM}_3(\Lambda_u)}+\frac{1}{2 \pi}
  b^{\rm SM}_3   \ln
  \left ( \frac{\Lambda_u}{\mu_R} \right) 
  \end{eqnarray}
\end{ceqn}  
where the parameter $b^{\rm SM}_3=-7$ in the Standard Model and $\mu_R$ is the renormalization scale. The QCD scale, i.e., the energy scale below which the SU(3) interactions are strong is defined by
\begin{ceqn}
\begin{eqnarray}
\Lambda_{\rm QCD}= \Lambda_u \exp\left(\frac{2 \pi}{\alpha_u}\right)^{1/{b^{\rm SM}_3}}\, .
\end{eqnarray}
The time variation of $\Lambda_{\rm QCD}$ is then determined by
\begin{eqnarray}
\frac{\dot \Lambda_{\rm QCD}}{\Lambda_{\rm QCD}}= -\frac{2\pi}{b^{\rm SM}_3} \frac{\dot \alpha_u}{\alpha_u^2}
+ \frac{\dot \Lambda_u}{\Lambda_u}\, ,
\end{eqnarray}
\end{ceqn}
and one can see that a time variation of the QCD scale could be due to either a time variation of the unification scale or of the unified coupling constant. For constant quark and electron masses this equation determines the ratio:
\begin{ceqn}
\begin{eqnarray} \label{result}
\frac{\dot \mu}{\mu}= \frac{2\pi}{b^{\rm SM}_3} \frac{\dot \alpha_u}{\alpha_u^2}
-\frac{\dot \Lambda_u}{\Lambda_u}\, .
\end{eqnarray}
\end{ceqn}
 
The running of the three coupling constants $\alpha_i$ of the Standard Model are given by  
\begin{ceqn}
\begin{eqnarray} \label{RGrun}
  \frac{1}{\alpha_i} \frac{\dot{\alpha}_i}{\alpha_i}=
  \frac{1}{\alpha_u} \frac{\dot{\alpha}_u}{\alpha_u} 
 - \frac{b_i}{2 \pi} \frac{\dot{\Lambda}_u}{\Lambda_u}
  \, ,
\end{eqnarray}
\end{ceqn}
where $b_i =(b^{{\rm SM}}_1, b^{{\rm SM}}_2, b^{{\rm SM}}_3)=(41/10,
-19/6, -7)$ are the coefficients of the renormalization group
equations for the Standard Model. This leads to the following relation for the fine structure constant  \cite{Calmet:2001nu,Calmet:2002ja,Calmet:2002jz}
\begin{ceqn}
\begin{eqnarray} \label{eq3A}
  \frac{1}{\alpha} \frac{\dot{\alpha}}{\alpha}= \frac{8}{3}
\frac{1}{\alpha_3} \frac{\dot{\alpha}_3}{\alpha_3} -\frac{1}{2 \pi}
\left(b_2+\frac{5}{3} b_1-\frac{8}{3} b_3\right)
\frac{\dot{\Lambda}_u}{\Lambda_u}\, .  
\end{eqnarray}
\end{ceqn}

In the supersymmetric extension of the Standard Model, the coefficients $b_i$ in Eqs. (\ref{RGrun}) need to be replaced by $b^{{S}}_i\!\!\!=(b^{{S}}_1, b^{{S}}_2, b^{{S}}_3)=(33/5, 1, -3)$ which are the coefficients of the renormalization group equations in the ${\cal N}=1$ supersymmetric case. 

SU(5) grand unification models require us to introduce supersymmetry between the weak scale and the unification scale to obtain a numerical unification of the gauge couplings of the Standard Model at the unification scale. 

One may consider different scenarios. First, we keep $\Lambda_u$ invariant and consider the case where $\alpha_u =\alpha_u (t)$ is time dependent. One then gets
\begin{ceqn}
\cite{Calmet:2001nu}
\begin{eqnarray}
\frac{\dot{\Lambda}_{\rm QCD}}{\Lambda_{\rm QCD}}= -\frac{3}{8} \frac{2 \pi}{b_3^{\rm SM}}
  \frac{1}{\alpha} \frac{\dot{\alpha}}{\alpha}= R\,
  \frac{\dot{\alpha}}{\alpha}\, .
\end{eqnarray}
\end{ceqn}
If we calculate $\dot{\Lambda}_{\rm QCD}/\Lambda_{\rm QCD}$ using the relation above in the case of 6 quark flavors, neglecting the masses of the quarks, we find $R \approx 46$. There are large theoretical uncertainties in $R$. Taking thresholds into account one gets $R=37.7\pm 2.3$ \cite{Calmet:2001nu}. The uncertainty in $R$ is given, according to $\Lambda_{\rm QCD} = 213^{+38}_{-35}~{\rm MeV}$, by the uncertainty in the ratio $\alpha/\alpha_s$, which is dominated by the uncertainty in $\alpha_s$.

We could alternatively consider the case where $\alpha_u$ is invariant but $\Lambda_u=\Lambda_u (t)$ is time-dependent. One gets \cite{Calmet:2002ja}
\begin{ceqn}
\begin{eqnarray}
\frac{\dot{\Lambda}_{\rm QCD}}{\Lambda_{\rm QCD}}=
\frac{b_3^{S}}{b_3^{\rm SM}} \left[ \frac{-2 \pi}{b_2^S+\frac{5}{3}
b_1^S}\right] \frac{1}{\alpha} \frac{\dot{\alpha}}{\alpha} \approx
-30.8\,\frac{\dot{\alpha}}{\alpha}\, . 
\end{eqnarray}
\end{ceqn}
It is interesting to note that the effects of a time variation of the unified coupling constant or of a time variation of the grand unified scale are going in opposite directions.  Clearly those are two extreme cases and a time variation of both parameters is conceivable. This could lead to cancellations between the different effects.
 
It should be clear that our results are strongly model-dependent. For
example, in SO(10) without supersymmetry, where a unification of the gauge couplings is possible due to threshold corrections, varying the grand unification scale, one finds \cite{Calmet:2006sc}:
\begin{ceqn}
\begin{eqnarray} 
\frac{\dot{\Lambda}_{\rm QCD} }{\Lambda_{\rm QCD}} 
= \left (\frac{-2 \pi}{b_2^{\rm SM} +\frac{5}{3} b_1^{\rm SM}}\right ) \frac{1}{\alpha}
\frac{\dot{\alpha}}{\alpha} = -234.8\,\frac{\dot \alpha }{\alpha}\, .
\end{eqnarray} 
\end{ceqn}
We thus see that simultaneous measurements of $\dot \mu/\mu$ and $\dot \alpha/\alpha$ would enable us to discriminate between grand unified theories with very low energy experiments. 

The model dependence in grand unification theories is what makes the detection of a possible time variation of the fundamental parameters so interesting. Indeed, QSNET could test grand unified theories without actually detecting any particle from a grand unified model. This is because QSNET will be sensitive to both variation of $\alpha$ and $\mu$, and if a variation is detected for either or both constants, it will be possible to discriminate between grand unified models. 

\subsubsection{Test of quantum gravity}

Quantum gravity will generate interactions between the Standard Model particles and any new particles of a hidden sector, e.g. dark matter particles  \cite{Calmet:2019jyz,Calmet:2019frv,Calmet:2020rpx,Holman:1992us,Barr:1992qq,Kallosh:1995hi}. It is easy to convince oneself that only non-perturbative quantum gravitational effects have the potential to be large enough to be relevant for clocks  as perturbative effects are generically very much smaller than non-perturbative ones \cite{Calmet:2019jyz,Calmet:2019frv}.

For a scalar field, quantum gravity will generate interactions of the type
\begin{ceqn}
\begin{equation}
	\mathcal{L}=\kappa^n  \phi^n \left (\frac{d^{(n)}_e}{4} F_{\mu\nu}F^{\mu\nu} -d^{(n)}_{m_e} m_e \bar \psi_e \psi_e \right )  + \kappa^n  \phi^n \left (\frac{d^{(n)}_g}{4} G_{\mu\nu}G^{\mu\nu} -d^{(n)}_{m_q} m_q \bar \psi_q \psi_q \right),
\end{equation}
\end{ceqn}
where $\kappa=\sqrt{4 \pi G_N}$ whether gauge interactions between the Standard Model and the hidden sector exist or not. The cases $n=1$ and $n=2$ correspond respectively to the linear and quadratic couplings discussed earlier. One expects $d^{(n)}_i\sim 1$ on very generic grounds \cite{Calmet:2019jyz,Calmet:2019frv} as these operators are normalised to the reduced Planck scale $M_P=1/\sqrt{8 \pi G_N}=2.4\times 10^{18}$ GeV which is the scale of quantum gravity. As these operators are generated via nonperturbative effects such as gravitational instantons, wormholes, or quantum black holes  \cite{Perry:1978fd,Gilbert:1989nq,Chen:2021jcb}, there is no further suppression to be expected such as factors of $(16 \pi^2)^{-k}$ due to $k$-loop factors. If the scale suppressing the operators is properly normalised, the Wilson coefficients $d^{(n)}_i$ must be of order one.

This reasoning leads quite generically to a bound on the mass of singlet scalar fields  \cite{Calmet:2019jyz,Calmet:2019frv}. Using data from the E\"ot-Wash experiment, we find $m<10^{-2}$ eV for a coupling $d^{(1)}\sim 1$. It follows that QSNET has the potential to probe quantum gravity: if a very light singlet scalar field was detected with $d^{(1)}\ll 1$, QSNET would have demonstrated that the linear operators discussed above are not generated by quantum gravity.

%%%%%%%%%%%%%%%%%%%%%%%%%%
\section{Summary and Conclusions}
\label{Sec:outlook}

Despite the fact that our understanding of physics strongly hinges on them, we know very little about the origin and the behaviour of fundamental constants. In particular, we do not know if they are true constants or rather feebly vary through space and time. The measurement of even the slightest variation would provide us with a clear research direction beyond the theories we have so far, which famously fail to explain the vast majority of the energy content of the universe. Indeed, we have shown that variations of fundamental constants could be linked to dark matter, dark energy, violations of fundamental symmetries of nature or could be evidence that the laws of physics undergo cosmological evolution. We have argued that a network of clocks provides us with a powerful opportunity to measure with unprecedented sensitivity and a high level of confidence variations of two fundamental constants: the fine structure constant, $\alpha$, and the electron-to-proton mass ratio, $\mu$. 

In this work we have introduced the QSNET project, which aims at realising such a network. We have described its first stage, that will include a range of atomic and molecular clocks in the UK with different sensitivities to variations of $\alpha$ and $\mu$. We have evaluated the performance that can be obtained by QSNET. As illustrated with a few examples, such performance will enable us to explore large uncharted territories of the dark sector, potentially discovering new physics and/or imposing new constraints over many models and theories, widening our understanding of the physics that governs the universe. More specifically, QSNET will be sensitive to
\begin{itemize}
    \item Drifts of $\alpha$ and $\mu$, with relevance for dark energy models and models that predict cosmological evolution of fundamental constants.
    \item Oscillations of $\alpha$ and $\mu$, that can be linked for example to %to violations of fundamental symmetries or
    virialised dark matter scalar fields.
    \item Transient events due to kinks or topological defects in dark-sector fields.
\end{itemize}
Additionally, QSNET will allow us to perform tests of quantum gravity, violations of fundamental symmetries and grand unification theories.

In its next stages, the QSNET network will be extended, ideally across the globe and with some nodes in space, including more clocks and allowing for novel and improved capabilities of detection of variations of fundamental constants. Also, as happened in the last five decades, the ongoing progress of clock technology will allow us to improve stability and accuracy at each node of the network, further pushing our abilities to explore the unknown 95\% of the universe.

\section*{Declarations}
\begin{backmatter}

\section*{Acknowledgements}
We are grateful to Geoffrey Barwood and Sean Mulholland for reading the manuscript.

\section*{Funding}%% if any
\justifying This work was supported by STFC and EPSRC under grants ST/T00598X/1, ST/T006048/1, ST/T006234/1, ST/T00603X/1,ST/T00102X/1,ST/S002227/1. We also acknowledge the support of the UK government department for Business, Energy and Industrial Strategy through the UK National Quantum Technologies Programme. The work of M.S.S. was supported by US NSF Grant No. PHY-2012068.  The work of Y.V.S.~was supported by the World Premier International Research Center Initiative (WPI), MEXT, Japan and by the JSPS KAKENHI Grant Number JP20K14460. A.V. acknowledges the support of the Royal Society and Wolfson Foundation.

\section*{Competing interests}
The authors declare that they have no competing interests.

\section*{Authors' contributions}
G.B., X.C., R.M.G., J.G., M.K., M.S.S.,  N.S., Y.V.S. and M.R.T. wrote the main manuscript text. All authors reviewed the manuscript.

\section*{Ethical Approval and Consent to participate}
Not applicable

\section*{Consent for publication}
Not applicable

\section*{Availability of supporting data}
Not applicable

%%  After submission of the .TEX file,                     %%
%%  you will be prompted to submit your .BBL file.         %%

% if your bibliography is in bibtex format, use those commands:
\bibliographystyle{bmc-mathphys} % Style BST file (bmc-mathphys, vancouver, spbasic).
\bibliography{QSNET}      % Bibliography file (usually '*.bib' )
% for author-year bibliography (bmc-mathphys or spbasic)
% a) write to bib file (bmc-mathphys only)
% @settings{label, options="nameyear"}
\end{backmatter}

\newpage
\appendix
\section{The QSNET clocks}
\label{Appendix:QSNET_clocks}

\subsection{Established frequency standards}
\label{Appendix:standards}
\subsubsection{$^{133}$Cs fountain clock}
\label{Appendix:Cs_clock}

Caesium fountain clocks have achieved inaccuracies as low as $1$--$2 \times 10^{-16}$~\cite{Weyers2018,Heavner2014,Guena2012,Szymaniec2016,Levi2014}. See, for example, the uncertainty budget for one of the $^{133}$Cs fountain clocks at NPL in Table~\ref{Cs_accuracy}.  There is no single systematic shift which significantly dominates the uncertainty budget so improvements would be required with respect to a number of effects in order to reduce the overall uncertainty.  Such improvements have become increasingly challenging as the experimental techniques associated with fountain clocks are already highly optimised.  Motivation for further improvements has also been hindered by the fact that optical clocks already offer significantly lower uncertainties.        

\begin{table}[!h]
\centerline{\begin{tabular}{|l |r| r|}
\hline
 $^{133}$Cs Effect & Shift /$10^{-16}$  & Uncertainty /$10^{-16}$         \\
\hline
\hline
Distributed cavity phase    & 0.2       & 1.1 \\
Blackbody radiation         & -163.5    & 1.0 \\
Second-order Zeeman         & 2475.7    & 0.8 \\
Microwave leakage           & 0.0       & 0.6 \\
Gravity                     & 13.0      & 0.5 \\
Cold collisions             & 2.0       & 0.4 \\
Background gas              & 0.0       & 0.3 \\
Microwave lensing           & 0.6       & 0.3 \\
Other                       & 0.0       & 0.3 \\
\hline
Total                       &           & 2.0 \\
\hline
\end{tabular}}
\caption{Fractional frequency shifts and uncertainties from different systematic effects in the $^{133}$Cs fountain clock, known as NPL-CsF2, at NPL~\cite{Szymaniec2016}.}
\label{Cs_accuracy}
\end{table}

%%%%%%%%%%%%%%%%%%%%
\begin{comment}

Sr, Cs, Yb$^+$ (overviews, energy levels/transitions, $K$ factors, accuracy table)
\newline \newline
State-of-the-art Sr accuracy at JILA~\cite{Bothwell2019}, $2.0 \times 10^{-18}$
\newline
State-of-the-art Yb$^+$ accuracy at PTB~\cite{Sanner2019}, $2.7 \times 10^{-18}$
\newline
State-of-the-art Cs accuracy, $1 - 2 \times 10^{-16}$
\newline PTB 1.7E-16~\cite{Weyers2018}
\newline NIST 1.1E-16~\cite{Heavner2014}.  [Value is contested, but it's not been officially revised].
\newline SYRTE 2.1E-16~\cite{Guena2012}
\newline NPL 2.0E-16~\cite{Szymaniec2016}
\newline INRIM 1.9E-16~\cite{Levi2014}

\end{comment}
%%%%%%%%%%%%%%%%%%%%

\subsubsection{$^{87}$Sr lattice clock}
\label{Appendix:Sr_clock}

The current state-of-the-art for the $^{87}$Sr optical lattice clock has an estimated fractional frequency uncertainty from systematic shifts of $2.0 \times 10^{-18}$~\cite{Bothwell2019}. The uncertainty budget is shown in Table~\ref{Sr_accuracy}, with the dominant contributions coming from the Stark shifts induced by blackbody radiation and the lattice trapping potential.  The uncertainty in the blackbody radiation shift can be broken down into two components: (i) the contribution from the environment due to uncertainty in the temperature at the atoms, and (ii) the contribution from uncertainty in the atomic polarisability coefficient at room temperature.  In the context of measuring \emph{variations} in the fundamental constants, however, it is not essential to know the exact value of the blackbody radiation shift, only to ensure that it is a reproducible, constant frequency offset from one measurement to the next.  
%% We can assume that the contribution from the polarisability coefficient remains constant over time, since any changes in fundamental constants that could perturb the value of the coefficient would add a negligible shift in the optical clock frequency as it would be a perturbation of a perturbation. 
Given that the polarisability coefficient is constant, it is only necessary to consider the contribution to the blackbody uncertainty arising from the temperature uncertainty.  As demonstrated in the room-temperature $^{87}$Sr optical lattice clock at JILA, it is already possible to characterise the temperature to within $2.9$~mK, leading to an achievable uncertainty of $2.0 \times 10^{-19}$ in the blackbody radiation shift.  Lower uncertainties are also being explored by operating experiments at cryogenic temperatures~\cite{Ushijima2015}, where the magnitude and corresponding uncertainty of the blackbody radiation shift become much smaller.  The NPL-Sr1 system operates at room temperature and is characterised by a total systematic uncertainty of $1.0 \times 10^{-17}$, dominated by the blackbody radiation contribution at $7 \times 10^{-18}$ resulting from a 350 mK temperature gradient across the science chamber~\cite{Hobson20}. Reducing this to 100 mK through careful control of the thermal environment would reduce its contribution to $2 \times 10^{-18}$.  

The uncertainty in the AC Stark shift from the laser light creating the lattice potential has several contributions, including the level of experimental control over the frequency, intensity, polarisation and spectral purity of the light.  Additionally, there is uncertainty in the contributions from higher-order effects such as hyperpolarisability and magnetic dipole and electric quadrupole terms.  However, mitigation strategies exist to reduce the uncertainty from these AC Stark shifts to the $10^{-19}$ level~\cite{Ushijima2018}.  

\begin{table}[!h]
\centerline{\begin{tabular}{|l |r| r|}
\hline
 $^{87}$Sr Effect & Shift /$10^{-18}$  & Uncertainty /$10^{-18}$         \\
\hline
\hline
Blackbody radiation (environment)   & -4974.1   & 0.2 \\
Blackbody radiation (coefficient)   & -         & 1.5 \\
Lattice AC Stark                  & -21.3     & 1.2 \\
Density                             & -12.3     & 0.4 \\
Background gas                      & -3.7      & 0.4 \\
d.c. Stark                          &  0.0      & 0.3 \\
Second-order Zeeman                 & -176.9    & 0.2 \\
Servo                               & 0.0       & 0.2 \\
\hline
Total                               &           & 2.0 \\
\hline
\end{tabular}}
\caption{Fractional frequency shifts and uncertainties from different systematic effects in the $^{87}$Sr optical lattice clock at JILA~\cite{Bothwell2019}.}
\label{Sr_accuracy}
\end{table}

\subsubsection{$^{171}$Yb$^+$ ion clock}
\label{Appendix:Yb_clock}
The current state-of-the-art for the E3 transition of $^{171}$Yb$^+$ has an estimated fractional frequency uncertainty from systematic shifts of $2.7 \times 10^{-18}$~\cite{Sanner2019}.  The uncertainty budget is shown in Table~\ref{Yb+_accuracy}, with the dominant contributions coming from the blackbody radiation and Doppler shifts.  At room temperature, there is a fractional frequency uncertainty of $1.3 \times 10^{-18}$ from an effective temperature uncertainty of $1.3$~K in the surroundings of the trapped ion, and there is a further contribution of $1.3 \times 10^{-18}$ from uncertainty in the polarisability coefficient.  Once again, when looking for \emph{variations} in fundamental constants, the uncertainty contribution from lack of knowledge of the polarisability coefficient can be ignored if we assume that the coefficient itself remains always constant.  The relevant contribution to the blackbody radiation shift uncertainty is then only that arising from temperature uncertainty. This could be reduced with better characterisation of the thermal environment, with uncertainties of $0.14$~K having already been demonstrated in the $^{171}$Yb$^+$ system at NPL~\cite{Nisbet-Jones2016}, leading to an uncertainty contribution of $1.3 \times 10^{-19}$ in the blackbody radiation shift.  The second-order Doppler shift has uncertainty contributions from the ion's thermal motion due to its 1~mK temperature, as well as from faster motional oscillations that are driven by radio-frequency (RF) electric fields when the ion is not perfectly centred within the trap.  Both of these contributions to the second-order Doppler shift could be reduced, with Raman-sideband cooling used to reduce thermal motion and with more stringent centering of the ion within the trap.            

\begin{table}[!h]
\centerline{\begin{tabular}{|l |r| r|}
\hline
 $^{171}$Yb$^+$ Effect & Shift /$10^{-18}$  & Uncertainty /$10^{-18}$         \\
\hline
\hline
Blackbody radiation (environment)     & -70.5   & 1.3 \\
Blackbody radiation (coefficient)     & -       & 1.3 \\
Second-order Doppler    & -2.3  & 1.5 \\
Probe light             & 0.0   & 0.8 \\
Quadratic d.c. Stark    & -0.8  & 0.6 \\
Quadrupole              & -5.7  & 0.5 \\
Background gas          & 0.0   & 0.5 \\
Second-order Zeeman     & -10.4 & 0.2 \\
Servo                   & 0.0   & 0.2 \\
\hline
Total                   &       & 2.7 \\
\hline
\end{tabular}}
\caption{Fractional frequency shifts and uncertainties from different systematic effects in the $^{171}$Yb$^+$ electric octupole clock at PTB~\cite{Sanner2019}.}
\label{Yb+_accuracy}
\end{table}

\subsection{CaF molecular lattice clock}
\label{Appendix:CaF_clock}

\begin{figure}[tb]
	\centering
		\includegraphics[width=0.5\textwidth]{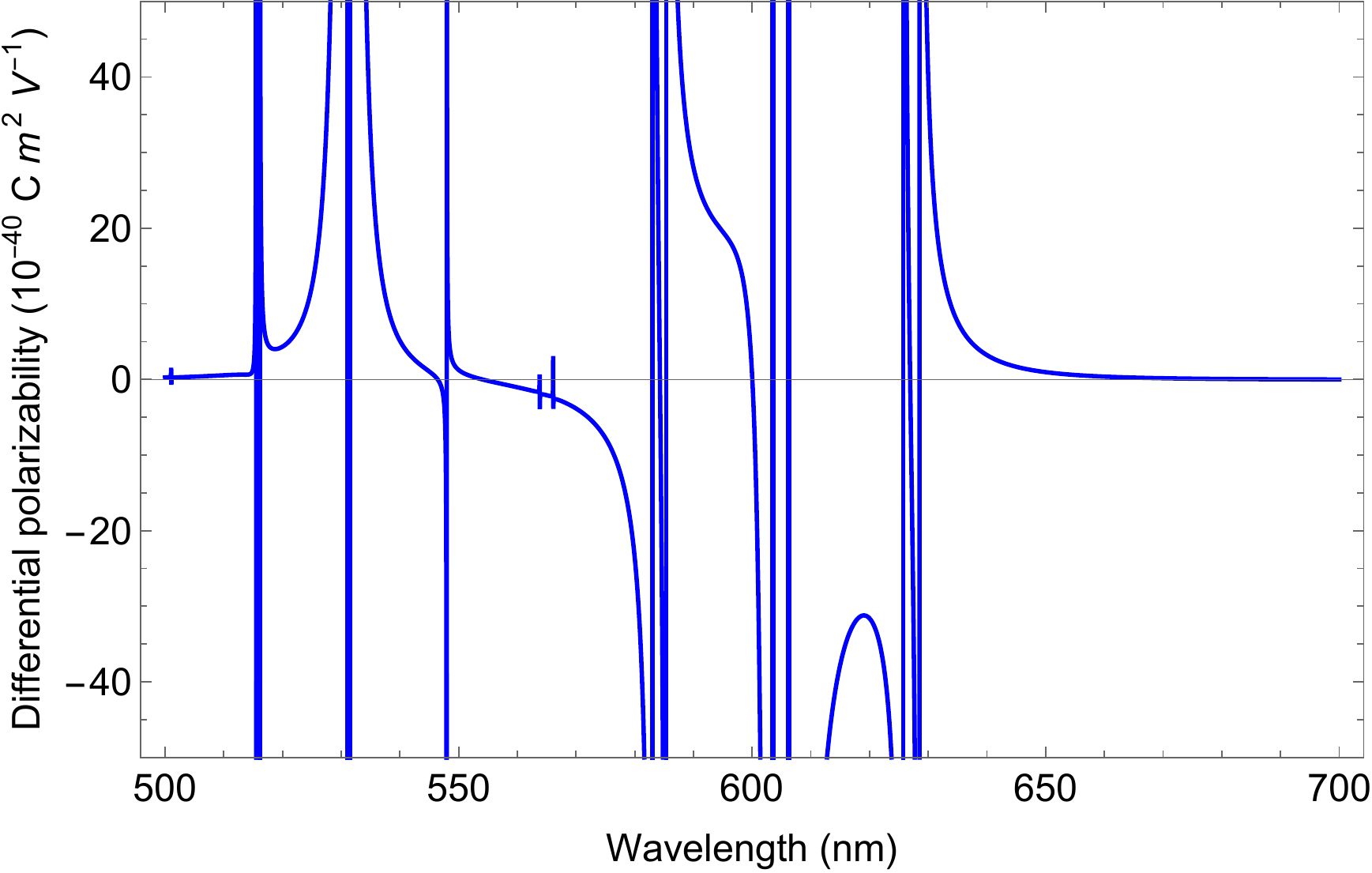}
	\caption{Differential scalar polarisability for the two vibrational states of CaF, plotted vs wavelength.}
\label{CaF_Polarizability}
\end{figure}

To make the most accurate measurement, the clock transition should be insensitive to the lattice intensity. This can be done by choosing a magic wavelength where the AC Stark shifts are equal for the upper and lower states. Figure \ref{CaF_Polarizability} shows the differential polarisability versus wavelength, calculated using data on the X, A, B, C, D and E electronic states. Transition strengths diminish rapidly as the change in vibrational quantum number increases, so only the low-lying vibrational levels of each electronic state are needed for the calculation. For the X, A, and B states, we construct Rydberg–Klein–Rees potentials, calculate the vibrational wavefunctions in these potentials, and use these to calculate the vibrational branching ratios for the A--X and B--X transitions~\cite{Fitch2021}. For the C--X, D--X and E--X transitions, we use the branching ratios given in \cite{Karthikeyan2017}. We find magic wavelengths near 554~nm, 600~nm and 696~nm. We note that these differ from the results presented in \cite{Kajita2018}, and that the magic wavelengths are very sensitive to the values of the vibrational branching ratios, especially for the shallow crossing at 696~nm. Table~\ref{CaF_lattices} gives the intensity needed to produce a lattice depth of 10~$\mu$K, and the sensitivity of the clock transition to lattice frequency at each of these wavelengths. From the set, the 696~nm lattice is particularly attractive because a frequency precision of 1~MHz leads to a fractional accuracy of 5 parts in $10^{19}$. Vector Stark shifts are also very small at this wavelength. Table~\ref{CaF_lattices} also shows the properties at a lattice wavelength of 1064~nm. Although this is not a magic wavelength, the AC Stark shifts of the two states differ by only 5 parts in $10^4$ and the clock transition is exceptionally insensitive to the lattice frequency. The dependence on lattice intensity limits the accuracy of the clock, but the intensity could be measured to high accuracy by using the large tensor Stark shift of the rotational transition. A measurement of the rotational frequency with a precision of $10^{-12}$, which is not too difficult~\cite{Blackmore2018,Caldwell2020}, determines the intensity to $10^{-6}$ which in turn results in a fractional inaccuracy of $5 \times 10^{-18}$ for the clock transition.

\begin{table}[!h]
\centerline{\begin{tabular}{|c |c| c| c| c|}
\hline
 Wavelength $\lambda_{\rm latt}$ & Trap sites & Intensity $I_{\rm latt}$  & $\frac{1}{f_0}\frac{d f_0}{d f_{\rm latt}}$ &  $\frac{1}{f_0}\frac{d f_0}{d I_{\rm latt}}$  \\
 
  [nm] & \ &  [W~cm$^{-2}$] & [MHz$^{-1}$] & [(W~cm$^{-2}$)$^{-1}$] \\
\hline
\hline
554.0 & Nodes & $1.8 \times 10^4$ & $6 \times 10^{-17}$ & 0 \\
600.1 & Nodes & $7.0 \times 10^2$ & $1 \times 10^{-16}$ & 0 \\
695.7 & Antinodes & $9.5 \times 10^3$ & $5 \times 10^{-19}$ & 0 \\
1064 & Antinodes & $2.5 \times 10^4$ & $3 \times 10^{-20}$ & $2 \times 10^{-16}$ \\
\hline
\end{tabular}}
\caption{Clock properties at three lattice wavelengths. The intensity is that needed for a depth of 10~$\mu$K, and the sensitivity to frequency is calculated at that intensity.}
\label{CaF_lattices}
\end{table}

Table \ref{CaF_accuracy} lists the uncertainties arising from systematic shifts of the clock transition. Here, we choose to use the 696~nm lattice. The first row gives the uncertainty arising from the lattice frequency, as discussed above. The components of the clock transition with $m=\pm 1$ have equal and opposite linear Zeeman shifts. Using the calculated vibrational dependence of the $g$-factors~\cite{Caldwell2020}, we find shifts of $\pm 0.05$~Hz~$\mu$T$^{-1}$. By controlling the magnetic field to 0.1~$\mu$T and ensuring that both components are driven equally to within 1\%, the fractional uncertainty is reduced to $3 \times 10^{-18}$. The transition with $m=0$ has a quadratic Zeeman  shift. Using the measured vibrational dependence of the hyperfine constants~\cite{Childs1981}, we estimate a shift of 0.014~Hz~$\mu$T$^{-2}$. At 0.1~$\mu$T, this translates into a fractional uncertainty of $8 \times 10^{-18}$. We note that controlling the magnetic field to 0.1~$\mu$T can be done by measuring the rotational frequency with a fractional precision of $7 \times 10^{-8}$, which is easily done. Using the vibrational dependence of both the dipole moment and the rotational constant, we estimate the DC Stark shift of the transition to be 1.2~Hz~(V/cm)$^{-2}$, so a 10~mV~cm$^{-1}$ uncertainty in the stray field translates to a fractional uncertainty of $7 \times 10^{-18}$. We calculate the blackbody shift of the clock transition to be $4 \times 10^{-18}$~K$^{-1}$ at 300~K, so measuring the temperature with an accuracy of 250~mK results in an uncertainty of $10^{-18}$. The first-order Doppler shift is eliminated in the lattice, and the second-order Doppler shift contributes a fractional uncertainty of about $4 \times 10^{-21}$. For driving the Raman transition, we find several wavelengths where the AC Stark shift due to the Raman lasers is eliminated. One interesting option is to use the lattice laser as one of the Raman lasers, with the other Raman laser at 725~nm. This laser produces a negligible AC Stark shift, contributing a fractional shift of only $9 \times 10^{-22}$. The uncertainty in this shift will be smaller still.

\begin{table}[!h]
\centerline{\begin{tabular}{|c |c| c|}
\hline
  Effect & Conditions & Fractional Uncertainty        \\
\hline
\hline
Lattice frequency & $\delta f_{\rm latt} < 1$~MHz  & $5 \times 10^{-19}$ \\
Zeeman shift & $B < 0.1$~$\mu$T, asymmetry $< 1$\% & $7 \times 10^{-18}$ \\
DC Stark shift & $E < 10$~mV~cm$^{-1}$ & $7 \times 10^{-18}$ \\
Blackbody shift & $\delta T < 250$~mK at 300~K & $10^{-18}$ \\
Doppler shift & $T=5$~$\mu$K & $4 \times 10^{-21}$ \\
Raman laser & 696~nm \& 725~nm & $< 9 \times 10^{-22}$ \\
\hline
\end{tabular}}
\caption{Projected systematic uncertainties for a CaF clock using a 696~nm lattice.}
\label{CaF_accuracy}
\end{table}

\subsection{N$_2^+$ molecular ion clock}
\label{Appendix:N2_clock}
%Molecular transitions have been proposed as ideal candidates for testing fundamental theories and for investigating spatio-temporal changed of fundamental constants. However, often transitions with high sensitivities are technically difficult to implement or have large systematic shifts. 
Recently, molecular nitrogen ions have been proposed as a candidate for precision spectroscopy \cite{Kajita}. The vibrational clock transition has a sensitivity of K$_\mu = 0.49$ and systematic shifts which are comparable with the currently best optical clocks and facilitate frequency measurements at an uncertainty below $10^{-18}$.

\begin{figure*}
	\centering
		\includegraphics[width=0.6\textwidth]{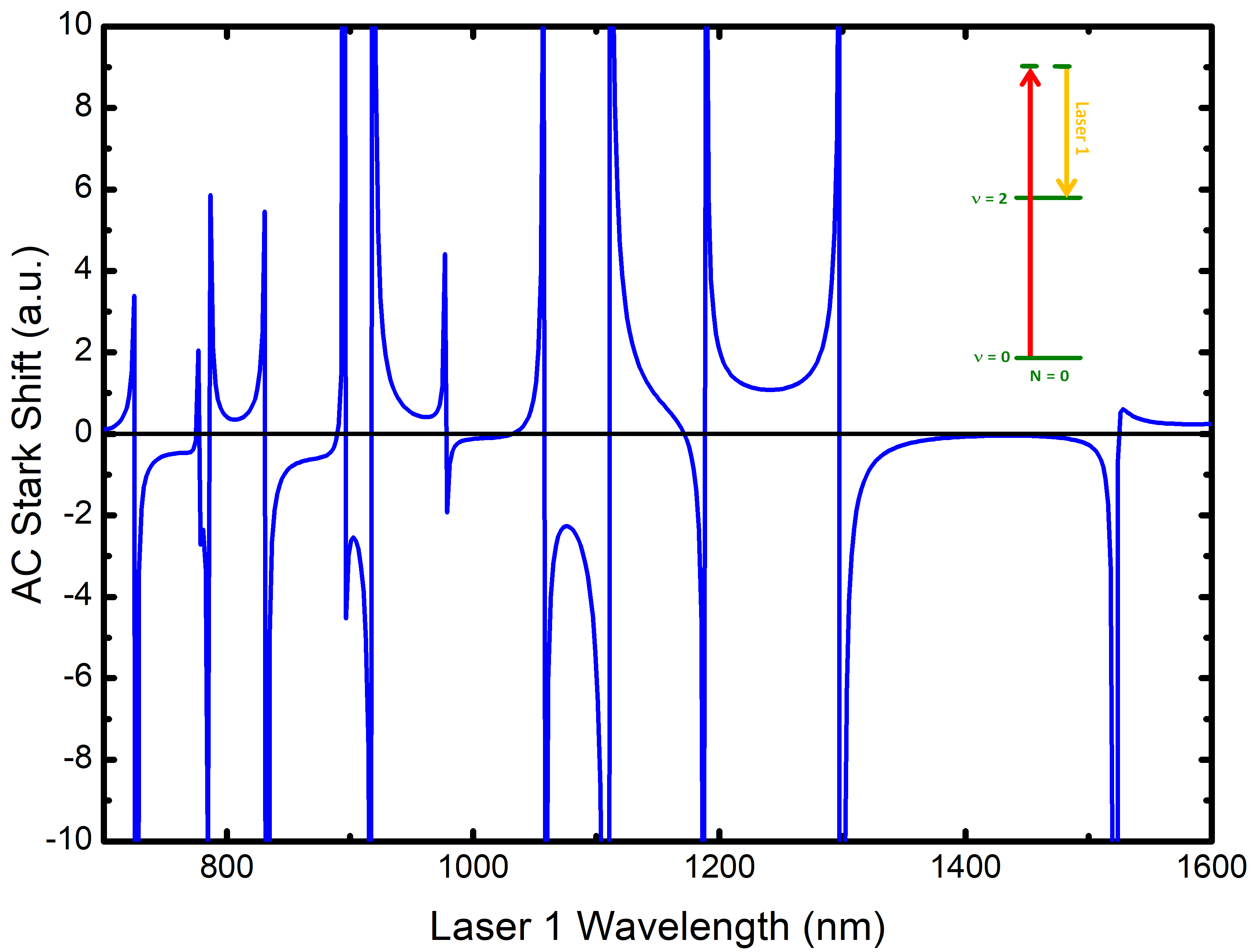}
	\caption{AC Stark shift of the N$_2^+$ clock transition as a function of the low frequency Raman laser wavelength.}
\label{figN2acStark}
\end{figure*}

With N$_2^+$ being an apolar molecule, there are no allowed purely rotational transitions. However, due to the coupling to the electronically excited states, the two clock states experience a second order fractional Stark shift of $4 \times 10^{-18}$~(V/cm)$^{-2}$. These shifts mostly cancel, leading to a Stark fractional shift of $8\times10^{-20}$~(V/cm)$^{-2}$, which leads to a systematic fractional Stark shift of $1\times10^{-20}$ for a typical stray electric field of 10 V/m. With all purely ro-vibrational transitions being dipole forbidden, and the first electronically excited state being 270 THz above the ground state and coupled to the clock states only through a very weak dipole moment, the effect of the blackbody radiation is also nearly cancelled, leading to a differential blackbody shift of $4\times10^{-18}$ at 300~K with an associated uncertainty of $3\times 10^{-21}$ for a temperature stability of 250~mK. The calculation of these systematic shifts is based on the available spectroscopic data for N$_2^+$. Even though N$_2^+$ is very well investigated and good spectroscopic data are available, there is still an uncertainty regarding the systematic effects of external fields on the clock transition. In particular, possible higher order couplings between high lying molecular states and the clock states remain to be investigated. 

In addition to systematic shifts due to external fields, the ion's motion can lead to Doppler shifts. With the molecular ion being co-trapped with a laser cooled atomic ion, the molecular ion's temperature is the same as that of the atomic ion. Using a calcium ion as a sympathetic coolant, the first-order Doppler shift is eliminated due to the Lamb-Dicke effect. The fractional second-order Doppler shift due to the ion's thermal motion is $-2\times 10^{-18}$ at the calcium ion's Doppler cooling limit and can be further reduced by employing sideband cooling, which is itself a prerequisite for quantum logic spectroscopy. In addition to the shift induced by the ion's thermal motion, excess micro-motion can lead to a motional shift. For typical micro-motion compensation, the associated shift is $-1\times 10^{-19}$. 

The $\nu=2, N=0 \longleftarrow \nu=0, N=0$ transition is even more forbidden, and is only accessible via a Raman transition, which can cause AC Stark shifts of both clock levels. By choosing the wavelengths of the Raman lasers correctly, the AC Stark shift of both clock levels due to the two lasers cancel. For the $\nu=2, N=0 \leftarrow\nu=0, N=0$ transition, the AC Stark shift of the clock transition is shown in figure \ref{figN2acStark}. One possible set of wavelengths is 714~nm and 1032~nm. These are directly accessible by diode lasers and are thus very convenient. Here the transition frequency is independent of the laser intensity if both Raman lasers have the same intensity. The clock transition fractional frequency shift for a differential intensity change is $-3.4\times 10^{-18}$~cm$^2$/W for the 714~nm laser and $3.4\times 10^{-18}$~cm$^2$/W for the Raman laser at 1032~nm. An intensity stability of 1\% corresponds to a fractional frequency shift of $\pm3.4\times 10^{-18}$. In order to eliminate a possible residual AC Stark shift, the hyper-Ramsey method can be employed \cite{Huntemann1} leading to a potential reduction by two orders of magnitude \cite{Huntemann2}.

In order to take advantage of the control possible in RF ion traps and to perform Raman spectroscopy, a single N$_2^+$ ion is trapped in an RF ion trap and sympathetically cooled with a co-trapped atomic ion. Narrow linewidth Raman lasers will be used to probe the transition with the state population change being monitored using the technique of state transfer to the calcium ion. The co-trapped calcium ion serves not only as a state read-out but also as a probe for the local environment within the ion trap. Using the calcium ion’s clock transition (S$_{1/2}$ -- D$_{5/2}$), the local magnetic field and blackbody shifts can be measured and used for correcting the results of the molecular spectroscopy.

\begin{table}
\centerline{\begin{tabular}{|c |c| c|}
\hline
  Effect & Conditions & Fractional Uncertainty        \\
\hline
\hline
Electric quadrupole shift &  & 0 \\
Zeeman shift &  &  0 \\
\multirow{2}{*}{DC Stark shift} & Typical electric fields in an ion trap  & $1 \times 10^{-20}$ \\
  & after micromotion compensation & \\
Blackbody shift & $\delta T < 250$~mK at 300~K & $3\times10^{-21}$ \\
Doppler shift & Doppler cooling limit of Ca$^+$ & $2\times10^{-18}$ \\
Raman laser & 1\% intensity stability & $3.4 \times 10^{-18}$ \\
\hline
\end{tabular}}
\caption{Projected systematic shifts for the N$_2^+$ clock.}
\label{N2accuracy}
\end{table}

With the lack of cycling transitions within molecules, the state read-out can’t be performed via electron shelving within the molecules but only through transfer of the molecular state to the auxiliary ion trapped alongside the molecular ion. With both ions simultaneously cooled through laser cooling of the calcium ion, the ions form a molecule-like structure with shared motional modes within the trapping potential. Using a state-selective dipole force, the molecular state can be mapped onto the joint motion of the ions and then transferred onto the calcium ion. Using Doppler velocimetry \cite{Hume, Sinhal} or quantum logic spectroscopy \cite{Wolf, Chou} the molecular state can then be read-out via the calcium ion’s fluorescence.

Due to the large number of molecular states, the nitrogen ion must be prepared in the ro-vibrational ground state of the electronic ground state. This can be accomplished by two-colour resonance-enhanced multi-photon ionisation (REMPI 2+1) of a skimmed, supersonic molecular beam. Employing the two-photon overtone transition from the molecular ground state X$^1\Sigma_g^+ \nu=0$ to the first electronically excited state a$^1\Pi_g^+ \nu=6$ at a laser wavelength of 255~nm, followed by an ionisation step at a laser wavelength of 212~nm, nitrogen ions can be prepared in a specific ro-vibrational state \cite{Gardner}.

\subsection{Cf highly charged ion clocks}
\label{Appendix:HCI_clock}

Recent experimental breakthroughs \cite{Schmoger1233,HCIRSI} allow  HCIs to be cooled and trapped to temperatures below 1~mK. In brief, the HCIs are initially produced and pre-cooled in an electron-beam ion trap (EBIT), then further cooled and guided using ion optics, and finally sympathetically cooled to mK and below using laser-cooled ion crystals \cite{HCIRSI,Schmoger1233}. HCIs can therefore be co-trapped with singly charged ions, allowing one to perform quantum logic spectroscopy \cite{Micke}.

Systematic effects for Cf clocks have been estimated in \cite{Cfclock} and the corresponding fractional accuracies are reported in table \ref{Cfaccuracy}. Shifts due to BBR can safely be ignored as the HCIs are trapped inside cryogenic systems that further suppress this effect. Electric quadrupole shifts cancel when $3M^2=F(F+1)$, a condition fulfilled by $|F=3,M=\pm2\rangle$ states. These are available in the ground state of both Cf$^{15+}$ and Cf$^{17+}$ and in the upper clock state of Cf$^{15+}$. The upper clock state of Cf$^{17+}$ has zero quadrupole moment. Supposing to operate in a magnetic field with stability of $\simeq 0.1$~mG, linear Zeeman effects can be cancelled out by averaging between $m_F=\pm2$ states, and the uncertainty due to the quadratic Zeeman effect can be reduced below 10$^{-18}$. For the $^{251}$Cf$^{15+}$ isotope, the quadratic Zeeman shift could be completely cancelled by averaging between $|F=2\rangle\rightarrow|F'=4\rangle$ and $|F=3\rangle\rightarrow|F'=5\rangle$ transitions, which are shifted by the same amount but with opposite signs. Within QSNET, we plan to use Ca$^+$ ions for sympathetically cooling the Cf HCIs. Similarly to N$^+_2$, the Doppler cooling limit of the Ca$^+$ ions leads to second-order Doppler shifts that for Cf are $\simeq10^{-19}$. With the information at hand, in principle it would be possible to reach fractional frequency uncertainty on the order of $10^{-19}$ for both ionization states. 

\begin{table}[!h]
\centerline{\begin{tabular}{|c | c| c| c|}
\hline
   & Conditions       & Cf$^{15+}$  &    Cf$^{17+}$     \\
\hline
\hline
\multirow{2}{*}{Electric quadrupole}  & $|F=3,m_F=\pm2\rangle\rightarrow|F=3,m_F=\pm2\rangle$  & 0 &      \\
                           & with $|F=3,m_F=\pm2\rangle$ ground state &  & 0     \\
BBR & $4.0\pm0.1$ K& $\ll 10^{-19}$  & $\ll 10^{-19}$  \\
Quadratic Zeeman &  $^{251}$Cf with $\delta B< 0.1$ mG & $5\times10^{-19}$ & $1\times10^{-18}$  \\
Second order Doppler  & Ca$^+$ Doppler limit & $1.5\times10^{-19}$ & $1.5\times10^{-19}$    \\
\hline
\end{tabular}}
\caption{Estimated fractional uncertainties for Cf HCI clocks \cite{Cfclock}.}
\label{Cfaccuracy}
\end{table}

\section{Dark matter}
\subsection{Feasible mass range for oscillating dark matter}
\label{Appendix:DM_mass_range}
The dark-matter field in Eq.~(\ref{oscillating_DM_field}) is classical, provided that $\gg 1$ bosons fit into the reduced de Broglie volume, $n_\phi [\lambda_\textrm{coh} / (2 \pi)]^3 \gg 1$ ($n_\phi$ is the boson number density), which for the local Galactic dark-matter energy density of $\rho_\textrm{DM,local} \approx 0.4~\textrm{GeV/cm}^3$ \cite{Zyla:2020PDG} is satisfied for $m_\phi \lesssim 1~\textrm{eV}$. 
Dark-matter particle masses less than $\sim 1~\textrm{keV}$ are precluded for fermionic particles, from the consideration of the available phase space density in dwarf galaxy haloes dictated by the Pauli exclusion principle; therefore, the classical dark-matter field under consideration must be bosonic. 
On the other hand, if very-low-mass bosons are to account for the observed dark matter, then their mass cannot be arbitrarily light.
Indeed, such bosonic fields would tend to suppress the formation of structures on length scales below the ground-state de Broglie wavelength of the bosons \cite{Khlopov:1985fuzzyDM,Gruzinov:2000fuzzyDM}. 
The reason is that the development of inhomogeneities in the bosonic density would be effectively inhibited on length scales below the ground-state de Broglie wavelength of the bosons, in accordance with the wave uncertainty principle. 
The de Broglie wavelength becomes astronomically large for sufficiently low-mass bosons, thereby precluding such bosons from saturating the observed dark matter abundance. 
The requirement that the ground-state de Broglie wavelength of the bosons fit inside the haloes of the smallest observed dwarf galaxies (which are $\sim 1~\textrm{kpc}$ in size and have a characteristic internal root-mean-square speed of $\sim 10~\textrm{km/s}$) gives a lower mass bound of $m_\phi \gtrsim 10^{-21}~\textrm{eV}$ if such bosons account for all of the cold dark matter. 
There are comparable lower mass bounds from the analysis of structures in Lyman-$\alpha$ forest data \cite{Viel:2017fuzzyDM,Viel:2018fuzzyDM}, as well as other astrophysical observations \cite{Marsh:2019fuzzyDM,Schutz:2020fuzzyDM}. 
For the boson masses $m_\phi \lesssim 10^{-21}~\textrm{eV}$, bosonic particles may account for up to $\mathcal{O}(10\%)$ of the dark matter. 

\subsection{Details on the exclusion plots}
\label{Appendix:DM_exclusion_plots}
In clock-based searches for dark-matter-induced oscillations, it is generally more favourable to perform more individual measurements with short averaging times (e.g., $\tau \sim 1~\textrm{s}$) when the individual measurements are statistically limited, instead of fewer individual measurements with long averaging times (e.g., $\tau \gtrsim 10^6~\textrm{s}$) when the individual measurements start to become limited by systematics. 
In other words, the benefits arise from lower \textit{fractional instabilities}. 
Clock-based measurements are best suited to search for signal frequencies greater than the inverse of the time span of the data set ($f_\textrm{min} = 1/T_\textrm{dataset}$), but less than the sampling frequency of the measurements, $f_\textrm{max}$. 
Nevertheless, frequencies below the inverse of the time span of the data set and above the sampling frequency can be sought with diminished sensitivity. 
In the first case, one simply fits the available data to the form of the expected signal in Eq.~(\ref{clock_DM_signal}), which is expected to be coherent for the relevant range of small dark-matter particle masses, at the cost of a loss in sensitivity by the factor of $(f_\textrm{signal}/f_\textrm{min})^2 < 1$. 
In the latter case, one can utilise the aliasing technique, at the cost of a loss in sensitivity by the factor of $f_\textrm{max}/f_\textrm{signal} < 1$. 
At higher signal frequencies, the use of cavities can be advantageous\cite{Stadnik:2015DM-LI,Stadnik:2016cavity}; for instance, when referencing a Sr clock against a cavity with a solid spacer between the mirrors, $K_\alpha (\textrm{Sr}) - K_\alpha (\textrm{cavity}) \approx +1$, compared with the relativistic correction factor of $+0.06$ for the Sr clock transition that is relevant when referencing a Sr clock against a different optical atomic clock.

Differences in heights between clocks at different nodes within the QSNET network will be accounted for by independently taken satellite clock and ground-based gravimeter data. 
The possible effect of the oscillating dark-matter field on the local gravitational acceleration $\boldsymbol{g}$ via temporal variations in the mass and radius of Earth is sub-leading compared to the direct effect of the oscillating dark-matter field on the clock energy levels, being parametrically suppressed by the factor $G M_\oplus / (R_\oplus c^2) \sim 10^{-9}$ compared with the latter, besides the additional common-mode suppression of local $\boldsymbol{g}$ variations when ``comparing'' ground-based clock and gravimeter data. 

Searches for the effects of an oscillating dark-matter field on atomic and molecular transition frequencies via apparent oscillations of the fundamental constants are strictly sensitive to combinations of parameters like $\sqrt{\rho_\phi} / \Lambda_X$ or $\sqrt{\rho_\phi} / \Lambda'_X$. 
In order to infer information about the new-physics energy scales $\Lambda_X$ or $\Lambda'_X$ with these types of measurements, one must make assumptions about the local value of $\rho_\phi$ during the course of the measurements. 
While the \textit{average} local Galactic dark-matter energy density is known with good certainty to be $\rho_\textrm{DM,local} \approx 0.4~\textrm{GeV/cm}^3$ \cite{Zyla:2020PDG}, the density and distribution of dark matter on length scales below $\sim 10 - 100~\textrm{pc}$ is poorly understood \cite{Bertone:2015DM,Karukes:2019DM}. 
Besides the possibility of cold dark matter naturally forming clumps and voids on length scales below $\sim 10$\,--$100~\textrm{pc}$, the amplitude of an ultra-low-mass bosonic dark-matter field is also expected to undergo stochastic fluctuations on sufficiently large temporal and spatial scales \cite{Derevianko:2018stochastic,Safdi:2018stochastic}. 
On time scales less than the coherence time, the scalar-field amplitude in Eq.~(\ref{oscillating_DM_field}) is expected to remain approximately constant in time and hence the resulting signal is expected to have the coherent form in Eq.~(\ref{clock_DM_signal}).
\begin{figure}
	\centering
		\includegraphics[width=0.8\textwidth]{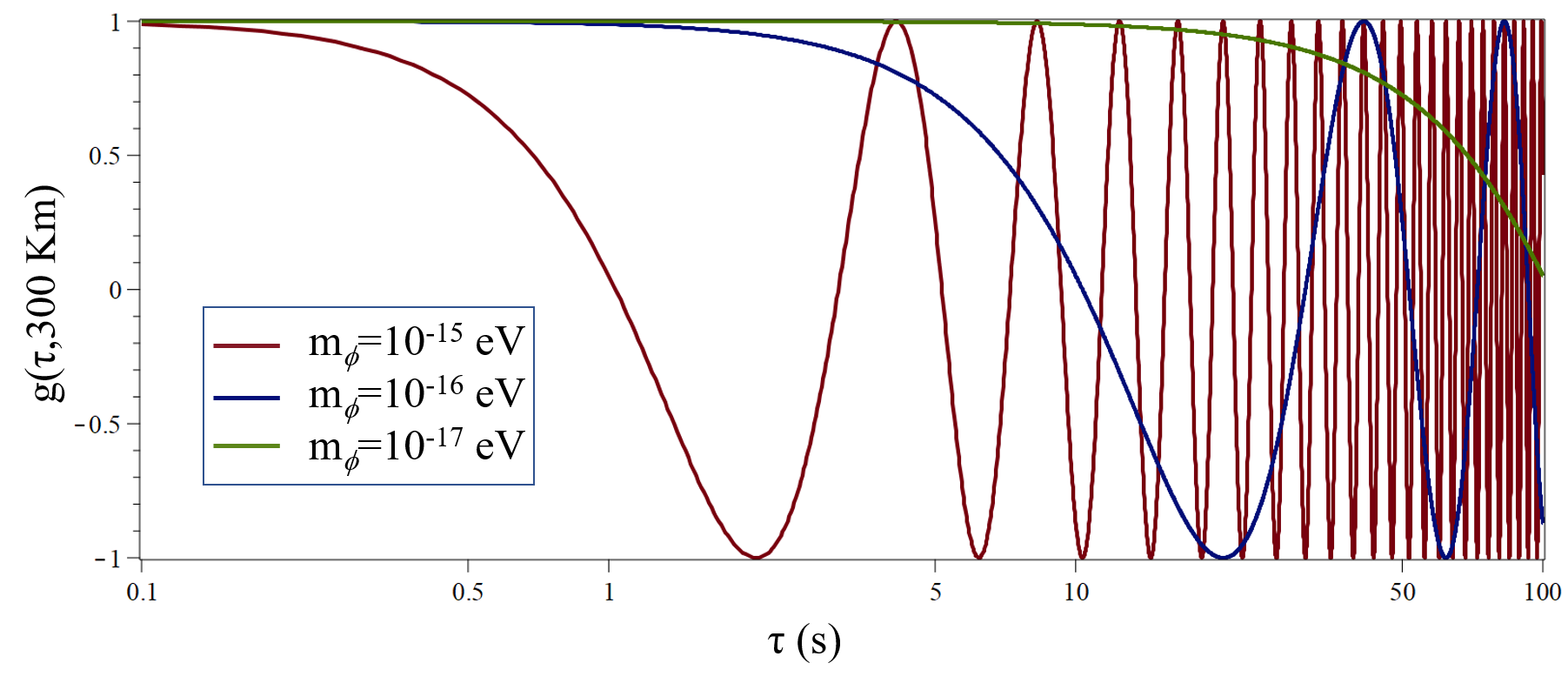}
	\caption{Normalised correlation function $g$ as a function of the delay time $\tau$ \cite{Derevianko:2018stochastic} calculated between two nodes separated by 300 km, corresponding to the longest extent of the QSNET network, for dark matter fields in the mass range $10^{-15}$--$10^{-17}$ eV.}
\label{Fig:correlations}
\end{figure}
In this case, the signal-to-noise ratio (SNR) improves with the integration time $t_{int}$ as $\textrm{SNR} \propto t_\textrm{int}^{1/2}$ and, if there are $N$ pairs of clocks located within a coherence length of one another, the SNR can further be improved by a factor of $N^{1/2}$. 
On the other hand, on time scales exceeding the coherence time, the scalar-field amplitude in (\ref{oscillating_DM_field}) is expected to fluctuate stochastically and the SNR is only expected to improve with the integration time as $\textrm{SNR} \propto t_\textrm{int}^{1/4} \tau_\textrm{coh}^{1/4}$. 
The use of multiple nodes within the QSNET network would allow one to probe the spatio-temporal correlation function for a range of dark-matter boson masses; see Fig.~\ref{Fig:correlations}.

\section{Fundamental symmetries}
\label{sec:fundsymms}
\subsection{Technical details regarding Lorentz-violating effects}
\label{ssec:fundsymmsdetails}
Terms in $S_{\text{LV}} = \int d^4 x\mathcal{L}_{\text{LV}}$ from Eq.~\eqref{SMEaction} may be expressed as a coordinate-independent contraction of an SME coefficient for Lorentz violation with a Lorentz-violating operator. A generic
term takes the form \cite{Kostelecky:2020hbb}
\begin{ceqn}
\begin{align}
\mathcal{L}_{\text{LV}} \supset k^{\mu\cdots\hphantom{\nu\hdots}a\cdots}_{\hphantom{\mu\cdots}\nu\cdots}(x)\mathcal{O}_{\mu\cdots\hphantom{\nu\cdots}a\cdots}^{\hphantom{\mu\cdots}\nu\cdots}(x),
\end{align}
\end{ceqn}
modulo possible derivatives of $k^{\mu\cdots\hphantom{\nu\hdots}a\cdots}_{\hphantom{\mu\cdots}\nu\cdots}(x)$. Here $\mathcal{O}_{\mu\cdots\hphantom{\nu\cdots}a\cdots}^{\hphantom{\mu\cdots}\nu\cdots}(x)$ is an $n$-dimensional Lorentz tensor and a product of field operators and gauge-covariant derivatives, and $k^{\mu\cdots\hphantom{\nu\hdots}a\cdots}_{\hphantom{\mu\cdots}\nu\cdots}(x)$ is a rank-$n$ SME coefficient. Both quantities can be functions of space-time position and may possess additional gauge-group indices. The coefficients may also depend on the underlying symmetry-breaking mechanism. They can be viewed as background tensor-valued quantities generating anisotropies in space-time that couple to the matter, gauge, and gravitational fields.
Existing experimental constraints suggest a perturbative treatment  around conventional physics is valid, thereby enabling a broad investigation of potential signals \cite{datatables}.

As discussed in Sec.~\ref{Sec:symmetries}, the leading Lorentz- and CPT-violating observables for clock-comparison experiments are expected to stem from modified free-fermion propagation terms \eqref{smeqed} at $d = 3, 4$. The covariant extension of these effects, including photon-sector modifications, reads
\begin{ceqn}
\begin{align}
\label{smeqedgeneral}
\mathcal{L} \supset &\tfrac{1}{2}\bar{\psi}\widehat{\Gamma}_{\nu}
i\overset{\text{\tiny$\leftrightarrow$}}{D^{\nu}}\psi - 
\bar{\psi}\widehat{M}\psi -\tfrac{1}{4}F_{\mu\nu}F^{\mu\nu} \nonumber\\
& -\tfrac{1}{4}F_{\kappa\lambda}\left(\widehat{k}_{F}\right)^{\kappa\lambda\mu\nu}\hspace{-1.5mm}F_{\mu\nu} + \tfrac{1}{2}
\epsilon_{\kappa\lambda\mu\nu}A^{\lambda}\left(\widehat{k}_{AF}\right)^{\kappa}\hspace{-1.5mm}F^{\mu\nu},
\end{align}
\end{ceqn}
where the gauge covariant derivative is $D^{\nu} = \partial^\nu + iqA^{\nu}$ with fermion charge $q$ and photon field $A^\nu$. The hat notation, e.g. $\widehat{\Gamma}^\nu$, indicates a generalization of the minimal operators \eqref{fermionterms} to operators of arbitrary mass dimension $d$. For example, $\widehat{M} \supset \widehat{b}^\mu\gamma_5\gamma_\mu$, where
\begin{ceqn}
\begin{equation}
\widehat{b}^{\mu} = \sum_{\begin{subarray}{l} d\ge 3 \\ d\;\text{odd}\end{subarray}
}b^{(d)\mu\alpha_1\cdots\alpha_{(d-3)}}(i\partial_{\alpha_1})\cdots(i\partial_{\alpha_{(d-3)}}).
\end{equation}
\end{ceqn}
Power counting shows the coefficients have mass dimension 
$4 - d$, where $d$ is the dimension of the associated operator. 
The coefficients within $\widehat{\Gamma}^\nu$ and $\widehat{M}$ affect the free propagation of fermions and the conventional QED interaction vertex through the covariant derivative. The coefficients $\left(\widehat{k}_{AF}\right)^{\kappa}$ and  $\left(\widehat{k}_{F}\right)^{\kappa\lambda\mu\nu}$ modify the propagation of the photon.

Electron- and nucleon-sector effects are particularly suitable for study in clock-comparison experiments. The prime reason is because the dominant Lorentz-violating corrections to atomic energy levels stem from free-fermion propagation as opposed to interaction terms, the latter of which involve the photon field and potentially powers of the fine structure constant $\alpha \ll 1$. For example, CMB polarisation measurements have placed especially stringent constraints $|\left(k_{AF}\right)^{\kappa}| \lesssim 10^{-43}$\;GeV on photon-sector coefficients, far exceeding the resolution provided by current clock-comparison tests. For these reasons, photon-sector coefficients are often set to zero for clock-comparison tests. However, not all effects can be naively ignored --- the dimensionless coefficients $\left(k_{F}\right)^{\kappa\lambda\mu\nu}$ demonstrate this point, since they mix with the fermion-sector $c$-type coefficients. Under a change of coordinates $x^\mu \rightarrow x^{\mu'} = x^\mu + \tfrac{1}{2}\left(k_F\right)^{\alpha \mu}_{\hphantom{\alpha\mu}\alpha \nu}x^\nu$, the coefficients $\left(k_{F}\right)^{\kappa\lambda\mu\nu}$ shift into the fermion-sector yielding an effective $c$-type coefficient $c'^{\mu\nu} \equiv c^{\mu\nu} - \tfrac{1}{2}(k_F)^{\alpha\mu\hphantom{\alpha}\nu}_{\hphantom{\alpha\mu}\alpha}$ \cite{Kostelecky:2010ze,Bailey:2004na}. Therefore, reported bounds on the $c$-type coefficients are technically bounds on the written combination of dimensionless fermion and photon coefficients.

Restricting attention to minimal effects, the resulting perturbed Hamiltonian in the nonrelativistic limit may be written as \cite{Kostelecky:1999mr}
\begin{ceqn}
\begin{align}
\label{lvh}
\delta h_{\text{LV}} &= (a_0 - m c_{00} - m e_0) +\left(-b_j + md_{j0} -
\tfrac{1}{2}m\epsilon_{jkl}g_{kl0} 
+ \frac{1}{2}\epsilon_{jkl}H_{kl}\right)\sigma^j 
\nonumber \\
& + \left[-a_j + m(c_{0j} + c_{j0}) + me_j \right]\frac{p_j}{m} \nonumber\\
& + \left[ b_0 \delta_{jk} - m(d_{kj}+ d_{00}\delta_{jk}) 
-m\epsilon_{klm}\left(\frac{1}{2}g_{mlj}+g_{m00}\delta_{jl}\right) 
-\epsilon_{jkl}H_{l0}\right]\frac{p_j}{m}\sigma^k \nonumber \\
& + \left[ m\left(-c_{jk} 
- \frac{1}{2}c_{00}\delta_{jk}\right)\right]\frac{p_jp_k}{m^2} \nonumber \\
& + \bigg\{\left[ m(d_{0j} + d_{j0}) 
- \frac{1}{2}\left(b_j + m d_{j0} + \frac{1}{2}m\epsilon_{jmn}g_{mn0} 
+ \frac{1}{2}\epsilon_{jmn}H_{mn}\right)\right]\delta_{kl} 
\nonumber \\
&+ \frac{1}{2}\left(b_l + \frac{1}{2}m \epsilon_{lmn}g_{mn0}\right)\delta_{jk} - m\epsilon_{jlm}(g_{m0k} + g_{mk0})\bigg\}\frac{p_j p_k}{m^2}\sigma^l \,,
\end{align}
\end{ceqn}
where $p_j$ is the three-momentum of the fermion, $\epsilon_{jkl}$ is
the three-dimensional Levi-Civita tensor, and the Pauli matrices satisfy the
algebra $[\sigma^j, \sigma^k] = 2i\epsilon_{jkl}\sigma^l$. This Hamiltonian has been the dominant quantity of study in the context of clock-comparison tests of Lorentz and CPT violation to date.

\subsection{The Sun-centred frame}
\label{ssec:fundsymmsSCF}
The SME coefficients are fixed by the choice of observer
frame. The energy shift~\eqref{deltaE} therefore
includes laboratory-frame coefficients.
Since the laboratory is typically a noninertial
frame rotating with the Earth,
nonzero coefficients as
viewed from the laboratory will oscillate
at harmonics of the Earth's sidereal frequency
$\omega_\oplus \simeq 2\pi/(23\;\text{h}\; 56\;\text{min})$.
For comparison between experiments, it is
useful to introduce an approximately inertial and
nonrotating frame where the SME coefficients
may be taken as constants to excellent approximation. It has become
convention to report constraints on the coefficients as they would
appear in the Sun-centred celestial-equatorial frame (SCF) \cite{Bluhm:2001rw,Kostelecky:2002hh,Bluhm:2003un}.
The coordinates of this frame are labeled by capital Cartesian indices $T$
and $J = X, Y, Z$.
The boundary condition $T=0$ is chosen as the year 2000 vernal equinox,
the $Z$ axis is aligned with the Earth's rotation axis,
the $X$ axis is defined to point from the Earth to the Sun at $T=0$,
and the $Y$ axis completes a right-handed coordinate system.
It is also useful to introduce the standard laboratory frame
with coordinates $x^j$ such that $x$ points to local south,
$y$ towards local east, and $z$ towards the local zenith.
The Sun-centred and standard laboratory frames are approximately related
by a rotation $x^j = \mathcal{R}^{j}_{\hphantom{j}J}X^J$ depending on
the laboratory colatitude $\chi$ where
$\widehat{z}\cdot \widehat{Z} = \cos\chi$
and the local sidereal angle $\omega_\oplus T_\oplus$~\cite{Kostelecky:2002hh},
\begin{ceqn}
\begin{align}
\label{scfrot}
\mathcal{R} =  \begin{pmatrix}\cos\chi\cos\omega_{\oplus} T_{\oplus} & \cos\chi\sin\omega_{\oplus} T_{\oplus} & -\sin\chi \\ -\sin\omega_{\oplus} T_{\oplus} & \cos\omega_{\oplus} T_{\oplus} & 0 \\ \sin\chi\cos\omega_{\oplus} T_{\oplus} & \sin\chi\sin\omega_{\oplus} T_{\oplus} & \cos\chi\end{pmatrix}.
\end{align}
\end{ceqn}
Note that $T_\oplus$ and $T$ are related by a linear shift depending
on the laboratory longitude \cite{Kostelecky:2016kkn}. After
performing the rotation~\eqref{scfrot} the coefficients
in the laboratory are linear combinations of the fixed SCF
coefficients and the laboratory colatitude and local sidereal angle. Furthermore, if the laboratory apparatus frame differs from the standard laboratory frame coordinates $x^j$, additional transformations must be performed.
This implies that every experiment is sensitive to a unique
combination of SME coefficients.
Comparing the resulting theoretical frequency shifts with
those extracted from experiment leads to constraints on SME coefficients in the SCF.

\end{document}